\documentclass[aps,prd,superscriptaddress,nofootinbib,11pt]{revtex4}
\usepackage[english]{babel}
\usepackage[utf8]{inputenc}
\usepackage{graphicx}   
\usepackage{slashed}
\usepackage{epstopdf}
\usepackage{verbatim}   
\usepackage{color}      
\usepackage{subfigure}  
\usepackage{multirow}
\usepackage{hyperref}   
\usepackage{float}
\usepackage{epsfig,rotating}
\usepackage{amsmath,amssymb}
\usepackage{dsfont}
\restylefloat{table}
\numberwithin{equation}{section}

\newcommand{\vx}{\vec{x}}
\newcommand{\vp}{\vec{p}}
\newcommand{\vq}{\vec{q}}
\newcommand{\vk}{\vec{k}}

\newcommand{\be}{\begin{equation}}
\newcommand{\ee}{\end{equation}}
\newcommand{\bea}{\begin{eqnarray}}
\newcommand{\eea}{\end{eqnarray}}

\newcommand{\ket}[1]{|#1\rangle}
\newcommand{\bra}[1]{\langle#1|}
\newcommand{\order}[1]{\mathcal{O}(#1)}

\begin{document}
\title{Cosmological decay of Higgs-like scalars into a fermion channel.}
\author{Daniel Boyanovsky}
\email{boyan@pitt.edu} \affiliation{Department of Physics and
Astronomy, University of Pittsburgh, Pittsburgh, PA 15260}
\author{Nathan Herring}
\email{nmh48@pitt.edu} \affiliation{Department of Physics and
Astronomy, University of Pittsburgh, Pittsburgh, PA 15260}

\date{\today}

\begin{abstract}
We study the   decay  of a Higgs-like scalar Yukawa coupled to massless fermions in post-inflationary cosmology, combining a non-perturbative method   with an adiabatic expansion. A dynamical renormalization is introduced to describe the formation of a renormalized (quasiparticle) state. The renormalized survival probability $\mathcal{P}_\Phi(t)$  is ultraviolet finite, independent of the cutoff and decays on much longer time scales.   During radiation domination, for a (quasi) particle ``born'' at time $t_b$ and decaying at rest in the comoving frame,   $\mathcal{P}_\Phi(t)  = \Big[\frac{t}{t_b}\Big]^{-\frac{Y^2}{8\pi^2}}~
e^{ \frac{Y^2}{4\pi^2}\,\big(t/t_b\big)^{1/4} }
\,e^{-\Gamma_0\,(t-t_b)}~  \mathcal{P}_\Phi(t_b) $, with  $\Gamma_0$ the decay rate at rest in Minkowski space-time. The power   with ``anomalous dimension'' and stretched exponential are remnants of the   formation of the quasiparticle and consequence of the cosmological redshift. For an  ultrarelativistic particle  we find $\mathcal{P}_\Phi(t)  =  e^{-\frac{2}{3}\Gamma_0\,t_{nr}\,(t/t_{nr})^{3/2}}~  \mathcal{P}_\Phi(t_b)$ before it becomes non-relativistic at a time $t_{nr}$ as a consequence of the cosmological redshift. For $t\gg t_{nr}$ we find  $\mathcal{P}_\Phi(t)  = \Big[\frac{t}{t_{nr}}\Big]^{-\frac{Y^2}{8\pi^2}}~
e^{ \frac{Y^2}{4\pi^2}\,\big(t/t_{nr}\big)^{1/4} }~\Big[\frac{t}{t_{nr}}\Big]^{\Gamma_0 t_{nr}/2}
\,e^{-\Gamma_0\,(t-t_{nr})}~  \mathcal{P}_\Phi(t_{nr})$. The extra power is a consequence of the memory   on the past history of the decay process.  We compare these results to an S-matrix inspired phenomenological Minkowski-like decay law modified by an instantaneous Lorentz factor to account for cosmological redshift. Such phenomenological description \emph{under estimates the lifetime of the particle}.  For very long lived, very weakly coupled particles, we   obtain an \emph{upper bound} for the survival probability as a function of redshift $z$ valid throughout the expansion history $\mathcal{P}_\Phi(z) \gtrsim  e^{-\frac{\Gamma_0}{H_0}\,\Upsilon(z,z_b)}\,\mathcal{P}_\Phi(z_b)$, where $\Upsilon(z,z_b)$ only depends on cosmological parameters and $t_{nr}$.

\end{abstract}


\maketitle

\section{Introduction}

The decay and scattering of particles are some of the most fundamental processes in particle physics, within and beyond the Standard Model, with profound impact in cosmology. These processes are ultimately responsible for establishing a state of local thermodynamic and chemical equilibrium and are fundamental ingredients in kinetic processes in the early universe\cite{kolb,linde,bernstein}. Particle decay is not only ubiquitous, but it plays an important role in big bang nucleosynthesis (BBN)\cite{kolb,steigman,sarkar,fields,serpicobbn,pospelovbbn,jeda}, and the generation of the baryon and lepton asymmetries\cite{trodden,kolba,lepgen,lepgen2}.  The  decay of long-lived dark matter particles is constrained by various cosmological and astrophysical probes\cite{zentner,dm1,dm2,dm3,dm4}, and recently it has been suggested that the two body decay of a  long lived dark matter particle  may relieve the tension between distance ladder and cosmic microwave background measurements of the Hubble constant\cite{loeb}.

 Most treatments of particle decay (and/or inverse decay) in cosmology implement  the S-matrix quantum field theory approach as in Minkowski space-time. In this framework, the unstable decaying state is prepared at a time far in the past ($t \rightarrow -\infty$), and one obtains the  transition amplitude to a given final state far in the future ($t\rightarrow \infty$). Taking the infinite time limit in the transition amplitude yields a total energy conserving delta function. Squaring this delta function to obtain the transition probability yields a total energy conservation delta function multiplied by the total elapsed time. Dividing by this large time and summing over all the final states for  a given decay channel gives the total transition probability per unit time, namely a decay rate. Energy conservation, \emph{a consequence of the infinite time limit}, yields kinematic constraints (thresholds) for decay and scattering processes.

In an expanding cosmology such an approach is at best approximate and at worst unreliable when the Hubble expansion rate is large  even during a post-inflationary early stage  of a radiation dominated cosmology, or if the lifetime of the particle is of the order of the Hubble time. In a spatially flat Friedmann-Robertson-Walker (FRW) cosmology there are three space-like Killing vectors associated with spatial translational invariance and spatial momentum conservation, however, as a consequence of cosmological expansion  there is no global time-like Killing vector, therefore particle energy is not manifestly conserved in scattering or decay processes.

A consistent formulation of dynamic processes in an expanding cosmology requires implementing methods of quantum field theory in curved space time\cite{parker,zelstaro,fulling,birford,bunch,birrell,fullbook,mukhabook,parkerbook}. Early studies revealed a wealth of novel phenomena such as particle production\cite{parker,birford,bunch} and processes that are forbidden in Minkowski space time as consequence of strict energy conservation.

S-matrix theory was extended to simple cosmological space times to study the decay of a massive particle into two massless particles conformally coupled to gravity in ref.\cite{spangdecay}. In references\cite{vilja,viljafer} these methods were adapted to calculate the decay of a massive bosonic particle \emph{at rest} into two massless bosonic particles conformally coupled to gravity and into massless fermions Yukawa coupled to a scalar.

More recently\cite{herring} the decay of bosonic particles into two other bosonic degrees of freedom during a radiation dominated era was studied by implementing a non-perturbative method.  This method was  adapted to quantum field theory  from the study of linewidths in quantum optics\cite{boyww,boycosww}, combined with a physically motivated adiabatic expansion. While the results of this reference agreed with those obtained in ref.\cite{vilja} for a particle decaying \textit{at rest} in the comoving frame \emph{in the long time limit}, they revealed new phenomena for highly relativistic decaying particles as a consequence of the cosmological redshift, and the \emph{relaxation of kinematic thresholds} as a consequence of energy uncertainties determined by the Hubble scale.

 Our study  in this article is a natural extension of that in ref.\cite{herring} focusing on decay of a heavy bosonic particle into fermions, a more relevant case for standard model physics (and probably beyond) since most of the fermionic degrees of freedom in the standard model (with the possible exception of neutrinos) are Yukawa coupled to the Higgs boson.

 \vspace{1mm}

 \textbf{Brief summary:}

 The study of fermionic degrees of freedom as decay products introduces several conceptually important distinctions with the bosonic case studied in refs.\cite{vilja,herring} that results in novel aspects of cosmological decay. First, fermionic degrees of freedom couple to the background gravitational field via the spin connection\cite{weinbergbook,parker,duncan,casta,birrell,parkerbook,fercurved1,fercurved2,fercurved3,feradia,feradia2,
feradia3,boydVS,baacke}. Secondly, fermions Yukawa coupled to a bosonic degree of freedom yield a renormalizable theory. Recently the decay of a bosonic particle Yukawa coupled to fermions was studied within a non-perturbative real time framework in Minkowski space-time\cite{zenoqft}. This study revealed novel transient dynamics associated with the \emph{dressing} of the decaying particle by fermion-antifermion pairs into a \emph{quasiparticle} state, which decays on a longer time scale. Such ``dressing'' leads to the necessity of an   ultraviolet divergent   renormalization of the decaying state and a detailed understanding of the various time scales to separate the many-particle dynamics of renormalization and dressing from that of the actual decay of the \emph{quasiparticle}. Such dynamical effects cannot be addressed within an S-matrix framework since these effects are not secular in time and their contribution vanishes when the transition probability is divided by the total time in the infinite time limit. The dynamics of dressing and quasiparticle formation have been recently addressed in ref.\cite{collins} for a consistent interpretation of the reduction formula  in  asymptotic quantum field theory.

We introduce a \emph{dynamical renormalization} that absorbs the ultraviolet divergences associated with fermion pairs into a renormalized survival probability at a renormalization  time scale $t_b$. The survival probability obeys a \emph{dynamical renormalization group equation} with respect to $t_b$.  The cosmological redshift encodes the memory of the transient dynamics of quasiparticle formation in the decay law not seen in Minkowski space-time. If the decaying particle is ultrarelativistic, the decay dynamics depends crucially on $t_{nr}$, the time scale at which it becomes non-relativistic as a consequence of the cosmological redshift.  An S-matrix inspired, phenomenologically motivated, Minkowski-like decay law is shown to  \emph{under estimate} the lifetime of the decaying state. Section (II) introduces the model and the adiabatic approximation, section (III) summarizes the non-perturbative framework to obtain the time evolution of the survival probability. In section (IV) we obtain the decay function for massless fermions during radiation domination, section (V) describes the dynamical renormalization method, section (VI) analyzes the decay dynamics of the renormalized survival probability during radiation domination, compares the results to an S-matrix inspired decay function, and introduces an upper bound  to the decay function for very long lived, very weakly coupled particles  valid all throughout the  expansion history. Section (VII) discusses the various results analyzing their regime of validity and highlighting several implications. Section (VIII) presents our conclusions summarizing the main results. Various appendices contain technical details, in particular appendix (\ref{app:Minko}) derives the decay law in Minkowski space time, highlighting the renormalization aspects to compare to the curved space-time case.

\section{The model:}\label{sec:model}

We consider   a Higgs-like scalar field  Yukawa coupled to one Dirac fermion in a spatially flat Friedmann-Robertson-Walker (FRW) cosmology with scale factor $a(t)$ in comoving time. Generalizing to include Majorana fermions and/or more fermionic species is straightforward.

In comoving
coordinates, the action is given by
\be
S   =  \int d^3x \; dt \;  \sqrt{-g} \,\Bigg\{
\frac{1}{2}{\dot{\phi}^2}-\frac{(\nabla
\phi)^2}{2a^2}-\frac{1}{2} \, \big[M^2 +\xi \,R\big]  \phi^2   +
\overline{\Psi}  \Big[i\,\gamma^\mu \;  \mathcal{D}_\mu -m_f-Y \phi \Big]\Psi     \Bigg\}\,, \label{lagrads}
\ee where
\be R= 6\Big[\frac{\ddot{a}}{a}+\Big(\frac{\dot{a}}{a} \Big)^2 \Big] \,,\label{ricci}\ee is the Ricci scalar,  and $\xi$ is the coupling  to gravity, with $\xi=0, 1/6$ corresponding to minimal or conformal coupling, respectively. Introducing  the vierbein field $e^\mu_a(x)$  defined as
$$
g^{\mu\,\nu}(x) =e^\mu_a (x)\;  e^\nu_b(x) \;  \eta^{a b} \; ,
$$
\noindent where $\eta_{a b}$ is the Minkowski space-time metric,
the curved space time Dirac gamma- matrices $\gamma^\mu(x)$
and the fermionic covariant derivative $\mathcal{D}_\mu$ are given
by\cite{weinbergbook,birrell,duncan,casta}
\be
\gamma^\mu(x) = \gamma^a e^\mu_a(x) \quad , \quad
\{\gamma^\mu(x),\gamma^\nu(x)\}=2 \; g^{\mu \nu}(x)  \; ,
\label{gamamtx}\ee where the $\gamma^a$ are the Minkowski space time Dirac matrices, chosen to be in the standard Dirac representation, and the covariant derivative $\mathcal{D}_\mu$ is given in terms of the spin connection by

\be
\mathcal{D}_\mu   =    \partial_\mu + \frac{1}{8} \;
[\gamma^c,\gamma^d] \;  e^\nu_c  \; \left(\partial_\mu e_{d \nu} -\Gamma^\lambda_{\mu
\nu} \;  e_{d \lambda} \right) \,,  \label{fermicovader}
\ee  where $\Gamma^\lambda_{\mu
\nu}$ are the usual Christoffel symbols.

For an (FRW)
  in conformal time $d\eta= dt/a(t)$,   the metric becomes
\be
g_{\mu\nu}= C^2(\eta) \;  \eta_{\mu\nu}
\quad , \quad C(\eta)\equiv a(t(\eta))\,,\label{gmunu}
\ee

\noindent where  $\eta_{\mu\nu}=\textrm{diag}(1,-1,-1,-1)$ is the flat
Minkowski space-time metric, and the vierbeins $e^\mu_a$ are given by
\be
 e^\mu_a = C^{-1}(\eta)\; \delta^\mu_a ~~;~~ e^a_\mu = C(\eta) \; \delta^a_\mu \,. \label{vierconf}\ee

The fermionic part of the action in conformal coordinates now becomes
\be S_f = \int d^3 x\,d\eta\,C^4(\eta)\,  \overline{\Psi}(\vec{x},\eta)\,\Bigg[ i \frac{\gamma^0}{C(\eta)}\,\Big(\frac{d}{d\eta} + 3 \frac{C^{'}(\eta)}{2C(\eta)}\Big) + i \,\frac{\gamma^i}{C(\eta)}\nabla_i - m_f - Y\,\phi\Bigg]\Psi(\vec{x},\eta)\,. \label{Sf}\ee

The Dirac Lagrangian density in conformal time simplifies to
\be
\sqrt{-g} \; \overline{\Psi}\Big(i \; \gamma^\mu \;  \mathcal{D}_\mu
\Psi -m_f-Y\phi \Big)\Psi  =
\big(C^{3/2}(\eta)\,\overline{\Psi}(\vec{x},\eta)\big) \;  \Big[i \;
{\not\!{\partial}}-(m_f+Y\phi) \; C(\eta) \Big]
\big(C^{3/2}(\eta)\,{\Psi}(\vec{x},\eta)\big)\,, \label{confferscal}
\ee
\noindent where $i {\not\!{\partial}}=\gamma^a \partial_a$ is the usual Dirac
differential operator in Minkowski space-time in terms of flat
space time $\gamma^a$ matrices. Introducing the conformally rescaled fields
\be C(\eta)\, \phi(\vx,t) = \chi(\vx,\eta)~~;~~ C^{\frac{3}{2}}(\eta)\,{\Psi(\vx,t)}= \psi(\vx,\eta)\,, \label{rescaledfields}\ee
and neglecting surface terms, the action becomes
   \be  S    =
  \int d^3x \; d\eta \, \Big\{\mathcal{L}_0[\chi]+\mathcal{L}_0[\psi]+\mathcal{L}_I[\chi,\psi] \Big\} \;, \label{rescalagds}\ee
  with
  \bea \mathcal{L}_0[\chi] & = & \frac12\left[
{\chi'}^2-(\nabla \chi)^2- \mathcal{M}^2 (\eta) \; \chi^2 \right] \,, \label{l0chi}\\
\mathcal{L}_0[\psi] & = & \overline{\psi} \;  \Big[i \;
{\not\!{\partial}}- M^2_f(\eta)   \Big]
 {\psi}  \,,\label{l0psi}\\ \mathcal{L}_I[\chi,\psi] & = & -Y\chi  \overline{\psi}\,\psi \; . \label{lI}\eea    The effective time dependent masses are given by
 \be
  \mathcal{M}^2(\eta) =m^2_\phi \,C^2(\eta)- \frac{C''(\eta)}{C(\eta)}\,(1-6\xi)  \,,  \label{massconformal} \ee  and
\be M_f(\eta) = m_f \,C(\eta) \,.  \label{masfer}\ee

In the non-interacting case, $Y =0$, the Heisenberg equations of motion for the spatial Fourier modes with comoving wavevector $\vec{k}$ for the conformally rescaled scalar field are
\be
  \chi''_{\vk}(\eta)+
\Big[k^2+\mathcal{M}^2(\eta)
\Big]\chi_{\vk}(\eta)  =   0\,.      \label{chimodes}\ee

The Heisenberg fields are quantized in a comoving volume $V$,   the real scalar field $\chi$ is  expanded as
\be
\chi(\vx,\eta)   =   \frac{1}{\sqrt{V}}\,\sum_{\vk} \Big[a_{\vk}\,g_k(\eta)\,e^{i\vk\cdot\vx}+ a^\dagger_{\vk}\,g^*_k(\eta)  \,e^{-i\vk\cdot\vx}\Big] \,,\label{chiex}  \ee where the mode functions $g_k(\eta)$ obey
\be  \Big[\frac{d^2}{d\eta^2}+k^2+\mathcal{M}^2(\eta)
\Big]g_k(\eta)  =   0\,.      \label{modebosons}\ee The mode functions are chosen to obey the Wronskian condition
\be g^{\,'}_k(\eta)g^*_k(\eta) - g^{*\,'}_k(\eta) g_k(\eta) = -i\,,  \label{wron}\ee
and  $a,a^\dagger$ obey the usual canonical commutation relations.

For Dirac fermions the field $ \psi({\vec x},\eta) $ is expanded  as
\be
\psi (\vec{x},\eta) =    \frac{1}{\sqrt{V}}
\sum_{\vec{k},\lambda=1,2}\,   \left[b_{\vec{k},\lambda}\, U_{\lambda}(\vec{k},\eta)\,e^{i \vec{k}\cdot
\vec{x}}+
d^{\dagger}_{\vec{k},\lambda}\, V_{\lambda}(\vec{k},\eta)\,e^{-i \vec{k}\cdot
\vec{x}}\right] \; ,
\label{psiex}
\ee
where the spinor mode functions $U,V$ obey the  Dirac equations\cite{fercurved1,fercurved2,fercurved3,feradia,feradia2,feradia3,boydVS,baacke}
\bea
\Bigg[i \; \gamma^0 \;  \partial_\eta - \vec{\gamma}\cdot \vec{k}
-M_f(\eta) \Bigg]U_\lambda(\vec{k},\eta) & = & 0 \,,\label{Uspinor} \\
\Bigg[i \; \gamma^0 \;  \partial_\eta + \vec{\gamma}\cdot \vec{k} -M_f(\eta)
\Bigg]V_\lambda(\vec{k},\eta) & = & 0 \,.\label{Vspinor}
\eea

These   equations become simpler by  writing
\bea
U_\lambda(\vec{k},\eta) & = & \Bigg[i \; \gamma^0 \;  \partial_\eta -
\vec{\gamma}\cdot \vec{k} +M_f(\eta)
\Bigg]f_k(\eta)\, \mathcal{U}_\lambda \,,\label{Us}\\
V_\lambda(\vec{k},\eta) & = & \Bigg[i \; \gamma^0 \;  \partial_\eta +
\vec{\gamma}\cdot \vec{k} +M_f( \eta)
\Bigg]h_k(\eta)\,\mathcal{V}_\lambda \,,\label{Vs}
\eea
\noindent with $\mathcal{U}_\lambda;\mathcal{V}_\lambda$ being
constant spinors\cite{boydVS,baacke} obeying
\be
\gamma^0 \; \mathcal{U}_\lambda  =  \mathcal{U}_\lambda
  \qquad , \qquad
\gamma^0 \;  \mathcal{V}_\lambda  =  -\mathcal{V}_\lambda\,. \label{Up}
\ee
Inserting (\ref{Us},\ref{Vs}) into the Dirac equations (\ref{Uspinor},\ref{Vspinor}) and using (\ref{Up}), it follows that the mode functions $f_k(\eta);h_k(\eta)$ obey the
equations
\bea \left[\frac{d^2}{d\eta^2} +
k^2+M^2_f(\eta)-i \; M'_f(\eta)\right]f_k(\eta) & = & 0 \,, \label{modefermionf}\\
\left[\frac{d^2}{d\eta^2} + k^2+M^2_f(\eta)+i \; M'_f(\eta)\right]h_k(\eta)
& = & 0 \,.\label{modefermionh}
\eea

Multiplying the Dirac equations on the left by $\gamma^0$, it is straightforward to confirm that
\be \frac{d}{d\eta} (U^\dagger_\lambda(q,\eta)\,U_\lambda(q,\eta)) =0 ~~;~~ \frac{d}{d\eta} (V^\dagger_\lambda(q,\eta)\,V_\lambda(q,\eta)) =0 \,.\label{constnorm}\ee We choose the normalizations
\be U^\dagger_\lambda(q,\eta)\,U_{\lambda'}(q,\eta) = V^\dagger_\lambda(q,\eta)\,V_{\lambda'}(q,\eta) = \delta_{\lambda,\lambda'}\,, \label{normas}\ee so that the operators $b,b^\dagger, d,d^\dagger$ obey the canonical anticommutation relations. Furthermore, we will choose particle-antiparticle boundary conditions so that $h_k(\eta) = f^*_k(\eta)$ (see below). We note that for $m_f =0$ the conformally rescaled fermi fields obey the same equations as in Minkowski space-time but in terms of conformal time, whereas this only occurs for bosons if they are conformally coupled to gravity, namely with $\xi = 1/6$, or for a radiation dominated cosmology (see below). The equivalence of massless fermions to those in Minkowski space-time will allow a direct comparison with the case of decay in flat space time studied in ref.\cite{zenoqft} and summarized in appendix (\ref{app:Minko}),  and to interpret the differences with the curved space-time case.

\subsection{Adiabatic approximation in post-inflationary cosmology: }

   The standard (post-inflation) cosmology is  described by  radiation (RD), matter (MD) and dark energy (DE) dominated stages, we take the latter to be described by a cosmological constant. Friedmann's equation in comoving time is
\be \Big(\frac{ \dot{a}}{a}\Big)^2 = H^2(t) = H^2_0\,\Bigg[ \frac{\Omega_M}{a^3(t)}+ \frac{\Omega_R}{a^4(t)}+ \Omega_\Lambda \Bigg] \,,\label{Hubble} \ee
  where    the scale factor is normalized to $a_0=a(t_0)=1$ today. We take as representative the following  values of the parameters \cite{wmap,spergel,planck}:
\be H_0 = 1.5\times 10^{-42}\,\mathrm{GeV}~~;~~ \Omega_M = 0.308~~;~~ \Omega_R = 5\times 10^{-5}~~;~~ \Omega_\Lambda = 0.692 \,.\label{cosmopars}\ee
 Passing  to   conformal time $\eta$ with $d\eta = dt/a(t)$, where the metric is given by (\ref{gmunu}) and $C(\eta) \equiv a(t(\eta))$, it follows that
\be \frac{d\,C(\eta)}{d\eta}   = H_0\,\sqrt{\Omega_M}\,\Big[a_{eq}+C(\eta)+s\,C^4(\eta)\Big]^{1/2}\,,\label{dera}\ee with
\be a_{eq}= \frac{\Omega_R}{\Omega_M} \simeq 1.66\,\times 10^{-4}~~;~~ s = \frac{\Omega_\Lambda}{\Omega_M} \simeq 2.25 \,,\label{rands}\ee $a_{eq}$ is the scale factor at matter-radiation equality.

Hence the different stages of cosmological evolution, namely   (RD),  (MD), and  (DE), are characterized by
\be C(\eta) \ll a_{eq} \Rightarrow \text{RD}~~;~~ a_{eq} \ll C(\eta) \lesssim 0.76 \Rightarrow \text{MD} ~~;~~ C(\eta) > 0.76 \Rightarrow \text{DE} \,. \label{stages}\ee We will begin by  studying  the dynamics of  particle decay during the (RD)     dominated era,  generalizing  afterwards to the case of a very long lived, very weakly coupled particle.    During (RD) and (MD) we find,

\be C(\eta)= H_R\,   \eta \,\Bigg[ 1 +    \frac{H_R  \,  \eta}{4\,a_{eq}} \Bigg] \,, \label{Crdmd}\ee where
\be H_R= H_0\,\sqrt{\Omega_R}\simeq 10^{-44}\,\mathrm{GeV}   \,,  \label{Hs}\ee and   conformal time in terms of the scale factor is given by
\be \eta(C) = \frac{2\, {a_{eq}}}{H_R}\,\Bigg[\sqrt{1 + \frac{C}{a_{eq}}}-1 \Bigg]\,. \label{etaofa}\ee

  During the (RD) stage
 \be C(\eta) \simeq H_R\,\eta \,,\label{RDcofeta}\ee and the relation between conformal and comoving time is given by \be \eta = \Big( \frac{2\,t}{H_R}\Big)^{\frac{1}{2}} \Rightarrow a(t) = \Big[ 2\,t H_R\Big]^{\frac{1}{2}}\,, \label{etaoft} \ee a result that will prove useful in the study of the decay law during this stage.

\textbf{Bosonic fields:}

Solving the mode equations (\ref{modebosons},\ref{modefermionf},\ref{modefermionh}) with the  cosmological scale factor (\ref{Crdmd}) is obviously very challenging, instead we implement a physically motivated adiabatic expansion. To highlight the nature of the expansion let us consider first the bosonic mode equation (\ref{modebosons}). The term proportional to $C''/C$ in (\ref{modebosons}) vanishes identically in a radiation dominated cosmology or for conformally coupled bosonic fields for which $\xi = 1/6$. We argue below that   we can consistently neglect this term to leading order in the adiabatic expansion all throughout the cosmological evolution during (RD) and (MD) (see eqn. (\ref{addot})). Neglecting this term, the mode equation  (\ref{modebosons}) becomes
\be \Big[\frac{d^2}{d\eta^2} + \omega^2_k(\eta)\Big]g_k(\eta) = 0 ~~;~~ \omega^2_k(\eta) = k^2 + m^2_\phi\, C^2(\eta)\,.   \label{modeboson2}\ee    We recognize that
\be \omega_k(\eta) = C(\eta)\, E_k(t)\,, \label{comovingE} \ee where
\be E_k(t) = \sqrt{k^2_p(t)+m^2_\phi} ~~;~~ k_p(t) = k/a(t)\,, \label{comoE} \ee is the local energy measured by a comoving observer, and $k_p(t)$ is the physical wavevector redshifting with the cosmological expansion.

Writing the solution of (\ref{modeboson2}) in the WKB form\cite{birrell,fullbook,mukhabook,parkerbook,birford}
\be g_k (\eta) = \frac{e^{-i\,\int^{\eta}_{\eta_i}\,W_k(\eta')\,d\eta'}}{\sqrt{2\,W_k(\eta)}} \,, \label{WKB}\ee and inserting this ansatz into (\ref{modeboson2}) it follows that $W_k(\eta)$ must be a solution of the equation\cite{birrell}
\be W^2_k(\eta)= \omega^2_k(\eta)- \frac{1}{2}\bigg[\frac{W^{''}_k(\eta)}{W_k(\eta)} - \frac{3}{2}\,\bigg(\frac{W^{'}_k(\eta)}{W_k(\eta)}\bigg)^2 \bigg]\,.  \label{WKBsol} \ee

This equation can be solved in an \emph{adiabatic expansion}
\be W^2_k(\eta)= \omega^2_k(\eta) \,\bigg[1 - \frac{1}{2}\,\frac{\omega^{''}_k(\eta)}{\omega^3_k(\eta)}+
\frac{3}{4}\,\bigg( \frac{\omega^{'}_k(\eta)}{\omega^2_k(\eta)}\bigg)^2 +\cdots  \bigg] \,.\label{adexp}\ee We refer to terms that feature $n$-derivatives of $\omega_k(\eta)$ as of n-th adiabatic order. The nature and reliability of the adiabatic expansion is revealed by considering the term of first adiabatic order
\be \frac{\omega^{'}_k(\eta)}{\omega^2_k(\eta)} = \frac{m^2_{\phi}\, C(\eta) C^{'}(\eta)}{\Big[ k^2 + m^2_{\phi}\,C^2(\eta) \Big]^{3/2}}\,, \label{firstordad}\ee this is most easily recognized in \emph{comoving} time $t$ in terms of the comoving local energy (\ref{comovingE},\ref{comoE})   and the Hubble expansion rate
 \be H(t) = \frac{\dot{a}(t)}{a(t)} = \frac{C^{'}(\eta)}{C^2(\eta)} \,. \label{Hub}\ee     In terms of these variables, the first order adiabatic ratio  (\ref{firstordad}) becomes\cite{herring}
 \be \frac{\omega^{'}_k(\eta)}{\omega^2_k(\eta)} = \frac{H(t)}{\gamma^2_k(t)\,E_k(t)}\,.  \label{adratio1}\ee where
 \be \gamma_k(t) = \frac{E_k(t)}{m_{\phi}} \,, \label{lorefac}\ee is the local Lorentz factor.

 The adiabatic approximation relies on the smallness of the (time dependent)   adiabatic ratio
 \be \frac{H(t)}{E_k(t)} \ll 1 \,, \label{adratio2}\ee    corresponding  to the physical wavelength $\propto 1/k_p(t)$ and/or the Compton wavelength   of the particle $  1/m_\phi$ being \emph{much smaller} than the size of the particle horizon $d_H(t) \propto 1/H(t)$  at a given time. During (RD)   the particle horizon grows as $a^2(t)$ and during  (MD) it grows as $a^{3/2}(t)$ whereas the physical wavelength grows as $a(t)$. Therefore, if at a given initial time the adiabatic approximation is valid and  $H(t)\ll E_k(t)$   the reliability of the adiabatic expansion \emph{improves} with the cosmological expansion.

 To understand the origin of this approximation consider that the decaying   particle is produced in the (RD) stage during which
 \be H(t) \simeq 1.66 \sqrt{g_\text{eff}}~ \frac{T^2(t)}{M_\text{Pl}}  \,,\label{Hrad}\ee where   $g_\text{eff} \lesssim  100$ is the number of ultrarelativistic degrees of freedom. Therefore,
 \be \frac{H(t)}{E_k(t)} \lesssim  \Bigg[\frac{T(t)}{E_k(t)}\Bigg]\, \Bigg[\frac{T(t)}{\mathrm{GeV}}\Bigg] \times 10^{-18}\,.  \label{HoverErd} \ee

 An \emph{upper bound} on this ratio is obtained by  considering  that the decaying particle is produced at the   scale of grand unification  with $T \simeq 10^{15}\,\mathrm{GeV}$, assuming that this scale describes the onset of the (RD) era. Taking a typical comoving energy   $E_k(t) \simeq T(t) $,   one finds that $H(t)/E_k(t) \lesssim 10^{-3}$ and diminishes with cosmological expansion and diminishing  temperature. This argument suggests that for typical particle physics processes  the adiabatic ratio $H(t)/E_k(t) \ll 1$   throughout the post-inflation thermal history.

In terms of this  adiabatic ratio, we find
\be
\frac{\omega^{''}_k(\eta)}{\omega_k^3(\eta)}  = \frac{1}{\gamma^2_k(t)}\Big(\frac{R(t)}{6 E^2_k(t)}+\frac{H^2(t)}{E_k^2(t)}\Big)-\frac{H^2(t)}{\gamma^4_k(t)E_k^2(t)}\,, \label{secordad}
\ee
where $R(t)$ is the Ricci scalar (\ref{ricci}). Furthermore, it is straightforward to find that
\be \frac{C''}{C\,\omega^2_k} = 2 \Big( \frac{\dot{H}}{2\,E^2_k} + \frac{H^2}{E^2_k} \Big) = \alpha\, \,\frac{H^2}{E^2_k} ~~;~~ \alpha\simeq 0 ~~~ (RD)~;~ \alpha \simeq \frac{1}{2}~~~(MD)\,, \label{addot} \ee therefore,  this ratio is of second adiabatic order and can be safely neglected to the leading adiabatic order pursued in this study, justifying the simplification of the mode equations to (\ref{modeboson2}) even for non-conformal coupling to gravity.

  In this study  we consider the zeroth-adiabatic order with the mode functions given by
\be g_k (\eta) = \frac{e^{-i\,\int^{\eta}_{\eta_i}\,\omega_k(\eta')\,d\eta'}}{\sqrt{2\,\omega_k(\eta)}} \,. \label{WKB0ord}\ee  Since the decay function is $\propto Y^2$, keeping the    zeroth adiabatic order yields the \emph{leading contribution} to the   decay law. Furthermore, as shown in detail in ref.\cite{herring} particle production as a consequence of cosmological expansion is an effect of  higher order in the adiabatic expansion, thus it can be safely neglected to leading order.

  The phase of the mode function has an immediate interpretation in terms of comoving time and the local comoving energy (\ref{comovingE},\ref{comoE}), namely
\be e^{-i\,\int^{\eta}_{\eta_i}\,\omega_k(\eta')\,d\eta'} = e^{-i\,\int^{t}_{t_i}\,E_k(t')\,dt'}\,,\label{phase}\ee which is a natural  generalization of the phase of \emph{positive frequency particle states in Minkowski space-time}.

During the (RD) era with $C(\eta)$ given by (\ref{RDcofeta}) we find that the criterion (\ref{adratio2})  for the validity  of the adiabatic approximation implies
\be \omega_k(\eta)\, \eta = \frac{E_k(t)}{H(t)} \gg 1 \,. \label{adprod}\ee

\textbf{Fermi fields:}
The adiabatic expansion is straightforwardly applied to the fermionic case and has been discussed in the literature\cite{fercurved2,fercurved3,feradia,feradia2,feradia3}. Beginning with the mode equations (\ref{modefermionf},\ref{modefermionh}) with $M'_f(\eta) = m_f \,C'(\eta)$ and, now with $\omega^2_k(\eta) = k^2 + M^2_f(\eta)$, it follows that
\be \frac{M'_f(\eta)}{\omega^2_k(\eta)}= \frac{H(t)}{\gamma_k(t)\,E_k(t)}\,, \label{mprimf}\ee therefore the purely imaginary term in these mode equations are of first adiabatic order and will be neglected to leading (zeroth) adiabatic order.   Hence, to leading order we find
\be f_k(\eta) = h^*_k(\eta) = \frac{e^{-i\,\int^{\eta}_{\eta_i}\,\omega_k(\eta')\,d\eta'}}{\sqrt{2\,\omega_k(\eta)}}\,.  \label{wkbfer} \ee In what follows we will refer to $\omega^2_k(\eta) = k^2 + M^2(\eta)$ for both bosonic and fermionic degrees of freedom with $M^2(\eta) = m^2 C^2(\eta)$ for either case. To leading (zeroth) order in the adiabatic expansion   the Dirac spinor solutions  in the standard Dirac representation and with the normalization conditions (\ref{normas})  are found to be
\be U_\lambda(\vk,\eta) =  \frac{e^{-i\,\int^{\eta}_{\eta_i}\,\omega_k(\eta')\,d\eta'}}{\sqrt{2\,\omega_k(\eta)}}\, \left(
                            \begin{array}{c}
                              \sqrt{\omega_k(\eta)+M_f(\eta)}\, \chi_\lambda \\
                              \frac{\vec{\sigma}\cdot \vec{k}}{\sqrt{\omega_k(\eta)+M_f(\eta)}}\, \chi_\lambda \\
                            \end{array}
                          \right) ~~;~~ \chi_1 = \left(
                                                   \begin{array}{c}
                                                     1 \\
                                                     0 \\
                                                   \end{array}
                                                 \right) \;; \; \chi_2 = \left(
                                                                       \begin{array}{c}
                                                                         0 \\
                                                                         1 \\
                                                                       \end{array}
                                                                     \right) \,,
 \label{Uspinorsol} \ee  and

 \be V_\lambda(\vk,\eta) =  \frac{e^{i\,\int^{\eta}_{\eta_i}\,\omega_k(\eta')\,d\eta'}}{\sqrt{2\,\omega_k(\eta)}}\,\, \left(
                            \begin{array}{c}
                               \frac{\vec{\sigma}\cdot \vec{k}}{\sqrt{\omega_k(\eta)+M_f(\eta)}}\, \varphi_\lambda \\
                               \sqrt{\omega_k(\eta)+M_f(\eta)}\, \varphi_\lambda  \\
                            \end{array}
                          \right) ~~;~~ \varphi_1 = \left(
                                                   \begin{array}{c}
                                                    0 \\
                                                     1 \\
                                                   \end{array}
                                                 \right) \;; \; \varphi_2 = -\left(
                                                                       \begin{array}{c}
                                                                         1 \\
                                                                         0 \\
                                                                       \end{array}
                                                                     \right) \,.
 \label{Vspinorsol} \ee

To leading adiabatic order these spinors satisfy the completeness relations
\bea \sum_{\lambda=1,2} U_{\lambda,a}(\vk,\eta)\,\overline{U}_{\lambda,b}(\vk,\eta') &  =  & \frac{e^{-i\int^\eta_{\eta'} \omega_k(\eta_1) \,d\eta_1}}{2\sqrt{\omega_k(\eta)\omega_k(\eta')}}\,\,
\Lambda^{+}_{\vk,ab}(\eta,\eta')   \nonumber \\  \sum_{\lambda=1,2} V_{\lambda,a}(\vk,\eta')\,\overline{V}_{\lambda,b}(\vk,\eta)   &  = &  \frac{e^{-i\int^\eta_{\eta'} \omega_k(\eta_1) \,d\eta_1}}{2\sqrt{\omega_k(\eta)\omega_k(\eta')}}\,\,
\Lambda^{-}_{\vk,ab}(\eta',\eta)  \,,\label{relaspinors}\eea where the projector operators at different times $\Lambda^{+}_k(\eta,\eta')~;~\Lambda^{-}_k(\eta',\eta)$ and their properties are given in appendix (\ref{app:pros}).

\section{Non-perturbative approach to the decay law:}

In Minkowski space-time, the decay rate of a particle is typically computed via S-matrix theory by obtaining the transition probability per unit time from an in-state prepared in the infinite past to an out-state in the infinite future. Obviously, such an approach -- taking the infinite time limit--  is not suitable in a time dependent gravitational background. An alternative approach in Minkowski space-time   considers the Dyson-resummed propagator in frequency space that includes radiative corrections through the self-energy. The imaginary part of the self-energy evaluated on the mass shell in frequency space is identified with the decay rate, and a Breit-Wigner approximation to the full propagator, namely approximating the self-energy near the  (complex) pole   yields the exponential decay law. Again such an approach is not available in an expanding cosmological background where the lack of a time-like Killing vector prevents Fourier transforms in time-frequency, and makes the self-energy     explicitly dependent on two time arguments, not only on the difference.

Instead we implement a   quantum field theory method that complements and extends the Wigner-Weisskopf theory of atomic linewidths, that is particularly suited to study time evolution in time dependent situations. This method is manifestly unitary and yields a non-perturbative description of transition amplitudes and probabilities directly in real time. We summarize below the main aspects of the method as it applies to this study, referring the reader to\cite{herring,boyww,boycosww} for details. The total Hamiltonian in conformal time is given by $H_0+H_I$ where $H_0$  is the free field Hamiltonian and
 \be H_I(\eta) = Y \int d^3x \, \chi(\vx,\eta) \,\overline{\psi}(\vx,\eta)\,\psi(\vx,\eta)\label{HI} \ee is the interaction Hamiltonian in the interaction picture. Passing to the interaction picture wherein a given state is expanded in the Fock states   associated with the creation and annihilation operators $a,a^\dagger, b,d$, etc. of the free theory, namely $\ket{\Phi(\eta)}_I = \sum_n \mathcal{C}_n(\eta)\ket{n}$, the amplitudes obey the coupled equations
\begin{equation}
i\frac{d}{d\eta} \mathcal{C}_n(\eta) = \sum_m \mathcal{C}_m(\eta)\bra{n}H_I(\eta)\ket{m}\,. \label{coupeqns}
\end{equation}

This is an infinite hierarchy of integro-differential equations for the coefficients $\mathcal{C}_n(\eta)$. Consider that initially the state is $\ket{\Phi}$ so that $\mathcal{C}_\Phi(\eta_i) =1\,;\,\mathcal{C}_{\kappa}(\eta_i) =0$ for $\ket{\kappa}\neq \ket{\Phi}$, and consider a  first order transition  process
$\ket{\Phi}\rightarrow\ket{\kappa}$  to    intermediate multiparticle states $\ket{\kappa}$ with transition matrix elements $\langle \kappa|H_I(\eta)|\Phi\rangle$. Obviously the state $\ket{\kappa}$ will be connected to other multiparticle states $\ket{\kappa'}$  different from $\ket{\Phi}$ via $H_I(\eta)$. Hence for example up to second order in the interaction, the  state $\ket{\Phi}\rightarrow \ket{\kappa}\rightarrow \ket{\kappa'}$. Restricting the hierarchy to \emph{first order transitions} from the initial state $\ket{\Phi} \leftrightarrow \ket{\kappa}$, and neglecting the contribution from vacuum diagrams which just yield a re-definition of the vacuum state\footnote{This is one of the main differences with the method used in references\cite{spangdecay,vilja,viljafer} where a disconnected vacuum diagram is \emph{also} included in the transition amplitude.} (see discussion in ref.\cite{herring})  results in the following  coupled equations
\bea
i\frac{d}{d\eta} \mathcal{C}_\Phi(\eta) &= &  \sum_\kappa \mathcal{C}_\kappa(\eta)\bra{\Phi}H_I(\eta)\ket{\kappa}\label{eqA}\\
i\frac{d}{d\eta} \mathcal{C}_\kappa(\eta) &= & \mathcal{C}_\Phi(\eta)\bra{\kappa}H_I(\eta)\ket{\Phi}~~;~~ \mathcal{C}_\Phi(\eta_i) =1\,;\,\mathcal{C}_{\kappa}(\eta_i) =0 \,. \label{eqK}
\eea

 Equation (\ref{eqK}) with $C_{\kappa}(\eta_i)=0$ is formally solved by
 \be \mathcal{C}_{\kappa}(\eta) = -i\int^{\eta}_{\eta_i}\,\bra{\kappa}H_I(\eta')\ket{\Phi}\,C_{\Phi}(\eta')\,d\eta' \,, \label{kappapop}\ee
 and inserting this  solution into equation (\ref{eqA}) we find
\begin{equation}
\frac{d}{d\eta}\, \mathcal{C}_\Phi(\eta) = -\int_{\eta_i}^\eta d\eta' \,
\Sigma_\Phi(\eta,\eta')~ \mathcal{C}_\Phi(\eta')\,,  \label{diffeqfi}
\end{equation} where we have introduced   the \emph{self-energy}
\be \Sigma_\Phi(\eta;\eta') =
\sum_\kappa \bra{\Phi}H_I(\eta)\ket{\kappa}
\bra{\kappa}H_I(\eta')\ket{\Phi} \,. \label{sigmaPhi}\ee

We study the decay of a single particle bosonic state into a fermion-anti-fermion pair to  leading order in the Yukawa coupling and the adiabatic approximation. Therefore the initial state is a single particle bosonic state with momentum $\vk$, namely $\ket{\Phi} \equiv \ket{1^{\chi}_{\vk}}$. The set of states $\ket{\kappa}$ with a non-vanishing matrix element of $H_I$ with this single particle state are $\ket{\kappa} \equiv \ket{1^f_{\vp,\lambda },\overline{1^f}_{\vq,\lambda'}}$ where $\lambda,\lambda'$ are the polarization of the fermion and antifermion states. The matrix elements entering in the evolution of the amplitudes are
\bea  \bra{1^{\chi}_{\vk}}H_I(\eta)\ket{1^f_{\vp,\lambda },\overline{1^f}_{\vq,\lambda'}}  & =  & \frac{V\,\delta_{\vk,\vp+\vq}}{V^{3/2}}\, \sum_{a} U_{\lambda,a}(\vp,\eta)\,\overline{V}_{\lambda',a}(\vq,\eta)\, g^*_k(\eta) \nonumber \\
\bra{1^f_{\vp,\lambda },\overline{1^f}_{\vq,\lambda'}}H_I(\eta') \ket{1^{\chi}_{\vk}} & = & \frac{V\,\delta_{\vk,\vp+\vq}}{V^{3/2}}\, \sum_{b}  {V}_{\lambda',b}(\vq,\eta')\,\overline{U}_{\lambda,b}(\vp,\eta')\,g_k(\eta')\,, \label{mtxeles}\eea and the self-energy (\ref{sigmaPhi}) to leading order in the adiabatic expansion becomes
\bea &&  \Sigma_{\chi}(k,\eta,\eta')   =   \sum_{\vp,\vq}\sum_{\lambda,\lambda'}\Big[\bra{1^{\chi}_{\vk}}H_I(\eta)\ket{1^f_{\vp,\lambda },\overline{1^f}_{\vq,\lambda'}}\bra{1^f_{\vp,\lambda },\overline{1^f}_{\vq,\lambda'}}H_I(\eta') \ket{1^{\chi}_{\vk}}\Big] = \nonumber \\
& & Y^2\,\frac{e^{i\int^\eta_{\eta'} \omega^\phi_k(\eta_1)d\eta_1} }{2\sqrt{\omega^{\phi}_k(\eta)\omega^{\phi}_k(\eta')}} \int \frac{d^3 p}{(2\pi)^3}\, \frac{e^{-i\int^\eta_{\eta'} (\omega^\psi_p(\eta_1)+\omega^\psi_q(\eta_1))d\eta_1} }{4\sqrt{\omega^{\psi}_p(\eta)\omega^{\psi}_p(\eta')}\, \sqrt{\omega^{\psi}_q(\eta)\omega^{\psi}_q(\eta')}} ~~\mathrm{Tr}\Big[\Lambda^+_{\vp}(\eta,\eta')\Lambda^-_{\vq}(\eta',\eta)\Big]\,, \label{selfie} \eea where $\vq = \vk -\vp$. This is the fermionic one-loop self energy in curved space time to leading order in the adiabatic expansion.

Obviously the differential equation (\ref{diffeqfi}) cannot be solved exactly with the above self-energy. In Minkowski space time the self-energy is a function of the time difference allowing a solution via Laplace transform\cite{boyww,boycosww}. However,  in a time dependent expanding cosmology such an approach is not available. This is a consequence of the lack of a global time-like Killing vector. Instead for weak coupling we resort to a \emph{Markov} approximation\cite{herring}. While details are available in ref.\cite{herring,boycosww,boyww} to which the reader is referred, we summarize here the main aspects of this approximation.

 We begin by introducing

\be
\mathcal{E}_\Phi(\eta,\eta')  \equiv \int_{\eta_i}^{\eta'} \Sigma_\Phi(\eta,\eta'')\,d\eta'' \,, \label{Evar}
\ee
such that
\be \frac{d}{d\eta'}\,\mathcal{E}_\Phi(\eta,\eta')=\Sigma_\Phi(\eta,\eta') \,,  \label{trick}\ee with the condition
\be   \mathcal{E}_\Phi(\eta,\eta_i)=0 \,. \label{condiEA}\ee

Then \eqref{diffeqfi} can be written as
\begin{equation}
\frac{d}{d\eta} \mathcal{C}_\Phi(\eta)  =  -\int_{\eta_i}^\eta d\eta' \,
\frac{d}{d\eta'}\mathcal{E}_\Phi(\eta,\eta')\,  \mathcal{C}_\Phi(\eta')\,,
\end{equation}
which can be integrated by parts to yield
\begin{equation}
\frac{d}{d\eta} \mathcal{C}_\Phi(\eta) = -\mathcal{E}_\Phi(\eta,\eta) C_\Phi(\eta) +\int_{\eta_i}^\eta d\eta' \,
\mathcal{E}_\Phi(\eta,\eta') \frac{d}{d\eta'}\mathcal{C}_\Phi(\eta') \label{eqcfi}\,.
\end{equation}
Since $\mathcal{E}_\Phi \propto \mathcal{O}(Y^2)$ the first term on the right hand side of (\ref{eqcfi})  is of order $Y^2$, whereas the second is $\order{Y^4}$ because $d\,\mathcal{C}_\Phi(\eta)/d\eta  \propto  Y^2$. Therefore  up to $\mathcal{O}(Y^2)$    the evolution equation for the amplitude $\mathcal{C}_\Phi$ becomes
\begin{equation}
\frac{d}{d\eta} \mathcal{C}_\Phi(\eta) = -\mathcal{E}_\Phi(\eta,\eta)\, \mathcal{C}_\Phi(\eta)  \,,\label{LOeq}
\end{equation} with   solution
\begin{equation}
\mathcal{C}_\Phi(\eta) = \exp\Bigl(-\int^\eta_{\eta_i}\mathcal{E}_\Phi(\eta',\eta')\,d\eta'\Bigr)\,\mathcal{C}_\Phi(\eta_i)\,. \label{Cfina}
\end{equation} This expression clearly highlights the non-perturbative nature of the Wigner-Weisskopf approximation. The imaginary part of the self energy $\Sigma_\Phi$ yields a \emph{renormalization} of the adiabatic frequencies and will not be addressed here\cite{boycosww,boyww}, whereas the real part determines the decay law
\be  \mathcal{P}_\Phi(\eta) \equiv |\mathcal{C}_\Phi(\eta)|^2 = e^{-\int^\eta_{\eta_i} \Gamma_\Phi(\eta') d\eta'} \,\mathcal{P}_\Phi(\eta_i) ~~;~~ \Gamma_\Phi(\eta) = 2 \int^\eta_{\eta_i}d\eta_1\,\mathrm{Re}\,[\Sigma_\Phi(\eta,\eta_1)]   \,, \label{probfi}\ee where we introduced the \emph{survival probability} $\mathcal{P}_\Phi(\eta)$ with $\mathcal{P}_\Phi(\eta_i)=|C_\Phi(\eta_i)|^2$. This final expression for the survival probability directly exhibits the non-perturbative nature of the method. The self-energy is given by (\ref{selfie}) to leading order in Yukawa coupling.


In references (\cite{herring},\cite{boyww},\cite{boycosww},\cite{zenoqft}) it has been established that this  non-perturbative framework  correctly describes the short, intermediate and long time dynamics in Minkowski space-time. It provides a real-time non-perturbative resummation of Feynman diagrams to a given order in the perturbative expansion and in Minkowski space-time is equivalent to the time evolution obtained from the inverse Fourier transform of the Dyson resummed propagator in momentum space. For a decaying particle, the propagator has a complex pole in the second (or higher) Riemann sheet, for weak coupling,  in the narrow width approximation  the long time behavior is completely determined by this pole. The Wigner-Weisskopf and Markov approximation yield exactly the same result including wave function renormalization and \emph{also} describes correctly the early time behavior\cite{zenoqft}.

 This equivalence has been discussed in greater detail in refs. (\cite{boyww},\cite{boycosww},\cite{zenoqft}).  The expansion yielding the Markov approximation (\ref{eqcfi}) can be systematically implemented (\cite{boyww},\cite{boycosww},\cite{zenoqft}) as an expansion in time derivatives of the \emph{amplitudes}, which in turn is an expansion in powers of the coupling (squared). This expansion relies on a \emph{separation of time scales}: the typical scale(s)  in the self-energy kernel is the inverse mass of the decaying particle $1/M$, whereas the typical scale of time evolution of the amplitude is $\propto 1/Y^2 M$, which determines the relaxation rate. This separation can be surmised from eqn. (\ref{LOeq}) which is tantamount to taking $\mathcal{C}_\phi$ outside the integral and evaluating it at $\eta'=\eta$. Namely, the amplitude varies very slowly on the time scale of variation of the self-energy.  For vanishing coupling, the amplitudes remain constant, thus vary slowly for weak coupling, as compared to the time variation of the self-energy.

 In the adiabatic approximation in an  expanding cosmology, the time scales in the self energy are completely determined by the adiabatic frequencies as explicitly shown by expression (\ref{selfie}). Therefore, even with expansion the time scales in the self-energy are much shorter than the relaxation time scale $\propto 1/Y^2$ of the decaying state. This separation, even during an expanding cosmology, but in the adiabatic approximation validates, the Markov approximation.


\section{Massless fermions:}\label{sec:masslessfer}

Our goal in this article is to  study  the decay of a heavy Higgs-like scalar field into much lighter fermions, neglecting the fermion masses. This is a suitable scenario for the standard model where the Higgs scalar can decay into all the charged leptons and quarks but for the top,  and the quark and lepton masses may be safely neglected. Such scenario also includes the possibility of decay into neutrinos in the case that  neutrino masses originate in Yukawa couplings to a Higgs-like scalar beyond the standard model. We postpone the study of decay into heavier fermionic degrees of freedom to a companion article.
Focusing on the case of massless fermions   allows a direct comparison with results in Minkowski space time, which are summarized in appendix (\ref{app:Minko}). Furthermore, understanding this simpler case  provides a pathway towards the more general case of massive fermions to be studied elsewhere.

 For massless fermions  $\omega^\psi_k(\eta) = k$, in this case  the projector operators $\Lambda^\pm$ in (\ref{selfie}) are given by eqn.  (\ref{Lambdasmassless}) in appendix (\ref{app:pros}), and the self-energy (\ref{selfie}) can be written in dispersive form as
\be \Sigma_{\chi}(k,\eta,\eta')   = Y^2\,\frac{e^{i\int^\eta_{\eta'} \omega^\phi_k(\eta_1)d\eta_1} }{2\sqrt{\omega^{\phi}_k(\eta)\omega^{\phi}_k(\eta')}}\, \int  \,\rho(k_0,k)\,e^{-ik_0(\eta-\eta')}\,\frac{dk_0}{2\pi} \,, \label{selfidisp}\ee where the spectral density is given by
\be \rho(k_0,k) = 8\pi\, \int \frac{d^3p}{(2\pi)^3}\,\frac{\delta(k_0 -p - |\vk-\vp|)}{4 p\,|\vk-\vp|}\Big[p\,|\vk-\vp|- \vp\cdot(\vk-\vp) \Big]\,, \label{rho} \ee with the result
\be \rho(k_0,k) = \frac{1}{4\pi}\,(k^2_0-k^2)\, \Theta(k_0-k) \,.\label{rhofin} \ee We carry out the $k_0$ integral in (\ref{selfidisp}) by introducing an upper (comoving) ultraviolet cutoff $\Lambda$ \emph{and} a short time convergence factor $\eta-\eta' \rightarrow \eta-\eta'-i\epsilon$ with $\epsilon \rightarrow 0^+$ and replacing   $k^2_0 \rightarrow -d^2/d\eta^{\,'\,2}$  yielding the final result for the self-energy
\be \Sigma_{\chi}(k,\eta,\eta')   = -i\frac{Y^2}{16\pi^2 }\,\frac{e^{i\int^\eta_{\eta'} \omega^\phi_k(\eta_1)d\eta_1} }{ \sqrt{\omega^{\phi}_k(\eta)\omega^{\phi}_k(\eta')}}\,\Bigg[\frac{d^2}{d\eta^{\,'\,2}}+k^2 \Bigg]\,
\Bigg[\frac{e^{-i\Lambda(\eta-\eta'-i\epsilon)}-e^{-ik(\eta-\eta'-i\epsilon)} }{(\eta-\eta'-i\epsilon)} \Bigg]\,.  \label{sigchifin}\ee  In our analysis we will keep $\Lambda$ fixed but large and take the limit $\epsilon \rightarrow 0^+$ first, clearly this is the correct limit when the theory is considered as an effective field theory valid below a cutoff $\Lambda$.  We note that the flat space time limit is obtained by replacing $\eta \rightarrow t$, and the frequency $\omega^\phi_k$ to be time independent (see appendix (\ref{app:Minko})).

It remains to perform the time integrals to obtain  $\Gamma_\Phi(\eta)$ and   $\int^\eta_{\eta_i} \Gamma_\Phi(\eta')\,d\eta'$  given by eqn.  (\ref{probfi}). The total time derivative in (\ref{sigchifin}) is  integrated by parts and  consistently with keeping the leading order in the adiabatic expansion,  terms of the form $\omega'/\omega^2$ are neglected since these yield higher order adiabatic corrections. In the limit $\epsilon \rightarrow 0^+$ for fixed $\Lambda$ we find the \emph{decay function}
 \be \int^\eta_{\eta_i}  \Gamma_\Phi(\eta')\,d\eta' = \frac{Y^2}{8\pi^2} \,I(\Lambda,k,\eta)~~;~~ I(\Lambda,k,\eta)\equiv \Big[I_1(\Lambda,k,\eta)+I_2(\Lambda,k,\eta)+I_3(\Lambda,k,\eta)\Big] \,, \label{intgama}\ee where
\be I_1(\Lambda,k,\eta)= \frac{\Lambda - k}{\omega^\phi_k(\eta_i)}\Bigg\{1 - \sqrt{\frac{\omega^\phi_k(\eta_i)}{ \omega^\phi_k(\eta)}}\, \Bigg[\frac{\sin\Big(\int^\eta_{\eta_i}
\big(\Lambda-\omega^\phi_k(\eta')\big)d\eta' \Big)}{(\Lambda - k)(\eta-\eta_i)} +\frac{\sin\Big(\int^\eta_{\eta_i}
\big(\omega^\phi_k(\eta')-k\big)d\eta' \Big)}{(\Lambda - k)(\eta-\eta_i)}\Bigg]\Bigg\}\,, \label{I1} \ee
\bea I_2(\Lambda,k,\eta) & = &  \int^\eta_{\eta_i}\Bigg[\sqrt{\frac{\omega^\phi_k(\eta')}{\omega^\phi_k(\eta_i)}}+
\sqrt{\frac{\omega^\phi_k(\eta_i)}{\omega^\phi_k(\eta')}} \Bigg]\,\times
\Bigg[\frac{1-\cos\Big(\int^{\eta'}_{\eta_i}
\big(\omega^\phi_k(\eta_1)-\Lambda\big)d\eta_1 \Big)}{\eta'-\eta_i} \nonumber \\ & - &\frac{1-\cos\Big(\int^{\eta'}_{\eta_i}
\big(\omega^\phi_k(\eta_1)-k\big)d\eta_1 \Big)}{\eta'-\eta_i} \Bigg]\,d\eta'  \equiv I_{2a}(\Lambda,k,\eta)+I_{2b}(k,\eta)\,,\label{I2} \eea
\bea I_3(\Lambda,k,\eta) & = &  m^2_\phi \int^\eta_{\eta_i} \frac{1}{\sqrt{\omega^\phi_k(\eta')}}\Bigg\{\,\int^{\eta'}_{\eta_i}\,
\frac{C^2(\eta_1)}{\sqrt{\omega^\phi_k(\eta_1)}}~ \Bigg[\frac{\sin\Big(\int^{\eta'}_{\eta_1}
\big(\Lambda-\omega^\phi_k(\eta_2)\big)d\eta_2 \Big)}{\eta'-\eta_1} \nonumber \\ & + & \frac{\sin\Big(\int^{\eta'}_{\eta_1}
\big(\omega^\phi_k(\eta_2)-k\big)d\eta_2 \Big)}{\eta'-\eta_1} \Bigg] d\eta_1 \Bigg\} \, \,d\eta' \equiv I_{3a}(\Lambda,k,\eta)+I_{3b}(k,\eta)\,. \label{I3} \eea

In obvious notation the contributions $I_{2b}(k,\eta),I_{3b}(k,\eta)$ are the $\Lambda$ independent terms in $I_{2,3}$ respectively. These three contributions are studied separately below, analyzing their cutoff dependent and independent terms   extracting the different physics of each term.

\subsection{Analysis of $I_{1,2,3}$:}
In the following analysis we will take the cutoff $\Lambda$ to be the largest of all scales, in particular $\Lambda \gg \omega_k(\eta)$ at all times.

\textbf{$I_1$:}  $I_1$ vanishes identically as $\eta \rightarrow \eta_i$ and the oscillatory terms become negligibly small for $\Lambda(\eta-\eta_i) \gg 1$, therefore $I_1$ \emph{grows} to its asymptotic value
\be I_1 =  \frac{\Lambda-k}{\omega^\phi_k(\eta_i)} \label{I1asy}\ee
 very rapidly, on a time scale $\eta-\eta_i \simeq 1/\Lambda$. This divergent contribution corresponds to a renormalization of the amplitude and is similar to a linearly divergent renormalization in Minkowski space time\cite{zenoqft} (see appendix (\ref{app:Minko})).

 \textbf{$I_2$:} The technical details of the analysis of $I_2$ are relegated to appendix (\ref{app:I2}). The main result is that for $\Lambda\,(\eta-\eta_i) \gg 1$

 \be I_2(\Lambda,k,\eta) = 2\Big[\ln\big[\Lambda\,(\eta-\eta_i) \big]+ \gamma_E \Big]+I_{2b}(k,\eta) \label{I2asy}\ee where $\gamma_E=0.577\cdots$ is Euler's constant and $I_{2b}(k,\eta)$ is given by equation (\ref{I2b}) in appendix (\ref{app:I2}) where this   contribution is analyzed in detail. We discuss this contribution in further detail in sections (\ref{sec:renor},\ref{sec:dynamics}) below.

\textbf{$I_3$:} With $\Lambda \gg \omega_k$  the argument of the sine function in  the first term in eqn.  (\ref{I3}), namely in $I_{3a}(\Lambda,k,\eta)$,  simplifies to $\Lambda\,(\eta'-\eta_1)$, therefore
\be I_{3a}(\Lambda,k,\eta)= m^2_\phi \, \int^\eta_{\eta_i} \frac{1}{\sqrt{\omega^\phi_k(\eta')}}\,\Bigg\{\,\int^{\eta'}_{\eta_i}\,
\frac{C^2(\eta_1)}{\sqrt{\omega^\phi_k(\eta_1)}}~ \frac{\sin\Big(\Lambda\,(\eta'-\eta_1) \Big)}{\eta'-\eta_1} \,d\eta_1 \Bigg\}\, d\eta' \,.\label{I3a} \ee Defining $ \sigma = \Lambda\,(\eta'-\eta_1)~~;~~ \sigma_f = \Lambda\,(\eta'-\eta_i)$, and  taking the limits $\Lambda \rightarrow \infty~~;~~ \sigma_f \rightarrow \infty$, the integral over $\eta_1$ in eqn.  (\ref{I3a}) becomes
\be  \int^{\infty}_{0}  \frac{C^2(\eta'-\sigma/\Lambda)}{\sqrt{\omega^\phi_k(\eta'-\sigma/\Lambda)}} ~~\frac{\sin\sigma}{\sigma} \,d\sigma ~~~ {}_{\overrightarrow{\Lambda \rightarrow \infty}} ~~~  \frac{\pi}{2} \, \frac{C^2(\eta')}{\sqrt{\omega^\phi_k(\eta')}}   \,, \label{limI3a} \ee therefore in this limit we find
\be I_{3a}(k,\eta) = \frac{\pi}{2} \, m^2_\phi\, \int^\eta_{\eta_i} \frac{C^2(\eta')}{ {\omega^\phi_k(\eta')}}\, d\eta' = \frac{\pi}{2} \, m_\phi \int^t_{t_i} \frac{1}{\gamma_k(t')}\, dt' \label{I3atime}\ee where we used $\omega^\phi_k(\eta) = C(\eta) E^\phi_k(t) =  m_\phi\,C(\eta) \, \gamma_k(t) $ and $C(\eta) d\eta' = dt'$, with $\gamma_k(t)= \sqrt{1+k^2_p(t)/m^2}$ being the   Lorentz factor whose time dependence is a consequence of the cosmological redshift.

In appendix (\ref{app:I3}) we provide the analysis for $I_{3b}$, gathering both terms we find that
\be I_3(k,\eta)=  \frac{\pi}{2} \, m^2_\phi\, \int^\eta_{\eta_i} \frac{C^2(\eta')}{ {\omega^\phi_k(\eta')}}\,\Big[1+\mathcal{S}(k,\eta')\Big] \,d\eta'\,, \label{I3tot}\ee where $\mathcal{S}(k,\eta')$ is given by (\ref{capSnewvars}) with asymptotic limit $\mathcal{S}(k,\eta')\rightarrow 1 $ for large $\eta'$. Therefore $I_3=I_{3a}+I_{3b}$ \emph{does not} depend on $\Lambda$ in the limit $\Lambda\rightarrow \infty$. This is similar to the case in Minkowski space time (see appendix (\ref{app:Minko}) where the equivalent term is called $T_3(k,t)$, eqn. (\ref{T3})).

\section{Renormalization: dynamics of ``dressing''}\label{sec:renor} The final result for the decay function in (\ref{intgama}), $I(\Lambda,k,\eta)$ is given by ,
\be I(\Lambda,k,\eta) = \frac{\Lambda-k}{\omega^\phi_k(\eta_i)}+  2\,\ln\Big[\Lambda\,\eta_i\,e^{\gamma_E} \Big] + I_{fin}(k,\eta)\,, \label{totIfin}\ee where $I_{fin}(k,\eta)$ is independent of the cutoff $\Lambda$ in the limit $\Lambda \rightarrow \infty$, and for $(\eta-\eta_i) \gg 1/\Lambda$ it is given by
\be I_{fin}(k,\eta) = 2\,\ln\Big[\frac{\eta}{\eta_i}-1 \Big]+I_{2b}(k,\eta)+I_3(k,\eta)\,.  \label{Ifinito}\ee The linear and logarithmic dependence on the cutoff $\Lambda$ are \emph{exactly} the same as in Minkowski space time\cite{zenoqft}, as obtained in the appendix (\ref{app:Minko}). This similarity is expected as the cutoff dependence arises from the short distance behavior of the self-energy correction which should be insensitive to the curvature of space time. As discussed in ref.\cite{zenoqft} the origin of this divergence is the ``dressing'' of the bare single particle state by a cloud of fermion-anti-fermion pairs into a renormalized \emph{quasiparticle} state. In a renormalizable theory the growth of the density of states at high energy implies that this cloud of excitations contains high energy states. The dynamical build-up of the cloud of excitations occurs on a time scale $\eta-\eta_i \simeq 1/\Lambda$ at which the divergent contributions to $I_{1,2}$ saturate, see eqn. (\ref{I1})  and the discussion in appendix (\ref{app:I2}).

 The ``dressing'' of the bare into the physical renormalized quasiparticle state is accounted for by the wave-function renormalization of the amplitude\cite{zenoqft}. For large cutoff scale $\Lambda$ and for a weakly coupled theory with $Y^2 \ll 1$  there is a wide separation betweeen the time scales of formation of the dressed renormalized state $\eta-\eta_i \simeq 1/\Lambda$,   the time scale of typical oscillations $\eta -\eta_i \simeq  1/\omega^\phi_k(\eta)$ and finally the decay time scale $\eta-\eta_i \propto 1/Y^2\omega^\phi_k(\eta)$, which for weak coupling is the longest scale.   Therefore, we can evolve the initial state in time up to an intermediate time scale $\eta_b$ with $(\eta_b-\eta_i) \gg  1/\Lambda$, but much smaller than the typical decay time scale $\propto 1/Y^2\omega^{\phi}_k(\eta_i)$, so that the initial state   had enough time to be ``dressed''   by fermion-antifermion pairs into the renormalized quasiparticle state, but did not have time to decay. For example, taking $\eta_b-\eta_i = 1/\omega^{\phi}_k(\eta_i)$  fulfills the conditions of time scale separation because $\omega^\phi_k \ll \Lambda$, and  because  for $Y^2 \ll 1$ there will be many oscillations of the field before it decays. Taking this renormalization scale is tantamount to an ``on-shell'' renormalization scheme. We identify $\eta_b$ as the time of formation -- or ``birth'' -- of the ``dressed'' or quasiparticle state\cite{zenoqft}, which after formation decays on a much longer time scale.

  The time evolution of  the ``bare'' single particle state until it is renormalized or ``dressed''  is implemented by the following procedure. Writing
 \be I(\Lambda,k,\eta) \equiv  I(\Lambda,k,\eta_b)+ I_S(k,\eta,\eta_b)~~;~~ I_S(k,\eta,\eta_b) = I(\Lambda,k,\eta)-I(\Lambda,k,\eta_b)\,, \label{Isubs}\ee
where, taking $(\eta_b-\eta_i) \gg  1/\Lambda$, the subtracted quantity
\be I_S(k,\eta,\eta_b) =  2\,\ln\Big[\frac{\eta-\eta_i}{\eta_b-\eta_i} \Big]+  I_{2b}(k,\eta,\eta_b)  +  I_{3S}(k,\eta,\eta_b)\,,  \label{ISubfin}\ee is independent of $\Lambda$  for  $\eta > \eta_b$ and  $\Lambda (\eta_b-\eta_i) \gg 1$.  The subtracted contributions $I_{2b}(k,\eta,\eta_b)~;~I_{3S}(k,\eta,\eta_b)$ are defined as follows
\be I_{2b}(k,\eta,\eta_b) \equiv I_{2b}(k,\eta)-I_{2b}(k,\eta_b)~~;~~I_{3S}(k,\eta,\eta_b) \equiv I_{3}(k,\eta)-I_{3}(k,\eta_b)\,, \label{subsI23}\ee and are obtained explicitly in appendices (\ref{app:I2},\ref{app:I3}) respectively. During (RD) we find  (see appendix (\ref{app:I2}) for definitions and eqn. (\ref{I2bfin}))
\be I_{2b}(k,\eta,\eta_b) = -\int^{\xi}_{ \xi_b}  \Bigg[\sqrt{W[\xi']}+\frac{1}{\sqrt{W[\xi']}} \Bigg]\, \, \Big[1- \cos[J(\xi')]\Big]  \,\,\frac{d\xi'}{\xi'}  \,, \label{I2bar}\ee with
   \bea \xi & = &  (\eta-\eta_i)/\eta_i~~;~~  \xi_b= (\eta_b-\eta_i)/\eta_i\nonumber \\
  W[\xi] & = &  \frac{1}{\gamma_i}\Bigg[(\gamma^2_i-1)+(1+\xi)^2 \Bigg]^{\frac{1}{2}} ~~;~~ \gamma_i\equiv\gamma(\eta_i)\,, \label{defa}\eea    $J(\xi')$ is given by eqn. (\ref{Jofz}) in appendix (\ref{app:I2}),
 and
\be I_{3S}(k,\eta,\eta_b) = \frac{\pi}{2}\,m_\phi \int^\eta_{\eta_b} \frac{C(\eta')}{\gamma_k(\eta')}\Big[ 1+ \mathcal{S}(\eta')\Big]\, d\eta'\,, \label{I3bar}\ee  where $\mathcal{S}(\eta)$ is given by eqn. (\ref{capSnewvars}) in appendix (\ref{app:I3}).  The contribution from $I(\Lambda,k,\eta_b)$ is absorbed into wave-function renormalization $Z$ as follows. Writing  equation (\ref{probfi}) as
 \be \mathcal{P}_\Phi(\eta)  = e^{-\int^\eta_{\eta_i} \Gamma_\Phi(\eta') d\eta'} \,\mathcal{P}_\Phi(\eta_i) \equiv
 e^{-\int^\eta_{\eta_b} \Gamma_\Phi(\eta') d\eta'} \,\mathcal{P}_{\Phi,r}(\eta_b)  \,, \label{C2ren}\ee where the renormalized probability is given by
 \be \mathcal{P}_{\Phi,r}(\eta_b) = Z(\eta_b)\,\mathcal{P}_{\Phi}(\eta_i) ~~;~~ Z(\eta_b) =  e^{-\int^{\eta_b}_{\eta_i} \Gamma_\Phi(\eta') d\eta'}\,.  \label{Zren}\ee The exponent in the wave function renormalization  $Z(\eta_b)$ is given by
  \be \int^{\eta_b}_{\eta_i} \Gamma_\Phi(\eta') d\eta' = \frac{Y^2}{8\pi^2} \,I(\Lambda,k,\eta_b) \,, \label{logZ}\ee yielding an  ultraviolet divergent wave function renormalization. The \emph{renormalized} probability obeys
  \be \mathcal{P}_{\Phi,r}(\eta)  = e^{-\int^\eta_{\eta_b} \Gamma_\Phi(\eta') d\eta'} \,\mathcal{P}_{\Phi,r}(\eta_b)\,.  \label{renoproba}\ee

  The decay function that describes the time evolution of the \emph{renormalized} survival probability is given by
 \be \int^\eta_{\eta_b} \Gamma_\Phi(\eta') d\eta' = \frac{Y^2}{8\pi^2} \,I_{S}(k,\eta,\eta_b) \,, \label{decayfun}\ee it is \emph{finite} and independent of   $\Lambda$ in the large cutoff limit. The time scale $\eta_b$ acts as a renormalization scale, obviously the survival probability $\mathcal{P}_{\Phi,r}(\eta)$ is \emph{independent} of this renormalization scale, hence it obeys a \emph{dynamical renormalization group} equation, namely
 \be \frac{\partial}{\partial\eta_b}\,\mathcal{P}_{\Phi,r}(\eta) =0 \,. \label{drg}\ee The solution of this equation is, obviously\footnote{Note the similarity with the usual renormalization group function associated with the running of the wave function renormalization that yields anomalous dimensions.},
 \be \mathcal{P}_{\Phi,r}(\eta_A) =    e^{-\int^{\eta_A}_{\eta_B} \Gamma_\Phi(\eta') d\eta'} \,\mathcal{P}_{\Phi,r}(\eta_B)  \,.  \label{drgsol}\ee

   $\mathcal{P}_{\Phi,r}(\eta_b)$ describes the probability of the renormalized \emph{quasiparticle} state. This ``dressed'' state decays with the finite and cutoff independent decay function $\int^\eta_{\eta_b} \Gamma_\Phi(\eta') d\eta'$ on time scales much longer than the ``dressing'' or renormalization scale $\eta_b$.

 In the following analysis we will drop the subscript $r$ from $\mathcal{P}_{\Phi,r}$ to simplify notation since we will be strictly dealing with the \emph{renormalized} survival probability.

The decay function (\ref{decayfun}) depends explicitly on the initial time $\eta_i$ (see explicit expressions in appendix (\ref{app:I2})). However,  $\mathcal{P}_{\Phi,r}(\eta_b) $ is defined at the renormalization scale $\eta_b$ and it is taken to be the initial probability of the fully renormalized  state after all the short time transient dynamics that result in the ``dressing'' of the bare into the renormalized quasiparticle state have subsided. Therefore, the dependence of the contributions (\ref{I2bar},\ref{I3bar}) on $\eta_i$ must be traded for a dependence on $\eta_b$.

 Let us write
 \be \eta_b-\eta_i = \frac{\beta}{\Lambda} \,,\label{beta}\ee with  $\beta \gg 1$ so that the $\Lambda$ dependent terms in $I_{1,2}$ reached their asymptotic behavior. For example, the ``on-shell'' renormalization scheme corresponds to $\beta \equiv \Lambda/\omega_k(\eta_i)$. Therefore, in terms of the Hubble rate and the physical cutoff $\Lambda_{ph}(\eta_i) = \Lambda/C(\eta_i)$ at the initial time $H(\eta_i)$    we find in (RD)
 \be \frac{\eta_b}{\eta_i} = 1 + \beta \,\frac{H(\eta_i)}{\Lambda_{ph}(\eta_i)}   \,. \label{rateta}\ee Since the  cutoff scale $\Lambda$ is taken to be much larger than any of the energy scales and the adiabatic condition requires that $H(\eta)/E_k(\eta)\ll 1$ at all times, it follows that $H(\eta_i)/\Lambda_{ph}(\eta_i) \ll H(\eta)/E_k(\eta) \ll  1$. Furthermore, we find that
 \be \omega_k(\eta_i) = \omega_k(\eta_b)\Big[1-\beta\, \frac{\omega^{'}_k(\eta_b)}{\omega^2_k(\eta_b)}\,
 \frac{\omega_k(\eta_b)}{\Lambda} +\cdots\Big]\,, \label{overome}\ee the second term in the bracket is at most of first adiabatic order, this is the case for the ``on-shell'' renormalization scheme for which $\beta \, \omega_k(\eta_b)/\Lambda=1$. Hence, to leading adiabatic order we can safely replace $\omega_k(\eta_i) \rightarrow \omega_k(\eta_b)$ in the expressions. Using the results of appendix (\ref{app:I2}) we find that similar arguments justify the replacement $\gamma_k(\eta_i) \rightarrow \gamma_k(\eta_b)$ along with $\eta_i \rightarrow \eta_b$  in all the quantities that enter in the decay function. In the limit of large cutoff $\Lambda$ the trade-off between the variables at the initial time and those at the renormalization scale $\eta_b$ \emph{does not} depend on the cutoff as it must be for a consistent effective field theory description well below the cutoff scale. We note that the adiabatic approximation plays an important role in this separation and is a necessary ingredient because the frequencies depend on time unlike in Minkowski space-time. In particular for the ``on-shell'' renormalization scheme
 \be \frac{\eta_b}{\eta_i}-1 = \frac{1}{\omega_k(\eta_i)\,\eta_i} \ll 1 \,, \label{OSscheme}\ee because the adiabatic condition (during (RD)) corresponds to $\omega_k(\eta_i)\,\eta_i \gg 1$ (see equation (\ref{adprod})).

\section{Dynamics of decay.}\label{sec:dynamics}

Once we have absorbed the ultraviolet divergences into a renormalization of the amplitude, we now proceed   to analyze the main physical aspects of the decay dynamics leveraging the adiabatic approximation.

\subsection{Decay during radiation domination:}

We assume that the decaying particle has been produced early during the (RD) stage by some (unspecified) particle physics process at a high energy/temperature scale, focusing  first on the dynamics of decay during this era. The   subtracted decay function $I_S(k,\eta,\eta_b)$ (\ref{ISubfin}) can be written in a
compact manner amenable to a numerical study as
 \be I_S(k,\eta,\eta_b) = I^R_S(k,\eta,\eta_b) + I_{3S}(k,\eta,\eta_b)\,, \label{ISfina}\ee  with
 \be I^R_S(k,\eta,\eta_b) =  2\,\ln\Big[\frac{\xi}{\xi_b} \Big]-\Big(F_1[\xi,\xi_b]-F_2[\xi,\xi_b]\Big)\,,\label{IRS} \ee  and
\be I_{3S}(k,\xi)= \frac{\pi}{2}\, \frac{\omega_i \eta_i}{\gamma_i}\,\int^{\xi}_{\xi_b} \frac{(1+\xi')^2\,\Big[1+\mathcal{S}(\xi') \Big]}{\sqrt{(\gamma^2_i-1)+(1+\xi')^2}} \, \,d\xi'\,, \label{I3ofxi2} \ee where $\xi,W[\xi]$ are defined in eqn. (\ref{defa}), and  we have introduced the following   functions  (see appendices \ref{app:I2},\ref{app:I3})

\bea F_1[\xi,\xi_b] & = & \int^{\xi}_{ \xi_b}  \Bigg[\sqrt{W[\xi']}+\frac{1}{\sqrt{W[\xi']}} \Bigg]\, \, \frac{d\xi'}{\xi'}\,, \label{F1v2} \\
F_2[\xi,\xi_b] & = & \int^{\xi}_{ \xi_b}  \Bigg[\sqrt{W[\xi']}+\frac{1}{\sqrt{W[\xi']}} \Bigg]\, \, \cos[J(\xi')]   \,\,\frac{d\xi'}{\xi'} \,, \label{F2v2} \eea where $J[\xi]$ is defined in eqn. (\ref{Jofz}) in appendix (\ref{app:I2}). To leading adiabatic order (see  appendix (\ref{app:I3}))
  \be \mathcal{S(\xi')} = \frac{2}{\pi} \,Si\big[\alpha(\xi')\big] ~~;~~ \alpha(\xi') = \frac{\omega_i\,\eta_i}{\gamma_i} \,\xi'\,\Bigg[ \sqrt{(\gamma^2_i-1)+(1+\xi')^2}-\sqrt{(\gamma^2_i-1)}\Bigg]\,,  \label{SiI3s}\ee  where  $Si[x]$ is the sine-integral function (see equation (\ref{Szeroad}) and   discussion  in appendix (\ref{app:I3})).

 We highlight that the contribution $I^R_S$  is a distinct feature of the renormalizable Yukawa interaction and  of the fermionic density of states, whereas $I_{3S}$  in (\ref{I3bar}) is very similar to the decay function  found  in the scalar case studied in ref.\cite{herring}.

 As discussed above, to leading adiabatic order we set $\eta_b=\eta_i$   in $I_{3S}$ and  obtain  (see appendix (\ref{app:I3}))
\be I_{3S}(k,\eta,\eta_b) = \frac{\pi}{4}\,  {\omega_i \eta_i} \,\Bigg\{ (1+\xi)\,W[\xi] -1 - \frac{(\gamma^2_i-1)}{\gamma_i}\,\ln \Bigg[\frac{\gamma_i\,W[\xi]+(1+\xi)}{1+\gamma_i} \Bigg]  \Bigg\} +  \widetilde{I}_{3S}(k,\eta,\eta_b)\,, \label{I3Sxivars}\ee
\be \widetilde{I}_{3S}(k,\eta,\eta_b) = \frac{\pi}{2}\, \frac{\omega_i \eta_i}{\gamma_i}\,\int^{\xi}_{\xi_b} \frac{(1+\xi')^2\, \mathcal{S}(\xi') }{\sqrt{(\gamma^2_i-1)+(1+\xi')^2}} \, \,d\xi'\,, \label{wideI3S} \ee where   $\widetilde{I}_{3S}(k,\eta,\eta_b)$ must be obtained numerically.

 However, before we engage in a numerical study we analyze the different contributions   to extract a physical picture of which terms dominate at different time scales. In order to analyze the behavior in the different regimes,  we write the Lorentz factor both in terms of the variable $\xi= \frac{\eta}{\eta_i}-1$     (see appendix (\ref{app:identities})) as well as  in terms of comoving time with the equivalence $1+\xi \equiv \sqrt{t/t_i}$  (see also appendix (\ref{app:identities})),
 \be \gamma(\xi) = \Bigg[ \frac{(\gamma^2_i-1)}{(1+\xi)^2}+1 \Bigg]^{\frac{1}{2}} \equiv  \Bigg[ \frac{(\gamma^2_i-1)}{\Big(\frac{t}{t_i}\Big)}+1 \Bigg]^{\frac{1}{2}} = \Bigg[ \frac{t_{nr}}{t}+1 \Bigg]^{\frac{1}{2}} \equiv \gamma(t) \,, \label{gamaofchi}\ee where $t_{nr}$ is the comoving  time scale at which the decaying particle becomes non-relativistic, given by
 \be  t_{nr} = t_i\,(\gamma^2_i-1) = \frac{k^2}{2 m^2_\phi H_R} \,. \label{tnr}\ee

  Whence the limits
 \bea (\gamma^2_i-1) \ll (1+\xi)^2  &     \Rightarrow &  ~ \mathrm{Non-relativistic} ~~;~~   (\gamma^2_i-1) \gg (1+\xi)^2  \Rightarrow ~ \mathrm{Ultra-relativistic} \nonumber \\ t_{nr} \ll   {t}    & \Rightarrow &  ~ \mathrm{Non-relativistic} ~~;~~   t_{nr} \gg t  \Rightarrow ~ \mathrm{Ultra-relativistic} \,.  \label{relimits}\eea

Let us   focus first on the contribution $I^R_S(k,\eta,\eta_b)$ given by  (\ref{IRS}).   In Minkowski space-time the frequencies are time independent, therefore $W[\xi'] =1$ and $J(\xi') = (\omega_k-k)\eta_i\,\xi'$. The analysis of appendix (\ref{app:Minko}) shows that in Minkowski space time  for $\xi \gg 1$ the second term in (\ref{IRS}), namely  $F_1-F_2$,  yields   $2\ln[\xi/\xi_b] + \mathrm{constant}$, thereby cancelling the logarithmic time dependence of the first term (see appendix (\ref{app:Minko})). \emph{Such cancellation only  occurs during a limited interval in time  in the expanding cosmology as a consequence of the time dependence of the frequencies}. This follows from  the  analysis of appendix (\ref{app:I2}) which shows that there are three distinct stages:

\vspace{1mm}

\textbf{i)  $\xi \lesssim \xi_m$:}   where $\xi_m $ given by (\ref{chimax1},\ref{chimax2})  is the time scale  at which $F_2[\xi,\xi_b]$ reaches a maximum. During this interval
$F_1 - F_2$ in  (\ref{IRS}) is negligible and $I^R_S \simeq 2 \ln[\xi/\xi_b]$.

\vspace{1mm}

\textbf{ii)}  $\xi_m  < \xi \lesssim  \gamma_i$: during this interval   the function  $F_1[\xi,\xi_b]$ continues to rise monotonically whereas $F_2[\xi,\xi_b]$   oscillates around its constant asymptotic value $F_2[\xi,\xi_b]\simeq F_2[\xi_m,\xi_b] \simeq 2 \ln[\xi_m/\xi_b]$, a behavior summarized by   figure (\ref{fig:f12g10}) and    equation (\ref{fif2behaviorur}) in appendix (\ref{app:I2}).

For $\omega_i \eta_i \gg 1$ the results  (\ref{chimax1},\ref{chimax2}) show that  $\xi_m \ll \gamma_i$ for all values of $\gamma_i \geq 1$. Therefore, for $\gamma_i \gg 1$, during the interval $\xi_m \leq \xi  < \gamma_i$   it follows that $W[\xi'] \simeq 1$ and $F_1 \simeq 2\,\ln[\xi/\xi_b]$ thereby (approximately) cancelling the logarithm from the first term in $I^R_S$, whereas $F_2\simeq 2\ln[\xi_m/\xi_b]$ remains constant, yielding a plateau in $I^R_S$. This approximate cancellation  is effective during a time  interval  that increases for $\gamma_i \gg 1$ (see discussion in appendix (\ref{app:I2})). According with eqn. (\ref{gamaofchi}) and the limits (\ref{relimits})  during this interval, wherein $I^R_S$ is approximately constant,   the decaying particle is in the \emph{ultrarelativistic regime}. In this stage the constancy of $I^R_S$ is expected because in the ultrarelativistic regime the frequencies are nearly time independent since $\omega_k(\eta) \simeq k \simeq \omega_i$. Therefore $W[\xi] \simeq 1$ yielding $F_1 \simeq 2\ln[\xi/\xi_b]$ thereby cancelling the logarithmic time dependence of the first term in (\ref{IRS}), similarly to Minkowski space-time.

If $\gamma_i \gg 1$ the decaying particle is ``born'' ultrarelativistically and there is a (long) time window $\xi_m < \xi < \gamma_i$ within which $\sqrt{W[\xi']} \simeq 1$ and $F_1[\xi,\xi_b] \simeq 2\ln[\xi/\xi_b]$    thereby approximately  cancelling the first term in  $I^R_S$ whereas $F_2[\xi,\xi_b]$ remains nearly constant. Therefore   for $\gamma_i \gg 1$   it follows that  $I^R_S(k,\eta,\eta_b)$ rises rapidly on a time scale $\simeq \xi_m$ reaching  a maximum and remaining nearly constant $I^R_S \simeq 2\,\ln[\xi_m/\xi_b]$ until $\xi \simeq \gamma_i$.

\vspace{1mm}

\textbf{iii)  $ \xi \gg \gamma_i$: } The cosmological redshift eventually makes the decaying particle to become non-relativistic  when $\xi \gg \gamma_i \gg 1$.  During this stage the particle is \emph{non-relativistic} as a consequence of the cosmological redshift. The time dependence of the frequency now yields $\sqrt{W[\xi']}+1/\sqrt{W[\xi']} \gg  2$, hence $F_1 > 2\,\ln[\xi]$. In this stage it follows that $W[\xi] \approx \xi/\gamma_i$, therefore for $\xi \gg \gamma_i \gg 1$ we find that $F_1[\xi] \simeq 2 \sqrt{\xi/\gamma_i}$ and $F_2[\xi,\xi_b] \simeq 2 \ln[\xi_m/\xi_b]$. For $\xi \gg \gamma_i$, the integral for $F_1[\xi,\xi_b]$ is estimated by splitting it into the stages $\xi_b \leq \xi \leq \gamma_i$ and $\xi > \gamma_i$. The first stage yields $2 \ln[\gamma_i/\xi_b]$ since during this (ultrarelativistic) stage $W[\xi']\simeq 1$, and the second yields (approximately) $2\sqrt{\xi/\gamma_i}$ since during this (non-relativistic) stage $W[\xi] \approx \xi/\gamma_i$.

  In summary,  for a particle that is ``born'' ultrarelativistically, namely with  $\gamma_i \gg 1$,   the contribution $I^R_S$ rises rapidly up to a value $\simeq 2\,\ln[\xi_m/\xi_b]$ on a time scale $\xi_m \ll \gamma_i$ given by (\ref{chimax2}), remains nearly constant up to a time scale $\xi \simeq \gamma_i$ at which the particle becomes non-relativistic, and begins to fall-off as $-2\sqrt{\xi/\gamma_i}$ for $\xi \gg \gamma_i$.

  In the opposite limit when $\gamma_i \simeq 1$ the decaying particle is non-relativistic already at the initial time and $\omega_k(\eta) \simeq m_\phi\,C(\eta)$. In this case $F_2[\eta,\eta_b]$ saturates rapidly, on a scale $\xi_m \simeq \pi/\omega_i\,\eta_i \ll 1$, and $F_1[\eta,\eta_b]$ grows \emph{faster} than logarithmically, hence $F_1-F_2$  becomes larger than the logarithm in the first term  of $I^R_S$ and \emph{negative}. This behavior leads  to an   early    \emph{suppression}  of  decay.

\begin{figure}[ht!]
\begin{center}
\includegraphics[height=4in,width=5in,keepaspectratio=true]{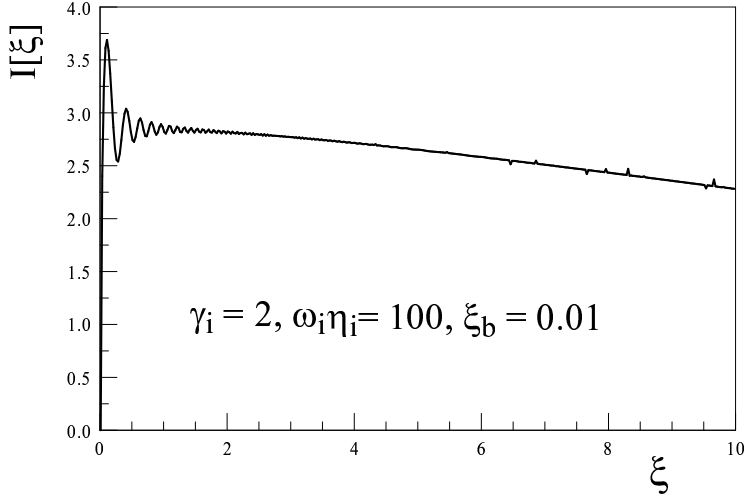}
\caption{The contribution $I^R_{S}$, eqn. (\ref{IRS}), for $\omega_i\eta_i=100, \xi_b = 0.01,\gamma_i=2$.}
\label{fig:sum2g2w100}
\end{center}
\end{figure}

\begin{figure}[ht!]
\begin{center}
\includegraphics[height=4in,width=5in,keepaspectratio=true]{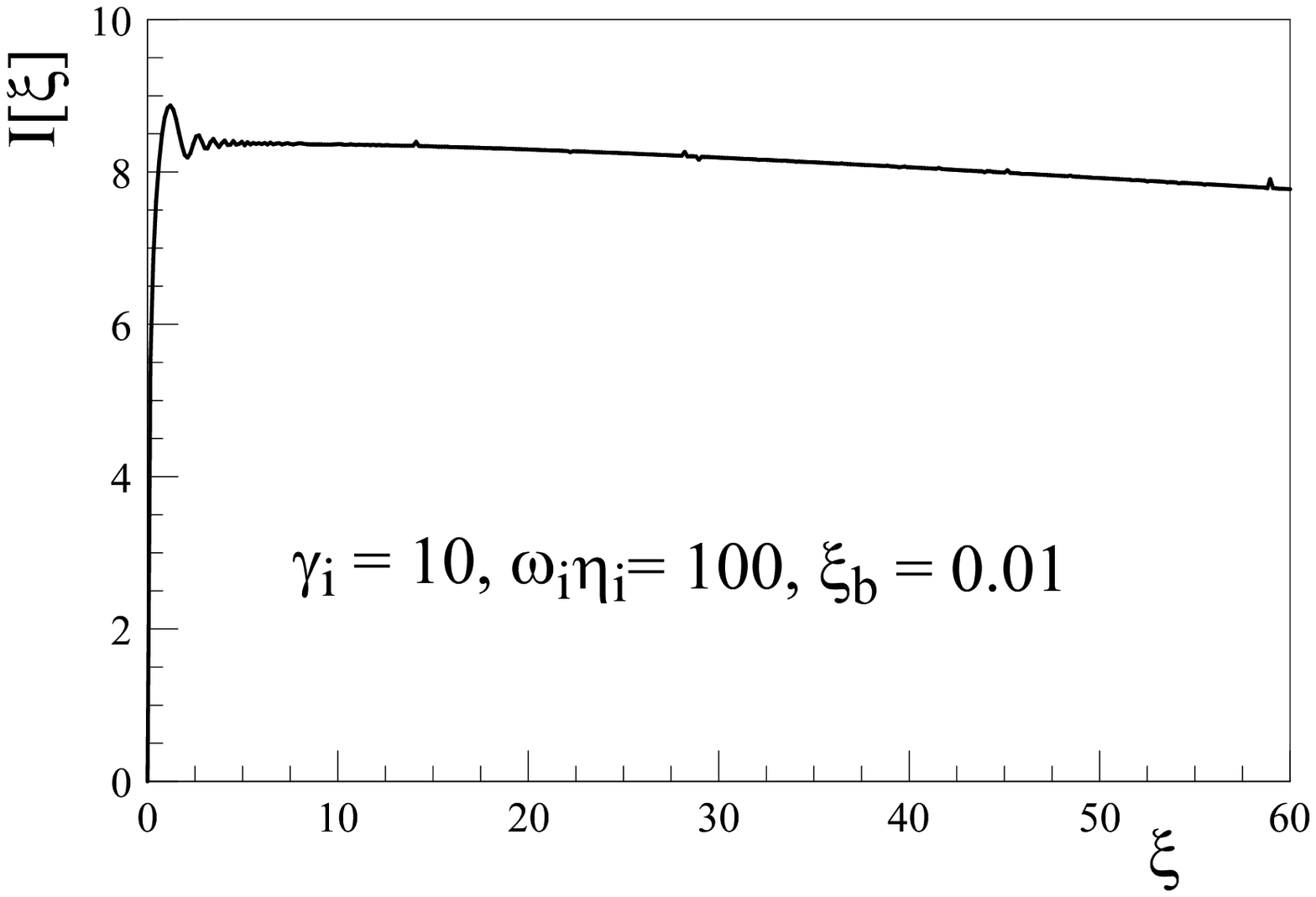}
\caption{The contribution $I^R_{S}$, eqn. (\ref{IRS}),  for $\omega_i\eta_i=100, \xi_b = 0.01,\gamma_i=10$.}
\label{fig:sum10g2w100}
\end{center}
\end{figure}

This analysis is approximately summarized   during the ultrarelativistic (UR) and non-relativistic (NR) regimes, by (see eqn. (\ref{fif2behaviorur}) in appendix (\ref{app:I2})),
\be I^R_S(k,\eta,\eta_b) \simeq  \Bigg\{ \begin{array}{ll}
                           2 \ln\big[\frac{\xi}{\xi_b}\big]\,\Theta(\xi_m-\xi) + 2 \ln\big[\frac{\xi_m}{\xi_b}\big]\,\Theta(\xi-\xi_m) ~~,~~\mathrm{for}~~ \gamma_i > \xi ~~(UR)\\
                            2 \ln\big[\frac{\xi_m}{\xi_b}\big]+2 \ln\big[\frac{\xi}{\gamma_i}\big]  -2\, \sqrt{\frac{\xi}{\gamma_i}}   ~~,~~\mathrm{for}~~\xi \gg \gamma_i > \xi_m ~~(NR)\,.
                         \end{array} \Bigg.   \label{IRSasy}\ee The main aspects of this analysis are
                         confirmed by a numerical study summarized  in figures (\ref{fig:sum2g2w100}) and (\ref{fig:sum10g2w100}) for $\gamma_i =2,10$ respectively. Notice the different scales in the figures highlighting the emergence of the plateau and the crossover to a diminishing (negative) square root behavior at a scale $\xi \simeq \gamma_i$.

\vspace{2mm}

\textbf{Decay at rest:} for a very massive particle ``born'' and decaying at rest in the comoving frame, namely for  $\gamma_i =1$, and $\omega_i \eta_i \gg 1$ we can provide an analytic form of the decay function for time scales $\xi \gg \xi_b \simeq 1/\omega_i \eta_i$ for on-shell renormalization. As discussed in appendices (\ref{app:I2},\ref{app:I3}), $F_2[\eta,\eta_b]$  reaches its asymptotic limit on a time scale $\xi \simeq \pi/2\omega_i\eta_i \ll 1$ (see equation (\ref{chimax1}) in appendix (\ref{app:I2})). Furthermore,  the function $\mathcal{S}(\xi')$ in (\ref{I3ofxi2}) reaches its asymptotic value $\mathcal{S} \simeq 1 $ at a time scale $\xi' \simeq \pi/\omega_i\eta_i\ll 1$. Therefore for $\xi' \gg 1/\omega_i\eta_i$ we can neglect the contribution from $F_2$ and   set $\mathcal{S}(\xi') =1$ in (\ref{I3ofxi2}), hence  $I_{3S}[k,\eta,\eta_b]$    is   given by the first term in  eqn. (\ref{I3Sxivars}) with $\gamma_i =1$ and multiplied by a factor $2$ to account for $\mathcal{S}=1$. Gathering all terms we find in this case ($\gamma_i =1~;~\omega_i\eta_i \gg 1~;~\xi \gg 1/\omega_i\eta_i$),
\be I_S(0,\eta,\eta_b) = 2\Bigg\{\ln[\xi]- \ln\Bigg[\frac{\sqrt{1+\xi}-1}{\sqrt{1+\xi}+1} \Bigg] - \sqrt{1+\xi} \Bigg\}+\frac{\pi}{2} \omega_i\eta_i \Bigg[\Big(1+\xi)^2 \Big]-1\Bigg] \,,\label{ISgama1}\ee where we have neglected a constant term of $\mathcal{O}(1)$. This expression displays all the features described above. Note that for $\xi \ll 1$ the logarithmic time dependence cancels out, but for $\xi \gg 1$ the first logarithm in (\ref{ISgama1}) continues to grow, however the \emph{negative} square root eventually dominates the contribution of the first terms within brackets. These are precisely the terms arising form the renormalization and their time dependence is a consequence of the time dependence of the frequencies.

 To compare to the decay law  in Minkowski space time it is convenient to cast the result (\ref{ISgama1}) in terms of comoving time, using $1+\xi= \eta/\eta_i$, with $\eta= \sqrt{2t/H_R}$ (see eqn. (\ref{etaoft}) valid in (RD)), and  the relation
 \be \frac{\omega_i \eta_i}{\gamma_i} = m_\phi H_R \, \eta^2_i = 2 \, m_\phi\,t_i\,. \label{relacoef}\ee Setting $\eta_i = \eta_b$ to leading adiabatic order,  we find for $\gamma_i=1$ and  $t \gg t_b$
\be I_S(0,t) \simeq  \ln\Big[\frac{t}{t_b} \Big]-2 \Bigg[\frac{t}{t_b}\Bigg]^{\frac{1}{4}}+ \pi \,m_\phi\,(t-t_b) \,,\label{ISoft} \ee leading to the survival probability for $t \gg t_b$
\be \mathcal{P}_\Phi(t)  = \Big[\frac{t}{t_b}\Big]^{-\frac{Y^2}{8\pi^2}}~
e^{ \frac{Y^2}{4\pi^2}\,\big(t/t_b\big)^{1/4} }
\,e^{-\Gamma_0\,(t-t_b)}~  \mathcal{P}_\Phi(t_b) ~~;~~ \Gamma_0 = \frac{Y^2}{8\pi}\,m_\phi\,.  \label{probarest}\ee This is one of the important results of this study. Remarkably $\Gamma_0$ is the same as the decay width at rest in Minkowski space time, however the power law with ``anomalous dimension'' $Y^2/8\pi^2$  and the \emph{stretched exponential} with the power law $(t/t_b)^{1/4}$ are a consequence of the renormalization and the time dependence of the frequencies,  manifestly a consequence of the expanding cosmology. The combined effect of these two terms yields a \emph{slowing down} of the decay as compared with the case of Minkowski space time with a concomitant enhancement of the lifetime of the decaying particle as compared to Minkowski space-time. This is a noteworthy result: as a consequence of the cosmological expansion the contribution from the renormalization and quasiparticle formation \emph{slows down the decay } leading to an enhancement of the lifetime of the initial state.

\vspace{2mm}

\textbf{Decay of   particles with $\gamma_i \gg 1$:} these are particles that are ``born'' ultrarelativistically. For $\gamma_i \gg 1$    the contribution from $I^R_S(k,\eta,\eta_b)$ has been  summarized by eqn (\ref{IRSasy}) and is  displayed in fig. (\ref{fig:sum10g2w100}): a rapid rise on a time scale $\xi_m \ll \gamma_i$ given by (\ref{chimax2}) up to $I^R_S \simeq 2\,\ln[\xi_m/\xi_b]$ followed by a near plateau  during the stage   while  $\xi \lesssim \gamma_i$. This contribution     falls off slowly as  $ -\sqrt{\xi/\gamma_i}$   during  the non-relativistic stage, $\xi \geq \gamma_i$ (see eqn. (\ref{IRSasy})).  While a quantitative analysis of $I_{3S}$ requires a numerical study, we can obtain a fairly accurate estimate as follows. The contribution from $\mathcal{S}$ to $I_{3S}$ (see equation (\ref{I3ofxi2})) is discussed in appendix (\ref{app:I3}), and can be approximately summarized as:  $\mathcal{S} \approx 0 $ for $\xi < \xi_s$ and $\mathcal{S}(\eta) \approx 1$ for $\xi > \xi_s$ with $\xi_s$ given by (\ref{solchimax},\ref{smallchim}).

With $\gamma_i \gg 1$, the  ultrarelativistic stage corresponds to  $\gamma_i \gg \xi$, during the stage  $\gamma_i \gg \xi_s \gg \xi$  it follows that $\mathcal{S} \approx 0$, using $1+\xi = \sqrt{t/t_i}$, and eqns. (\ref{tnr},\ref{relacoef}), during this stage $I_{3S}$  is given in comoving time  $t$ by
\be I_{3S}(t) = \frac{\pi}{2}\,m_\phi\,t_{nr}\Bigg[ G\Big[\frac{t}{t_{nr}}\Big]- G\Big[\frac{t_b}{t_{nr}}\Big] \Bigg] \,, \label{I3scomot}\ee where
\be G[x] = \Big[x(1+x) \Big]^{1/2} - \ln\Big[\sqrt{1+x}-\sqrt{x} \Big]\,, \label{Gofx}\ee also describes the decay function in the case of a scalar field decaying into two massless scalars\cite{herring}. During this stage for $t\ll t_{nr}$ we find
\be I_{3S}(t) =    \frac{\pi}{3} \, {m_\phi\,t_{nr}}   \, \Big(\frac{t}{t_{nr}}\Big)^{\frac{3}{2}}\,\Bigg[1-\Big(\frac{t_b}{t}\Big)^{\frac{3}{2}}  + \cdots  \Bigg]\,. \label{I3uroft}\ee

For $\gamma_i \gg \xi \gg \xi_s$   it follows that  $\mathcal{S} \simeq 1$, therefore the above result is multiplied by a factor 2. Hence, during the ultrarelativistic stage with $\gamma(t)\gg 1$, or $t\ll t_{nr}$,   and $\mathcal{S}  =1$ in (\ref{I3ofxi2}), it follows that
\be I_{3S}(t) \simeq  \frac{2\,\pi}{3} \, {m_\phi\,t_{nr}}\,\Big(\frac{t}{t_{nr}}\Big)^{3/2}\,\Bigg[1-\Big(\frac{t_b}{t}\Big)^{\frac{3}{2}}  + \cdots  \Bigg]  \, , \label{I3uroft2}\ee which when combined with the result (\ref{IRSasy}) yields in this ultrarelativistic regime, for $\gamma_i \gg \xi \gg \xi_s,\xi_m$
\be I_S(t) \simeq 2 \ln\Big[\frac{\xi_m}{\xi_b} \Big] + \frac{2\,\pi}{3} \, {m_\phi\,t_{nr}}\,\Big(\frac{t}{t_{nr}} \Big)^{3/2}\,\Bigg[1-\Big(\frac{t_b}{t}\Big)^{\frac{3}{2}}  + \cdots  \Bigg]  \,.  \label{ISurtot} \ee Neglecting the perturbatively small non-secular constant in the decay function from the first term in (\ref{ISurtot})\footnote{Or absorbing it in a \emph{finite} perturbatively small time independent wave function renormalization of $\mathcal{P}_\Phi$.} , we find in the time interval  for $t_{nr} \gg t \gg t_b$  during which the decaying particle is ultrarelativistic and the  transient dynamics   of   quasiparticle formation has saturated \be \mathcal{P}_\Phi(t)  =  e^{-\frac{2}{3}\Gamma_0\,t_{nr}\,(t/t_{nr})^{3/2}}~  \mathcal{P}_\Phi(t_b) \,.   \label{proba32}\ee

We can now use the property (\ref{drgsol}) and write for $t> t_{nr}$
\be \mathcal{P}_\Phi(t)= e^{-\int^\eta_{\eta_{nr}}\Gamma_\Phi(\eta')\,d\eta'}\,\mathcal{P}_\Phi(t_{nr})\,,\label{evoltnr}\ee where
\be \int^\eta_{\eta_{nr}}\Gamma_\Phi(\eta')\,d\eta' = \frac{Y^2}{8\pi^2} \,\Big[ I_{S}(k,\eta,\eta_b)-I_{S}(k,\eta_{nr},\eta_b)\Big] \,. \label{decfundife}\ee

After the decaying particle becomes non-relativistic for $\xi \gg \gamma_i$ or $t \gg t_{nr}$ when $\gamma(t) \simeq 1$, the contribution $\mathcal{S} \simeq 1$ and $I_{3S}(\xi)-I_{3S}(\xi_{nr})$ becomes
\be  I_{3S}(t)-I_{3S}(t_{nr}) = \pi \, m_\phi\,t   \Bigg[1- \frac{t_{nr}}{t}- \frac{t_{nr}}{2\,t}\,\ln\Big[\frac{t}{t_{nr}} \Big] + \cdots \Bigg]  \,, \label{I3oftnr}\ee   the dots in the above expression stand for terms of higher order in the ratio $t_{nr}/t$.

 Finally, combining with the result given by eqn. (\ref{IRSasy}),  the total decay function  after the particle has become non relativistic $\xi \gg \gamma_i$ (or $t\gg t_{nr}\gg t_b$)  is given in comoving  time by
\be I_S(t)-I_S(t_{nr}) \simeq \ln\Big[\frac{t}{t_{nr}} \Big]-2 \Bigg[\frac{t}{t_{nr}}\Bigg]^{\frac{1}{4}}+ \pi \, m_\phi\,t   \Bigg[1- \frac{t_{nr}}{t}- \frac{t_{nr}}{2\,t}\,\ln\Big[\frac{t}{t_{nr}} \Big] + \cdots \Bigg]\,, \label{ISnrtot} \ee where we have neglected a perturbatively small constant term and approximated $t_i \gamma^2_i \simeq t_{nr}$ for $\gamma_i \gg 1$. Hence for $t \gg t_{nr} \gg t_b$ we find
\be \mathcal{P}_\Phi(t)  = \Big[\frac{t}{t_{nr}}\Big]^{-\frac{Y^2}{8\pi^2}}~
e^{ \frac{Y^2}{4\pi^2}\,\big(t/t_{nr}\big)^{1/4} }~\Big[\frac{t}{t_{nr}}\Big]^{\Gamma_0 t_{nr}/2}
\,e^{-\Gamma_0\,(t-t_{nr})}~  \mathcal{P}_\Phi(t_{nr}) \,.   \label{probaurnr}\ee

It would be expected that after $t_{nr}$ when the particle has become non-relativistic as a consequence of the cosmological redshift, the time evolution of the survival probability would be similar to that of a particle born and decaying at rest. However, the result (\ref{probaurnr}) features an extra power law with exponent $\Gamma_0 t_{nr}/2$ as compared to the decay function for the particle born at rest, eqn. (\ref{probarest}). This difference  reflects the memory   of the past evolution in the form of the  integral (\ref{decfundife}).

 We can provide a measure of the impact of curved space time effects on the decay function by   comparing the results above to a phenomenological, S-matrix inspired   Minkowski decay law allowing for a \emph{local time dilation factor} to account for the cosmological redshift, namely
 \be \mathcal{P}^{(M)}_\Phi(t) = e^{-\frac{\Gamma_0}{\gamma(t)}\,(t-t_i)}\, \mathcal{P}_\Phi^{(M)}(t_i)\,, \label{Minkodecay}\ee where  $\Gamma_0 = \frac{Y^2\,m_\phi}{8\pi}$ is  the decay width at rest in Minkowski space time, and $\gamma(t)$ the local Lorentz factor (\ref{gamaofchi}). The comparison to  the cutoff independent subtracted decay function  (\ref{ISfina})  is facilitated by introducing
  \be I_M(t) = \frac{\pi\,m_\phi\,t}{\gamma(t)}\,\Big[1-\frac{t_i}{t} \Big] \,, \label{minkI}\ee  so that the Minkowski-like decay function is given by
   \be \frac{\Gamma_0}{\gamma(t)}\,(t-t_i) \equiv \frac{Y^2}{8\pi^2}\,I_M(t)\,, \label{IMink} \ee where a factor is included in (\ref{IMink}) to ensure that $I_M(t_i=t_b) =0$ consistently with the subtraction definining (\ref{ISfina}). For $t\gg t_i$ this \emph{phenomenological} decay function is interpreted as that of Minkowski space-time but with the    instantaneous Lorentz time dilation factor. For $t\gg t_i$ it provides a ``benchmark'' to compare the results obtained above for the decay function to an S-matrix inspired \emph{instantaneous} Minkowski decay law.

    Before we engage in a numerical comparison, it is illuminating to analyze the cases discussed above.

   \vspace{2mm}

   \textbf{Non-relativistic: $\gamma(t) =1$.} For this case the $I_S(t)$ is given by (\ref{ISoft}), the last term of which is precisely $I_M(t)$ for $\gamma(t)=1$. The first two terms in (\ref{ISoft}) yield a \emph{negative} contribution for $t \gg t_b =t_i$, therefore the cosmological decay function is \emph{smaller} in this case than the phenomenological Minkowski function, leading to a longer lifetime.

   \vspace{2mm}

   \textbf{Ultra-relativistic: $\gamma_i \gg 1$.} During the ultrarelativistic regime $\gamma(t) \gg 1$ ($t \ll t_{nr}$),  taking the time large enough so that the transient build-up of $\mathcal{S}$ in eqn. (\ref{I3ofxi2}) has saturated, the cosmological decay function is given by (\ref{ISurtot}) whereas $I_M(t) \simeq \pi m_\phi t  \,\big(t/t_{nr}\big)^{1/2}$. The  logarithmic term in (\ref{ISurtot}) could be fairly large for large $\gamma_i$ thereby yielding $I_S(t) > I_M(t)$ during a   time interval.  This can be understood from the following argument.

   As discussed above and in appendix (\ref{app:I2}), for $\gamma_i \gg 1$ the contribution  $I^R_S$ (see eqn. (\ref{IRS}))  rises on a time scale $\xi_m \simeq (3\pi \gamma^2_i/\omega_i\eta_i)^{1/3}$ up to a maximum $\simeq 2 \ln(\xi_m/\xi_b)$ after which it remains nearly constant up to $\xi \simeq \gamma_i$   yielding the logarithmic term in (\ref{ISurtot}).  For example, for $\gamma_i \simeq 200~,~ \omega_i \eta_i \simeq 100$ and ``on-shell'' renormalization with $\xi_b = 1/\omega_i \eta_i$, the contribution from  $I^R_S$ rises up to a value $\simeq \frac{4}{3} \ln[\sqrt{3\pi}\gamma_i \omega_i \eta_i] \simeq 14.7$ on a comoving time scale $t_m/t_i \approx 240$.  Since the Hubble time scale $1/H(t) = 2 t$ during (RD), it follows that   $I^R_S$ rises   up to the plateau  over $\simeq 240$ Hubble times, with the possibility that during this time $I_S(t) > I_M(t)$.  However, after the particle has become non-relativistic, namely for $t\gg t_{nr}$, the cosmological decay function $I_S(t)$ is given by (\ref{ISnrtot}) whereas    \be I_M(t) \simeq \pi m_\phi\,t\,\Big[ 1- \frac{t_{nr}}{2t} +\cdots \Big] \,, \label{IMnr}\ee  showing that  $I_S(t)  \ll I_M(t)$  for $t \gg t_{nr}$. This suggests a crossover behavior  for very large values of $\gamma_i$:  there is an early  time window during the ultrarelativistic stage wherein the cosmological decay function may be  larger than the Minkowski one, however as the decaying particle eventually becomes non-relativistic the latter will ultimately dominate. This behavior is borne out by a detailed numerical study.

    \begin{figure}[ht!]
\begin{center}
\includegraphics[height=4in,width=5in,keepaspectratio=true]{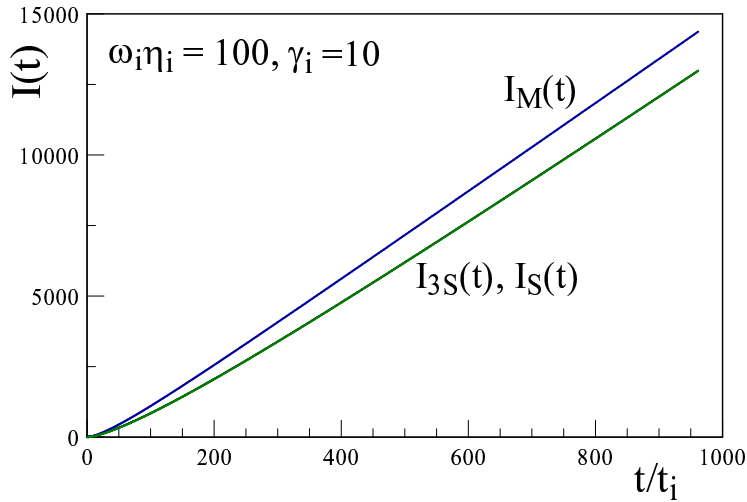}
\caption{Comparison between $I_M,I_S,I_{3S}$ for on-shell subtraction with $\omega_i \eta_i = 100,\gamma_i = 10$, $t_{nr}/t_i = 99$.}
\label{fig:compg10w100}
\end{center}
\end{figure}

    \begin{figure}[ht!]
\begin{center}
\includegraphics[height=4in,width=5in,keepaspectratio=true]{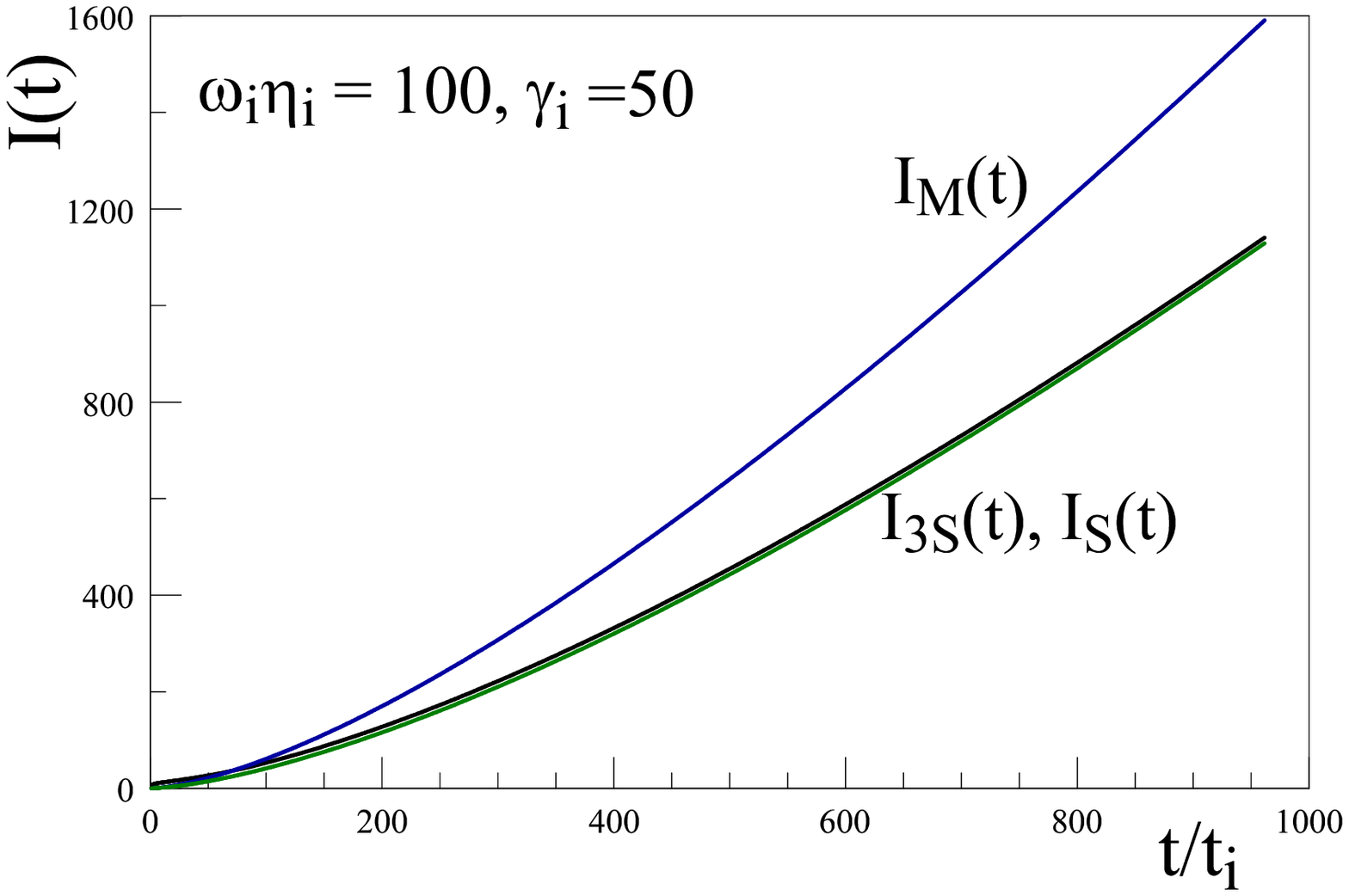}
\caption{Comparison between $I_M,I_S,I_{3S}$ for on-shell subtraction with $\omega_i \eta_i = 100,\gamma_i = 50$,$t_{nr}/t_i = 2499$.}
\label{fig:compg50w100}
\end{center}
\end{figure}

    \begin{figure}[ht!]
\begin{center}
\includegraphics[height=4in,width=5in,keepaspectratio=true]{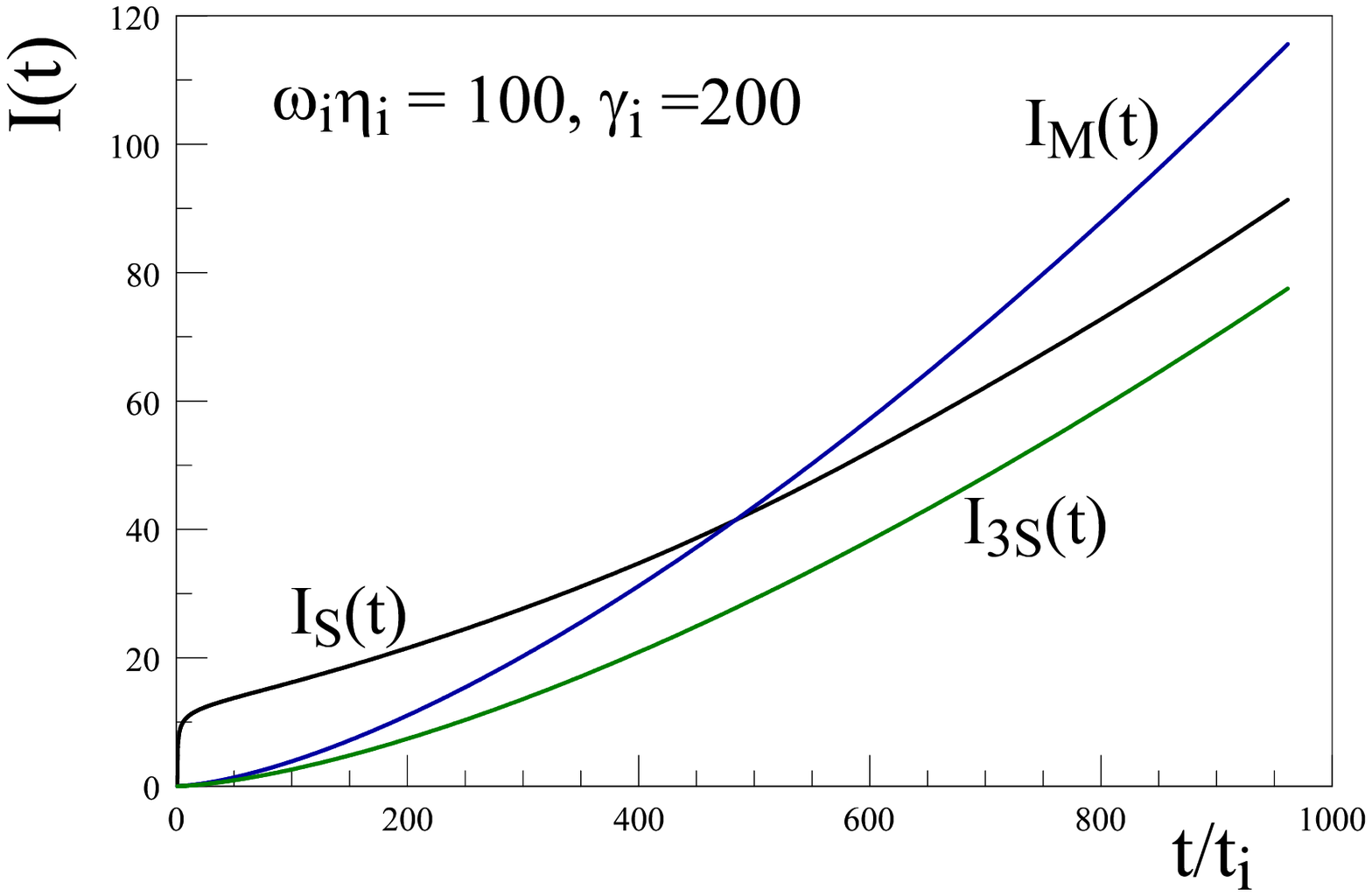}
\caption{Comparison between $I_M,I_S,I_{3S}$ for on-shell subtraction with $\omega_i \eta_i = 100,\gamma_i = 200$, $t_{nr}/t_i \simeq 4\times 10^4$.}
\label{fig:compg200w100}
\end{center}
\end{figure}

Figures (\ref{fig:compg10w100},\ref{fig:compg50w100},\ref{fig:compg200w100}) show a comparison between the phenomenological Minkowski decay function (\ref{minkI}), the total contribution $I_S$ (\ref{ISfina}) along with $I_{3S}$ (\ref{I3ofxi2}) for on-shell renormalization with $\omega_i \eta_i =100$ and $\gamma_i = 10,50,200$ respectively. For these values the transition time to the non-relativistic behavior is $t_{nr}/t_i \simeq  10^2, 2.5\times 10^3, 4\times 10^4$ respectively.   For $\gamma_i = 10, 50$ figs. (\ref{fig:compg10w100},\ref{fig:compg50w100}) show that $I_S$ and $I_{3S}$ are nearly indistinguishable, namely $I^R_S$ (\ref{IRS}) is subleading in these cases, and that the phenomenological $I_M$ is always \emph{larger} than $I_S$.

However, for $\gamma_i = 200$ fig. ( \ref{fig:compg200w100}) shows that the contribution from $I^R_S$ dominates at early time, rising on a time scale $t/t_i \simeq 100$.  In this case $I_M$ is smaller than $I_S$ during a substantial time window, $\approx 500$ Hubble times from the ``birth'' of the quasiparticle, before  crossing over to becoming the largest decay function.

 Therefore we conclude that in the ultrarelativistic case, for very large values of $\gamma_i$, the decay function is \emph{larger} than the phenomenological Minkowski one within a substantial time interval but eventually becomes smaller at a time scale that depends on the various parameters. In either case, at long time the decaying particle lives \emph{longer} than predicted by a Minkowski decay law extrapolated to the expanding cosmology. This is a generic result: after an intermediate time scale that depends on $\gamma_i$, the cosmological decay function is \emph{smaller} than the phenomenological Minkowski-like one. Therefore the S-matrix inspired phenomenological Minkowski decay law \emph{under estimates} the lifetime of the decaying particle.

\subsection{Long lived particles: decay during matter   domination or beyond.}\label{subsec:MD}

The discussion above focused on decay during the radiation dominated era that lasts until $C(\eta) = a_{eq} \simeq 10^{-4}$ corresponding to an ambient temperature $T\simeq  \mathrm{eV}$ at a time $t_{eq} \approx 10^{12}\,\mathrm{secs}$. If the decaying particle is very long lived as would befit a dark matter candidate, it would continue to decay during the matter and perhaps dark energy dominated eras. This case corresponds to an extremely small Yukawa coupling, which allows to safely neglect early transient effects that saturate at early times.
The general form of the decay function after renormalization is given by eqns. (\ref{decayfun},\ref{ISubfin}). Under the assumption of very weak Yukawa coupling  we can neglect the contribution from the cosine term in $I_{2b}$, eqn. (\ref{I2bar}) (the contribution $F_2$ in eqn. (\ref{F2v2})) and we can set $\mathcal{S} =1$ in eqn. (\ref{I3bar}). This is because both terms saturate on short time scales therefore they yield perturbatively small corrections to the decay function for very weak Yukawa coupling as compared to the terms that continue to grow in time. Hence, neglecting these perturbatively small   transient contributions     for very weak Yukawa couplings, the decay function simplifies to
\be \int^\eta_{\eta_b}\,\Gamma_\Phi(\eta')\,d\eta' = \frac{Y^2}{8\pi^2}\,\Bigg[ 2\,\ln\Big[\frac{\xi}{\xi_b} \Big]- F_1[\xi,\xi_b]  +  {\pi} \,m_\phi \int^\eta_{\eta_b} \frac{C(\eta')}{\gamma_k(\eta')} \,d\eta' \, \Bigg]+\cdots \,,\label{decfunmd} \ee where $\xi = (\eta-\eta_i)/\eta_i$ and $F_1$ is given by (\ref{F1v2}) and the dots in (\ref{decfunmd}) stand for constant terms that   are of $\mathcal{O}(Y^2)$.

For general scale factor  $W[\xi]$ is  given by
\be  W[\xi]   =    \frac{1}{\gamma_i}\Bigg[(\gamma^2_i-1)+
\frac{C^2(\eta)}{C^2(\eta_i)}  \Bigg]^{\frac{1}{2}}\,. \label{WgenC}\ee

Let us analyze each term separately in order to understand their behavior at long time during the (MD) era, taking as an upper bound $C(\eta) \simeq  \mathcal{O}(1)$, or, upon using eqn. (\ref{etaofa}) $\eta \simeq \sqrt{a_{eq}}/H_R$. With ``on-shell'' renormalization ($\xi_b = 1/\omega_i \eta_i$), we find
\be \ln\Big[\frac{\xi}{\xi_b} \Big] \simeq \ln\Big[\frac{\omega_i\,\sqrt{a_{eq}}}{H_R} \Big]\simeq \ln[10^{42}\,C(\eta_i)\Big]+ \ln\Bigg[\gamma_i\,\Big(\frac{m_\phi}{\mathrm{GeV}} \Big) \Bigg]\,. \label{logterm}\ee Taking the initial time to correspond to an initial temperature $10^{15}\,\mathrm{GeV}$ yields $C(\eta_i) \simeq 10^{-28}$ therefore the logarithm contribution to the decay function for $\eta \simeq \sqrt{a_{eq}}/H_R$ yields
\be \frac{Y^2}{4\pi^2}\,\ln\Big[\frac{\xi}{\xi_b} \Big] \simeq 0.82\,Y^2+ \frac{Y^2}{4\pi^2}\,\ln\Bigg[\gamma_i\,\Big(\frac{m_\phi}{\mathrm{GeV}} \Big) \Bigg]\,. \label{logairre}\ee

Obtaining the contribution from $F_1$ over the whole history from early (RD) into (MD) can be done numerically, although this is a rather challenging task because of the enormous dynamic range with the scale factor varying over twenty four orders of magnitude. However, we can provide a simple estimate of the remaining two terms of the decay function   at long time during the (MD) era and/or beyond. If the particle remains ultrarelativistic, then as discussed in the previous sections the contribution from $F_1$ cancels the logarithmic time dependence of the first term, hence the combination of the first two terms saturates  (this is the plateau in fig. (\ref{fig:sum10g2w100})) and yields a perturbatively small time independent contribution to the decay function. Hence during this ultrarelativistic stage   the last term in (\ref{decfunmd}) dominates the decay function.

After  the particle has become non-relativistic then $W[\xi] \simeq C(\eta)/\gamma_i C(\eta_i) \gg 1 $ and
\be F_1[\xi,\xi_b] \simeq \frac{1}{\sqrt{\gamma_i C(\eta_i)}}\,\int^\eta \frac{\sqrt{C(\eta')}}{\eta'}\,d\eta' \,,\label{F1mdap}\ee during (MD) using eqn. (\ref{Crdmd}) and taking as an upper bound $\eta \simeq \sqrt{a_{eq}}/H_R$ we find
\be F_1[\xi,\xi_b] \simeq \frac{1}{2\sqrt{\gamma_i C(\eta_i)} } \simeq \frac{10^{14}}{\sqrt{\gamma_i}} \,.\label{F1mdsub}\ee

Finally, we can estimate the last term in (\ref{decfunmd}) during the stage when the particle is non-relativistic and (MD) dominated, taking $\gamma(\eta') \simeq 1$ and taking $\eta \simeq \sqrt{a_{eq}}/H_R$, we find
\be m_\phi \int^\eta  \frac{C(\eta')}{\gamma_k(\eta')} \,d\eta' \simeq 10^{42}\,\Bigg(\frac{m_\phi}{\mathrm{GeV}} \Bigg) \,.\label{lasterm} \ee Since during the ultrarelativistic stage the time dependence of the first and second term cancel out and the last term dominates the decay dynamics,  we conclude that the last term in (\ref{decfunmd}) dominates the decay dynamics of  a very long-lived particle with very weak Yukawa coupling, \emph{all throughout the time evolution}. Since the first (logarithmic) term is always subdominant, and   the second term is negative, larger in magnitude than the logarithmic term but also subdominant at late time, the last term in (\ref{decfunmd}) yields  an \emph{upper bound} to the decay function throughout all the expansion history. It can be written  as a function of the redshift by    recalling   that $C(\eta) \,d\eta = dt$,  and using $dt = da/(a\,H(a))$ with $H(a)$ the Hubble expansion rate given by eqn. (\ref{Hubble}).  Writing the local Lorentz factor as  $\gamma(a(t)))= \Big[\frac{a^2_{nr}}{a^2(t)}+1\Big]^{1/2}~;~a_{nr} \equiv k/m_{\phi}$,   we find that the upper bound to the decay function at redshift $z$ is given by
\be \int^\eta_{\eta_b}\,\Gamma_\Phi(\eta')\,d\eta' \simeq \frac{\Gamma_0}{H_0}\,\Upsilon(z,z_b)\,,  \label{Upsiz} \ee where $\Gamma_0 = Y^2 m_\phi/8\pi$ is the decay rate at rest in Minkowski space time, and
\be \Upsilon(z,z_b) = \int^{1/(1+z)}_{1/z_b} \frac{da}{\sqrt{a^2_{nr}+a^2}\,\Big[ \frac{\Omega_M}{a^3}+ \frac{\Omega_R}{a^4}+ \Omega_\Lambda \Big]^{1/2}  } \,,  \label{upsfunc}\ee depends solely on the cosmological parameters and $a_{nr}=k/m_{\phi}$ the scale factor at which the decaying particle transitions from ultrarelativistic to non-relativistic, and we have taken $z_b \gg 1$.  In particular this result for the decay function is insensitive to the early transient dynamics.

 The redshift evolution of the survival probability all throughout the expansion history is summarized concisely as
\be \mathcal{P}_\Phi(z) \gtrsim  e^{-\frac{\Gamma_0}{H_0}\,\Upsilon(z,z_b)}\,\mathcal{P}_\Phi(z_b)\,. \label{probaz}\ee
  The inequality in eqn. (\ref{probaz}) reflects that eqn. (\ref{Upsiz}) yields an \emph{upper bound} to the decay function.  For $a_{nr}=0$, namely when the decaying particle is ``born'' at rest,  it follows that $\Upsilon(z,z_b) = H_0 (t-t_b)$ independently of the cosmology,
  and we can compare the result (\ref{Upsiz}) for $a_{nr}=0$  to the case of the particle decaying at rest given by eqn. (\ref{ISoft}) valid during the (RD) era. The  discussion on dominant terms above clarifies that the last term in (\ref{ISoft}) dominates the decay dynamics, whereas the first two terms combine into  a   negative contribution which becomes   subleading at long time  for very weak Yukawa couplings. Hence it is clear that for very weak Yukawa coupling and long time  (\ref{Upsiz})  becomes the leading contribution and yields an upper bound  to the decay function for long-lived particles decaying at rest. Furthermore, for $a\ll a_{nr}$, namely when the decaying particle is ultrarelativistic and taking this regime to be during the (RD) era with $a\propto t^{1/2}$ it follows that
\be \Upsilon(z,z_b) \propto t^{3/2}\,, \label{t32}\ee in agreement with the decay law (\ref{proba32}) during the ultrarelativistic regime in (RD). This analysis confirms the validity of the decay law (\ref{probaz}) with (\ref{upsfunc})  as an  upper bound to describe the evolution of the survival probability for very weakly coupled, long lived particles all throughout the cosmological evolution, under the assumption that the fermionic decay products can be considered massless in the decay process.

\section{Discussion}

The final form of the renormalized decay function, eqn. (\ref{ISfina})  describing the time evolution of the survival probability of the \emph{quasiparticle} state is amenable to a straightforward numerical study. The  analysis of section (\ref{sec:dynamics}) reveals a very rich dynamical evolution with various different time scales. The shortest time scales describe the build-up of the quasiparticle; this early transient dynamics is absorbed into a wave function renormalization of the quasiparticle survival probability at a time scale $t_b$. After this short time transient there remain the time scales over which $F_2$ (\ref{F2v2}) saturates at a constant value  and $\mathcal{S}$ (\ref{SiI3s}) rapidly approaches $\mathcal{S} \simeq 1$. The detailed dynamics over these scales  was studied analytically and numerically in appendices (\ref{app:I2}) and (\ref{app:I3})) respectively.  The evolution of the survival probability   on the intermediate and long time scales    becomes simpler and can be summarized succinctly. Furthermore, because the short time transients saturate to constant values, for weak Yukawa coupling  the largest contributions to the decay dynamics arises from   terms that are secular (grow in time) over  the intermediate and long time scales.

\vspace{2mm}

\textbf{Decay at rest in the comoving frame ($\gamma_i=1$):} The time evolution of the survival probability is given by

\be \mathcal{P}_\Phi(t)  = \Big[\frac{t}{t_b}\Big]^{-\frac{Y^2}{8\pi^2}}~
e^{ \frac{Y^2}{4\pi^2}\,\big(t/t_b\big)^{1/4} }
\,e^{-\Gamma_0\,(t-t_b)}~  \mathcal{P}_\Phi(t_b) \,, \label{probarestdos}  \ee where $ \Gamma_0 = \frac{Y^2}{8\pi}\,m_\phi\,$ is the decay width of a particle at rest in Minkowski space-time. The power law and stretched exponentials are both a remnant of the renormalization, or ``dressing''  of the bare into the quasiparticle state and a distinct consequence of the cosmological redshift. Indeed, in Minkowski space-time  the terms that give rise to these contribution become time independent after the transient dynamics, whereas, in curved space-time, the origin of these contributions    is the time dependence of the frequencies via the cosmological redshift.

The methods that we implemented in this study, a non-perturbative formulation combined with a physically motivated adiabatic expansion including a consistent treatment of renormalization, are very different from those implemented in ref.\cite{viljafer}.   The decay law of a particle at rest (\ref{probarestdos}) is also very different from that reported in ref.\cite{viljafer}. The origin of the discrepancy is not clear to us. However,    since the power law and stretched exponentials originate precisely from the contributions to the renormalization of the survival probability, we suspect that the discrepancy originates in the treatment of the ultraviolet divergences. These are of the same form as in Minkowski space-time (see appendix (\ref{app:Minko}) and ref.\cite{zenoqft}) as expected since these are short distance divergences,  but have not been discussed or addressed in ref.\cite{viljafer}. As explained above, the time dependence of the frequency yields an unexpected contribution to the decay law on longer time scales that originates in the dynamics of quasiparticle formation.

\vspace{2mm}

\textbf{Born ultrarelativistically:} if the particle is ``born'' or produced ultrarelativistically, namely with $\gamma_i \gg 1$  during (RD),  an important time scale is  $t_{nr} = \frac{k^2}{2 m^2_\phi H_0\sqrt{\Omega_R}}$, which determines when the particle transitions from being ultrarelativistic ($\gamma(t) \gg 1$ or $t \ll t_{nr}$) to non-relativistic ($\gamma(t) \simeq 1$ or $ t\gg t_{nr}$) as a consequence of the cosmological redshift. The  dynamical evolution of the survival probability is different in these stages.  \textbf{a)}  ultrarelativistic  stage:   ($\gamma(t) \gg 1$, or $t_b\ll t \ll t_{nr}$ )
\be \mathcal{P}_\Phi(t)  =  e^{-\frac{2}{3}\Gamma_0\,t_{nr}\,(t/t_{nr})^{3/2}}~  \mathcal{P}_\Phi(t_b) \,.   \label{probaurdos}\ee \textbf{b)}  non-relativistic stage ($t\gg t_{nr}$ or $\gamma(t) \simeq 1$),
\be \mathcal{P}_\Phi(t)  = \Big[\frac{t}{t_{nr}}\Big]^{-\frac{Y^2}{8\pi^2}}~
e^{ \frac{Y^2}{4\pi^2}\,\big(t/t_{nr}\big)^{1/4} }~\Big[\frac{t}{t_{nr}}\Big]^{\Gamma_0 t_{nr}/2}
\,e^{-\Gamma_0\,(t-t_{nr})}~  \mathcal{P}_\Phi(t_{nr}) \,.   \label{probaurnrdos}\ee Although for $t\gg t_{nr}$ the particle has become non-relativistic because of the cosmological redshift, as compared to the case of decay at rest (\ref{probarestdos}), this decay law features a new power with exponent $\Gamma_0 t_{nr}/2$. Its origin is the \emph{memory} of the decay function manifest in the form of the integral of the cosmological redshift in (\ref{I3bar}) over the whole history of the decay process. Therefore, even well after the decaying particle has become non-relativistic, the survival probability features an \emph{enhancement} factor that ``knows'' about the past history when the particle was ultrarelativistic. The dynamics during the  transition from the ultrarelativistic to the non-relativistic behavior must be studied numerically, and the previous section shows such study for several values of the parameters.

\vspace{2mm}

\textbf{Massless fermions vs massless bosons:} Ref.\cite{herring} studied the decay of a scalar into two massless scalars, therefore we can now compare the results of that study to those obtained here for the case of scalar decay into  massless fermions. The main difference is in the contribution $I^R_S$ in eqn.(\ref{ISfina}) which is given by eqn. (\ref{IRS}). The contribution from $I_{3S}$ to the decay function is the same for fermions and bosons, for example the function $G[x]$ is the same that enters in scalar decay\cite{herring}. The extra contribution, namely $I^R_S$ has the same origin as the ultraviolet divergent contributions that are absorbed in wave function renormalization. This is also the case in Minkowski space-time\cite{zenoqft} as shown in appendix (\ref{app:Minko}).  Whereas in Minkowski space time this contribution becomes time independent after a short time transient and is absorbed into wave function renormalization, in a FRW cosmology, it is time dependent as a consequence of the cosmological redshift and becomes important for non-relativistic particles. Namely, $I^R_S$ is a \emph{remnant} of the physical process of quasiparticle formation. There is no such contribution in the case of decay into two scalars because the theory in this case is superrenormalizable, hence there is no equivalent of the $I^R_S$ term. This contribution \emph{suppresses} the decay function at long time, thereby enhancing the lifetime of the decaying particle. This behavior is yet another source of discrepancy with  the results of ref.\cite{viljafer}, which finds a \emph{larger} rate in the fermionic case. The source of this discrepancy are precisely the ``anomalous'' power and stretched exponential which are a consequence of the quasiparticle formation and wave function renormalization. Although the decay probability requires an ultraviolet divergent wave function renormalization even in Minkowski space-time, this seems to be an aspect missing  in the treatment of ref.\cite{viljafer}.  The cumulative effect of these differences results in that a  meaningful comparison to our study has eluded us.

\vspace{2mm}

\textbf{``Benchmarking'' the decay law:} The decay laws obtained above are very different from the usual exponential decay familiar in Minkowski space time, one of the   reasons for the difference being the cosmological redshift. Thus a natural question arises: would an S-matrix inspired,  \emph{phenomenologically} motivated exponential decay law  with a  time dependent Lorentz factor to account  for the cosmological redshift describe even approximately the decay of the particle?. This motivates the comparison of the previous results to the following Minkowski-like decay law (in (RD))
\be \mathcal{P}^{(M)}_\Phi(t) = e^{-\frac{\Gamma_0}{\gamma(t)}\,(t-t_i)}\, \mathcal{P}_\Phi^{(M)}(t_i)~~;~~ \gamma(t) = \Big[\frac{t_{nr}}{t} +1 \Big]^{1/2}\,. \label{Minkodecaydos} \ee For decay at rest $\gamma(t)=1$, this decay law misses the power with anomalous dimension and the stretched exponential, whose combination is \emph{negative}. Therefore the Minkowski-like decay law \emph{over-estimates} the suppression of the survival probability in the case of decay at rest. For a particle that is produced ultrarelativistically, during the stage wherein $\gamma(t) \gg 1$, namely $t \ll t_{nr}$ one finds
\be \mathcal{P}^{(M)}_\Phi(t) = e^{- \Gamma_0 t_{nr} (t/t_{nr})^{3/2} }\, \mathcal{P}_\Phi^{(M)}(t_i) \,, \label{Minkodecayurdos} \ee which is \emph{smaller} than (\ref{probaurdos}). For $t \gg t_{nr}$ when the decaying particle has become non-relativistic
\be \mathcal{P}^{(M)}_\Phi(t) = e^{- \Gamma_0 \big(t-t_{nr}/2\big)  }\, \mathcal{P}_\Phi^{(M)}(t_i) \,. \label{Minkodecaynrdos} \ee Comparing this result with (\ref{probaurnrdos}) clearly shows that the phenomenological Minkowski decay law including the instantaneous Lorentz factor over-estimates the suppression of the survival probability, namely \emph{under estimates} the lifetime of the decaying state. The discrepancies with the cosmological decay law, both the factor $2/3$ in (\ref{probaurdos}) along with the powers and stretched exponential in (\ref{probaurnrdos}) are traced to i) the memory of quasiparticle formation, ii) the memory of the past evolution in the integral of the time dilation factor. None of these can be captured by a phenomenological Minkowski-like decay law including an instantaneous Lorentz factor as such description has no memory of the past evolution.
We draw two important conclusions from this comparison: i) a phenomenological, S-matrix inspired  Minkowski decay law \emph{under estimates} the lifetime of the decaying particle since it \emph{over estimates} the suppression of the survival probability, ii) describing particle decay in cosmology in terms of a decay \emph{rate}, even one that includes the cosmological redshift in the time dilation factor,  is not only not useful but is misleading insofar as missing  important physical processes and yielding a substantial under estimate of the lifetime of the decaying particle.

\vspace{2mm}

\textbf{On initial conditions:} we have taken the initial state to correspond to  a single particle state of a given momentum, to compare to the usual case in S-matrix theory. The calculation of a decay rate in S-matrix theory considers the transition amplitude from  ``in'' single particle state (prepared at time $-\infty$) to an ``out'' multiparticle state at time $+\infty$. Our main point is that such calculation is not meaningful in an expanding cosmology, motivating the study of the previous sections. Thus the chosen initial condition allows us to directly compare to what would be expected from S-matrix theory, appended with an exponential decay law with the rate calculated with S-matrix.

Alternative initial conditions may be considered but all imply not only technical complexities, but also conceptual aspects: a single particle but spatially localized wave packet will not only decay via the decay of the different single particle components for different momenta, but its amplitude will also decay as a consequence of dispersion and spreading. Spatially narrow wave packets will decay the fastest and  separating systematically the different physical processes is, in general, not only technically daunting but implies some ambiguity as to how to extract a ``decay''. Another physically motivated initial condition would be to take the initial state to emerge from the decay of a heavier particle. Obviously, such choice would have inherent ambiguities from the choice of the parent particle and its decay kinematics. These aspects notwithstanding, the framework developed in the previous sections can be simply adapted  to alternative initial conditions.

\vspace{2mm}

\textbf{Modifications to BBN?} Although the results obtained in this study do not apply directly to neutron decay, since we focused on scalar decay Yukawa coupled to  massless fermions, and the small phase space available for three body neutron decay is a result of the small neutron-proton mass difference, let us explore the consequences of the results  on this process, with all these caveats. First: the neutron is ``born'' after the QCD phase transition at $T_{QCD} \simeq 150\,\mathrm{MeV}$ at a time $t_b \simeq 10^{-5}\,\mathrm{secs}$, because the neutron mass $M_N \simeq \mathrm{GeV} \gg T_{QCD}$ it is ``born'' at rest in the plasma. Let us identify the dimensionless coupling $Y^2/8\pi \equiv \Gamma_N/M_N$ where $\Gamma_N \simeq 10^{-3}\,\mathrm{secs}^{-1}$ is the neutron's lifetime. Hence  $Y^2/8\pi \simeq 10^{-21}$, and taking  $t/t_b \simeq 1/\Gamma_N t_b \simeq 10^{8}$ we see that the power law with ``anomalous'' dimension and the stretched exponential correction to the usual exponential decay law in eqn. (\ref{probarest}) for decay at rest are all but negligibly small and would \emph{not} affect the dynamics of neutron decay during (BBN). Of course, there are the above mentioned caveats to this conclusion which should only be taken as an extrapolation and as a gross estimate of the effects. This analysis also suggests that the corrections to the decay law are more important for particles ``born'' very early during (RD) and very long lived a situation that befits most descriptions of a dark matter candidate.

\vspace{2mm}

\textbf{Emergence of  a local decay law with constant S-matrix decay rate.}  If  a measurement of the time evolution of the survival probability is carried out during a    \emph{ sufficiently short} time interval $\Delta t = t_f-ti$ and sufficiently long after the transient dynamics has subsided, we would expect that the decay law would be \emph{nearly} exponential with a nearly constant decay rate. Namely we would expect that \emph{locally} during such a short time interval the
survival probability is given by
\be \mathcal{P}[t_f] = e^{-\overline{\Gamma}_\Phi(k) (t_f-t_i)} \,\mathcal{P}[t_i]  \,,  \label{Smtxinsp}\ee where $\overline{\Gamma}_\Phi(k)$ is a constant, related to the S-matrix rate.  This law cannot emerge during the transient stage dominated by the power laws in the survival probabilities (\ref{proba32}), (\ref{probaurnr}). However,  in fact, it emerges naturally at longer time scales after the transient dynamics becomes negligible, from the   expression (\ref{Upsiz}) when considered during time intervals $\Delta t = t_f-t_i \ll 1/H(t_i)$, with $H(t_i)$ the   Hubble expansion rate at the time $t_i$. This is seen as follows: keeping only the last term in eqn. (\ref{decfunmd}) (neglecting transients), and passing to comoving time with $C(\eta')d\eta' = dt'$, it follows that
\begin{equation}
\int^{\eta_f}_{\eta_i}\,\Gamma_\Phi(\eta')\,d\eta' = \Gamma_0 \int^{t_f}_{t_i} \frac{dt'}{\gamma(t')} \,.\label{Smatins}
\end{equation} We now expand $\gamma(t')$ around $t_i$, $\gamma(t')=\gamma(t_i) + \gamma'(t_i) (t'-t_i) + \cdots$ and integrate to obtain,
\begin{equation}
 \Gamma_0 \int^{t_f}_{t_i} \frac{dt'}{\gamma(t')} = \frac{\Gamma_0}{\gamma_k(t_i)}\, \Delta t \,\Big[ 1+ \frac{1}{2} \beta^2_k \, H(t_i) \Delta t + \cdots \Big]~~;~~ \beta_k = \frac{k_p}{E_k}\,, \label{correcS}
\end{equation} therefore we clearly see that for time intervals $\Delta t \ll 1/H(t_i)$ the decay law features small departures from the exponential S-matrix inspired one, with corrections of order $\Delta t H(t_i)\ll 1$. This is expected on physical grounds: on very short time scales, much shorter than the Hubble time, a local Minkowski space-time approximation is warranted by the equivalence principle. For example, accelerator experiments \emph{today}  measure the   lifetime of standard model particles, these experiments are obviously insensitive to the Hubble time today $\simeq 13.5\,\mathrm{Gyr}$. Therefore the S-matrix-inspired calculation in Minkowski space time is warranted as it describes the measurement of lifetimes much smaller than $1/H_0$. In the early stages of a radiation dominated cosmology during rapid expansion, or for lifetimes comparable to the Hubble time, such   S-matrix inspired approximation \emph{does not} describe correctly the dynamics of decay, as discussed in detail in the previous sections.

\textbf{Caveats:} We have focused on studying scalar decay into massless fermion pairs, a situation that approximates most of the fermionic decay channels of a Higgs scalar in the standard model. An important aspect of this decay process is that it does not feature thresholds. Including the mass for the decay products introduces kinematic thresholds, a consequence of strict energy-momentum conservation. In ref.\cite{herring} it was argued that the Hubble rate of expansion introduces a natural energy uncertainty leading to a relaxation of the kinematic thresholds, thereby allowing processes that are forbidden in Minkowski space-time by energy conservation. Furthermore, ref.\cite{zenoqft} has shown that energy uncertainties associated with transient non-equilibrium aspects of the decay allow decay into heavier particles during a time interval. In an expanding cosmology these effects may combine with the energy uncertainty from Hubble expansion to \emph{enhance} the decay by opening up novel channels that would be otherwise forbidden by strict energy conservation. These aspects associated with the masses of the decay products will be the subject of further study.

The inclusion of masses for the decay products becomes a more pressing issue in the case of decay of very long lived particles studied in section (\ref{subsec:MD}) where we have extended the results obtained for the (RD) era to provide an \emph{upper bound} on the decay function all throughout the expansion history. Therefore,  the decay law (\ref{probaz}) with the decay function (\ref{upsfunc}) must  be understood within the context of decay of a heavy particle into massless or nearly massless fermionic channels with the caveat that such an approximation may be of limited validity during the (MD) or (DE) eras and should be interpreted as indicative of the decay dynamics.

In this study we have neglected finite temperature corrections to decay vertices and masses, their inclusion requires studying the time evolution of an initial \emph{density matrix}. Furthermore, if the decay products thermalize with the medium, their population build-up will lead to Pauli blocking factors thereby suppressing the decay of the parent particle.  These effects remain to be studied but are beyond the scope and goals of this article.

\vspace{2mm}

\textbf{Possible implications:} The time dependence of the decay function reveals non-equilibrium aspects that have not been previously recognized, not only from the transient build-up of the quasiparticle but also the memory effects that yield the unexpected power laws and stretched exponentials. These novel non-equilibrium effects may lead to interesting and perhaps important dynamics   relevant to baryogenesis and leptogenesis. In particular, we envisage corrections to quantum kinetic processes for particle production and their inverse processes. Typically   quantum kinetics inputs transition rates perhaps with finite temperature contributions but ultimately obtained from S-matrix theory. Namely, such transition rates are obtained in the infinite time limit and the forward and backward probabilities input strict energy conservation, and as a consequence,   obey detailed balance  . The richer time dependence of the decay function revealed by this study, with the hitherto unexplored novel non-equilibrium aspects,  suggests that similar dynamical processes \emph{may} enter in a modified quantum kinetic description in the early universe. We expect to report on these and other related issues in future studies.

\section{Summary, conclusions and further questions. }
In this article we studied the   decay of a   bosonic particle into massless fermions  via a Yukawa coupling in post-inflation cosmology. The approximation of massless fermions is warranted for a heavy Higgs-like scalar within or beyond the standard model decaying into most charged leptons or quarks (but for the top) of the standard model. We implemented  a non-perturbative method that yields the time evolution of the survival probability $\mathcal{P}_{\Phi}(t)$  combined with a physically motivated adiabatic expansion. This expansion is justified when $H(t)/E_k(t) \ll 1$ where $H(t)$ is the Hubble rate and $E_k(t)$ the local energy of the particle as measured by a comoving observer. We have argued that this approximation is valid for  typical particle physics processes during the radiation dominated era and beyond. In a standard cosmology the reliability of this approximation improves with the cosmological expansion, therefore if the adiabatic condition is fulfilled at the initial time when the decaying particle is produced, its reliability  improves along the expansion history.

Particle decay into fermionic channels introduces novel phenomena associated with ultraviolet divergences requiring renormalization that result into two different physical processes: i) the build-up of a \emph{quasiparticle} state out of the bare initial state by dressing with fermion-antifermion pairs, ii) the decay of this quasiparticle state via the emission of fermion pairs. These two different processes occur on widely separated time scales. We introduced a dynamical renormalization method that allows to separate the dynamics of formation of the quasiparticle from its decay on longer time scales.  It relies on introducing a renormalization time scale $t_b$ to absorb the transient dynamics of formation into the wave function renormalization of the quasiparticle state. The survival probability obeys a \emph{dynamical renormalization group equation} with respect to $t_b$. The decay function of this renormalized state is ultraviolet finite and cutoff independent.

We carried out a detailed analytic and numerical study of the decay function during the radiation dominated era. The dynamics of decay depends crucially on whether the particle is non-relativistic or relativistic.  For a particle that is ``born'' at rest in the comoving frame during (RD) we find that after   short time transients,  the survival probability is given by
\be  \mathcal{P}_\Phi(t)  = \Big[\frac{t}{t_b}\Big]^{-\frac{Y^2}{8\pi^2}}~
e^{ \frac{Y^2}{4\pi^2}\,\big(t/t_b\big)^{1/4} }
\,e^{-\Gamma_0\,(t-t_b)}~  \mathcal{P}_\Phi(t_b) ~~;~~ \Gamma_0 = \frac{Y^2}{8\pi}\,m_\phi\,. \label{restP}  \ee where $Y$ is the Yukawa coupling and $\Gamma_0$ is the decay rate at rest in Minkowski space-time. The scale $t_b$ is an intermediate time scale that describes the build-up of the quasiparticle state, and $\mathcal{P}(t_b)$ is the renormalized probability of such state.  The power of $t/t_b$  with ``anomalous'' dimension and the stretched exponential with power $1/4$ are both a remnant of the formation of the quasiparticle on long time scales as a consequence of the \emph{cosmological redshift}.

For the case in which the decaying particle is ``born'' ultrarelativistically the time evolution over the whole history during (RD) must be obtained numerically. Different   regimes emerge depending on whether the particle is ultrarelativistic for $t\ll t_{nr}$ or non-relativistic for $t\gg t_{nr}$ where $t_{nr}= k^2/(2m^2_\phi H_0\sqrt{\Omega_R}) $ is the time scale at which the decaying particle of mass $m_\phi$ that is born ultrarelativistically with comoving momentum $k$ transitions to being non-relativistic  as a consequence of the cosmological redshift. During the ultrarelativistic regime ($t\ll t_{nr}$) we find for $t \gg t_b$ that the decay function is a stretched exponential
 \be  \mathcal{P}_\Phi(t)  =  e^{-\frac{2}{3}\Gamma_0\,t_{nr}\,(t/t_{nr})^{3/2}}~  \mathcal{P}_\Phi(t_b) \nonumber \ee whereas  for $t\gg t_b$ and after the particle has become non-relativistic  ($t \gg t_{nr}$) we find
 \be \mathcal{P}_\Phi(t)  = \Big[\frac{t}{t_{nr}}\Big]^{-\frac{Y^2}{8\pi^2}}~
e^{ \frac{Y^2}{4\pi^2}\,\big(t/t_{nr}\big)^{1/4} }~\Big[\frac{t}{t_{nr}}\Big]^{\Gamma_0 t_{nr}/2}
\,e^{-\Gamma_0\,(t-t_{nr})}~  \mathcal{P}_\Phi(t_{nr}) \,.  \nonumber \ee The extra power of $t/t_{nr}$ as compared to the  case when the particle is born at rest  (see eqn. (\ref{restP})) is a consequence of the  \emph{memory} of the decay function on the past history during the ultrarelativistic stage.

The cosmological decay law is compared to a \emph{phenomenological} Minkowski-like, S-matrix inspired  decay law with an instantaneous Lorentz time dilation factor
\be   \mathcal{P}^{(M)}_\Phi(t) = e^{-\frac{\Gamma_0}{\gamma(t)}\,(t-t_i)}\, \mathcal{P}_\Phi^{(M)}(t_i)\,,   \ee we found that this phenomenological law describes at long times a \emph{much faster decay} thereby \emph{under estimating} the lifetime of the decaying particle.

The decay dynamics revealed by this study during (RD) allows us to extrapolate to the case of very long lived, i.e. very weakly coupled particles. We obtain a decay function that yields an \emph{upper bound} to the survival probability \emph{all throughout the expansion history} under the assumption of two body decay into a massless fermions, it is given by
\be \mathcal{P}_\Phi(z) \gtrsim  e^{-\frac{\Gamma_0}{H_0}\,\Upsilon(z,z_b)}\,\mathcal{P}_\Phi(z_b)\,, \ee where $\Upsilon(z,z_b)$  is given by (\ref{Upsiz}) and  depends \emph{only} on the cosmological parameters and the scale factor at which the particle transitions from ultrarelativistic to non-relativistic.

One important conclusion from these results is that using a \emph{decay rate} as measure of the decay dynamics is not a   useful concept and misses the correct dynamical evolution. An S-matrix calculation of transition amplitudes or probabilities, where the time interval is taken to infinity not only it does not capture the various different dynamical scales and temporal behaviour of the survival probability, but substantially \emph{under estimates}    the lifetime of the decaying state.

An important corollary of this study is that the S-matrix approach to describe quantum decay in the cosmological setting is in general inadequate, while it may yield a good approximation for processes of decay at rest for weakly coupled particles late in the cosmological history,  it misses important non-equilibrium dynamics.  The non-equilibrium effects revealed by our study, from the transient dynamics of the formation to the quasiparticle, to the memory of the decay function on the past history of the decaying particle \emph{could} be relevant in the quantum kinetics of processes in the very early universe. These could have potential impact in CP-violating non-equilibrium dynamics, baryogenesis and leptogenesis and merit further study.

\acknowledgements N.H. acknowledges discussions with A. Zentner. D. B. thanks John Collins for illuminating exchanges on aspects related to renormalization and quasiparticle formation.

\appendix
\section{Projectors} \label{app:pros}
Introducing the notation
\be  \Omega_k(\eta) \equiv \sqrt{\omega^\psi_k(\eta) + M_f(\eta)}~~; ~~\omega^\psi_k(\eta) = \sqrt{k^2+M^2_f(\eta)} \,,  \label{Omegdef}\ee  with the zeroth-adiabatic order spinors (\ref{Uspinorsol},\ref{Vspinorsol}), the projector operators are given by equations (\ref{relaspinors}). We find

 \be \Lambda^+_{\vk}(\eta,\eta') =  \left(
       \begin{array}{cc}
         \Omega_k(\eta)\Omega_k(\eta') \,\mathbb{I} ~ & ~ -\vec{\sigma}\cdot\vk\, \frac{\Omega_k(\eta)}{\Omega_k(\eta')} \\
         \vec{\sigma}\cdot\vk\, \frac{\Omega_k(\eta')}{\Omega_k(\eta)} ~ & ~ -\frac{k^2}{\Omega_k(\eta)\Omega_k(\eta')}\,\mathbb{I} \\
       \end{array}
     \right)\,, \label{Lplu}
  \ee

 \be \Lambda^-_{\vk}(\eta',\eta) =  \left(
       \begin{array}{cc}
        \frac{k^2}{\Omega_k(\eta)\Omega_k(\eta')}\,\mathbb{I}    ~ & ~ -\vec{\sigma}\cdot\vk\, \frac{\Omega_k(\eta)}{\Omega_k(\eta')} \\
         \vec{\sigma}\cdot\vk\, \frac{\Omega_k(\eta')}{\Omega_k(\eta)} ~ & ~ -\Omega_k(\eta)\Omega_k(\eta') \,\mathbb{I} \\
       \end{array}
     \right)\,, \label{Lmin}
  \ee  where $\mathbb{I} $ is the $2\times 2$ identity matrix. These expressions can be written more compactly introducing the following functions (suppressing the momentum and conformal time arguments),
  \bea \lambda_0 & = &  \frac{1}{2} \Big( \Omega_k(\eta)\Omega_k(\eta')+ \frac{k^2}{\Omega_k(\eta)\Omega_k(\eta')}\Big) \label{lambda0}\\
\lambda_1 & = &  \frac{1}{2}\Big( \frac{\Omega_k(\eta)}{\Omega_k(\eta')} + \frac{\Omega_k(\eta')}{\Omega_k(\eta)} \Big) \label{lambda1} \\
\lambda_2 & = &  \frac{1}{2}\Big( \frac{\Omega_k(\eta)}{\Omega_k(\eta')} - \frac{\Omega_k(\eta')}{\Omega_k(\eta)} \Big) \label{lambda2}\\
\lambda_3 & = &   \frac{1}{2} \Big( \Omega_k(\eta)\Omega_k(\eta')- \frac{k^2}{\Omega_k(\eta)\Omega_k(\eta')}\Big)    \label{lambda3}\,, \eea  as
\bea  \Lambda^+_{\vk}(\eta,\eta')  & = & \gamma^0\,\lambda_0 -\vec{\gamma}\cdot \vk \,\lambda_1 +  \vec{\gamma}\cdot \vk\,\gamma^0 \,\lambda_2 + \lambda_3 \label{Lplus} \\
\Lambda^-_{\vk}(\eta',\eta)  & = & \gamma^0\,\lambda_0 -\vec{\gamma}\cdot \vk \,\lambda_1 +  \vec{\gamma}\cdot \vk\,\gamma^0 \,\lambda_2 - \lambda_3 \label{Lmini}\,.  \eea

\vspace{2mm}

\textbf{Two relevant cases:} 1:) Equal time, $\eta= \eta'$,
\bea  \Lambda^+_{\vk}(\eta,\eta) & = & \gamma^0 \omega^\psi_k(\eta) - \vec{\gamma}\cdot \vk + M_f(\eta) \nonumber \\ & = & a(t) \Big[ {\not\!{K}}(t) + m_f\Big] ~~;~~ K_\mu(t)= (E^\psi_k(t), - \vec{k}_p(t)) \label{Lpluseqeta}\,,\eea

 \bea  \Lambda^-_{\vk}(\eta,\eta) & = & \gamma^0 \omega^\psi_k(\eta) - \vec{\gamma}\cdot \vk - M_f(\eta) \nonumber \\ & = & a(t) \Big[ {\not\!{K}}(t) - m_f\Big]  \label{Lmineqeta}\,. \eea
 2:) Massless fermions,
 \be \Lambda^+_{\vk} = \Lambda^-_{\vk} =   \gamma^0 k - \vec{\gamma}\cdot \vk \,. \label{Lambdasmassless}\ee

 \section{Minkowski space-time: $m_\psi =0$} \label{app:Minko}
The Minkowski space-time limit is obtained by replacing $\eta \rightarrow t$ and the frequencies are time independent. The self-energy in this case becomes\cite{zenoqft}
\be \Sigma_{\chi}(k,t,t') = \frac{Y^2}{16\pi^2}\,\frac{e^{i\omega^\phi_k(t-t')}}{\omega^\phi_k}\int dk_0 \rho(k_0,k) e^{-ik_0(t-t')}~~;~~ \rho(k_0,k) = (k^2_0-k^2)\,\Theta(k_0-k)\,. \label{sigmin}\ee Replacing $k^2_0 \rightarrow -d^2/dt^{\,'\,2}$, and introducing a  convergence factor $\epsilon \rightarrow 0^+$  yields

\be \Sigma_{\chi}(k,t,t')   = -i\frac{Y^2}{16\pi^2}\,\,\frac{e^{i\omega^\phi_k\,(t-t')}}{\,\omega^\phi_k}\,\Bigg[\frac{d^2}{dt^{\,'\,2}}+k^2 \Bigg]\,
\Bigg[\frac{e^{-i\Lambda(t-t'-i\epsilon)}-e^{-ik(t-t'-i\epsilon)} }{(t-t'-i\epsilon)} \Bigg]\,,  \label{sigchifinminko}\ee and the decay function
\be \int^t_0 \Gamma_k(t')\, dt' = 2 \int^t_0 \Bigg\{\int^{t'}_0 \mathrm{Re}\Big[ \Sigma_{\chi}(k,t',t^{''}\Big] dt^{''}\Bigg\}\,dt' \,. \label{decminko} \ee Integrating by parts twice the derivative term in (\ref{sigchifinminko}) we find
\be \int^t_0 \Gamma^\phi_k(t')\, dt' = \frac{Y^2}{8\pi^2\,\omega^\phi_k} \Bigg[T_1(\Lambda,k,t)+T_2(\Lambda,k,t)+T_3(k,t) \Bigg]\,,\label{3Ts}\ee where
\be T_1(\Lambda,k,t)= \frac{1}{\epsilon}\Big(e^{(\omega^\phi_k-k)\epsilon}-e^{(\omega^\phi_k-\Lambda)\epsilon} \Big) - \frac{\sin\Big((\Lambda-\omega^\phi_k)t\Big)}{t}- \frac{\sin\Big((\omega^\phi_k-k)t\Big)}{t}\,,\label{T1} \ee
\be T_2(\Lambda,k,t) =   2\,\omega^\phi_k \int^t_0 \Bigg[\frac{1-\cos\Big((\Lambda-\omega^\phi_k)t'\Big)}{t'}-\frac{1-\cos\Big((k-\omega^\phi_k)t'\Big)}{t'} \Bigg] \,dt' \label{T2} \ee

\be T_3(k,t) =  m^2_\phi \int^t_0 \Bigg[Si\Big((\Lambda-\omega^\phi_k)\,t' \Big)+ Si\Big(( \omega^\phi_k-k)\,t' \Big) \Bigg] dt'\,,\label{T3} \ee where $Si(t) = \int^t_0 \sin(x)/x dx$. Taking $\epsilon \rightarrow 0^+$ and keeping $\Lambda$ large but finite yields
\be T_1(\Lambda,k,t) = (\Lambda-k)\Bigg[ 1 - \frac{\sin\Big((\Lambda-\omega^\phi_k)\,t\Big)}{(\Lambda-k)\,t}- \frac{\sin\Big((\omega^\phi_k-k)\,t\Big)}{(\Lambda-k)\,t}\Bigg]\,, \label{T1fina}\ee
\be T_2(\Lambda,k,t) =  2\,\omega^\phi_k \Bigg[\ln\Big( \frac{\Lambda-\omega^\phi_k}{\omega^\phi_k-k}\Big) - Ci\big[(\Lambda-\omega^\phi_k)\,t \big] + Ci\big[(\omega^\phi_k-k)\,t \big]\Bigg]\,,\label{T2fina}\ee and
\be T_3(k,t)= m^2_\phi\Bigg\{t\Big[Si\big[(\Lambda-\omega^\phi_k)\,t \big]+Si\big[(\omega^\phi_k-k)\,t \big] \Big]- \frac{\Big[1-\cos\big[(\Lambda-\omega^\phi_k)\,t  \big] \Big] }{(\Lambda-\omega^\phi_k)} - \frac{\Big[1-\cos\big[(\omega^\phi_k-k)\,t  \big] \Big] }{(\omega^\phi_k-k)}  \Bigg\} \,, \label{T3fina}\ee where for $\Lambda t \rightarrow \infty$ it follows that $Si\big[(\Lambda-\omega^\phi_k)\,t \big]\rightarrow \pi/2$, and $Ci\big[(\Lambda-\omega^\phi_k)\,t \big]\rightarrow 0$. Taking   the limit $\Lambda \rightarrow \infty$ yields
\bea \int^t_0 \Gamma^\phi_k(t') dt'  & = &   \frac{Y^2}{8\pi^2\,\omega^\phi_k} \Bigg\{ \Lambda-k +  2\,\omega^\phi_k  \ln\Bigg[ \frac{\Lambda}{\omega^\phi_k-k}\Bigg]    + m^2_\phi \,t \Big[ \frac{\pi}{2} + Si\big[(\omega^\phi_k-k)\,t \big]\Big] + Ci\big[(\omega^\phi_k-k)\,t \big] \nonumber \\ & - &  \frac{\Big[1-\cos\big[(\omega^\phi_k-k)\,t  \big] \Big]}{(\omega^\phi_k-k)}+ \mathcal{O}(1/t)  \Bigg\}\,. \label{decminfina}  \eea
This is exactly the same result as obtained in ref.\cite{zenoqft} integrating in $k_0$ first.

 \section{Useful identities:}\label{app:identities}
In this appendix we gather some useful identities valid during the radiation dominated stage (see also appendix (\ref{app:I2})).
\bea \omega(\eta) & = &  \Big[k^2+ m^2 \, H^2_R\,\eta^2 \Big]^{1/2} = \Big[k^2+ m^2 \, H^2_R\,\eta^2_i + m^2\,H^2_R\,(\eta^2-\eta^2_i) \Big]^{1/2} \equiv \frac{\omega_i}{\gamma_i}\,\Big[\gamma^2_i -1 + \frac{\eta^2}{\eta^2_i}\Big]^{1/2} \nonumber \\ \omega_i  & = &  \omega(\eta_i)~~;~~ \gamma_i = \gamma(\eta_i) \,.  \label{ome2id}\eea The local Lorentz factor in conformal time is given by

\bea \gamma(\eta) &  =  & \Bigg[\frac{(\gamma^2_i-1)}{\Big(\frac{\eta}{\eta_i}\Big)^2} + 1  \Bigg]^{1/2}= \Bigg[\frac{k^2}{m^2\,H^2_R\,\eta^2} + 1  \Bigg]^{1/2} \equiv  \Big[\frac{\eta^2_{nr}}{\eta^2} + 1  \Big]^{1/2} = \Big(\frac{\eta_i}{\eta}\Big) \,\Big[\gamma^2_i -1 +  \frac{\eta^2}{\eta^2_i} \Big]^{1/2}  \nonumber \\ \gamma^2_i & = &  1+ \frac{\eta^2_{nr}}{\eta^2_i}~~;~~ \eta_{nr} = \frac{k}{m\,H_R} = \eta_i\,\sqrt{\gamma^2_i -1}\,,  \label{gamma2id} \eea  yielding the identity
\be \gamma^2(\eta)-1 = \Big(\frac{\eta_i}{\eta}\Big)^2\,(\gamma^2_i-1)\,.  \label{gamaid}\ee

  The relationship with comoving time $t$ is obtained via eqn. (\ref{etaoft}), namely
\be \gamma(\eta(t)) = \Bigg[\frac{(\gamma^2_i-1)}{\big(\frac{t}{t_i}\big)} + 1  \Bigg]^{1/2} \equiv \Bigg[\frac{t_{nr}}{t} + 1  \Bigg]^{1/2}\,.  \label{gamaoft}\ee The conformal and comoving time scales $\eta_{nr}\,, t_{nr}$ respectively,  determine  the scale at which the decaying particle transitions from relativistic with $\gamma(\eta) \gg 1$ for $ \eta \ll \eta_{nr}$ or $t \ll t_{nr}$  to non-relativistic with $\gamma(\eta) \simeq 1 $ for $\eta \geq  \eta_{nr}  $ or $t \geq t_{nr}$.
In terms of $\eta,\eta_i$ we find,

\be \omega(\eta)\,\eta = \frac{\omega_i\,\eta_i}{\gamma_i}\,\Big(\frac{\eta}{\eta_i} \Big) \, \Big[\gamma^2_i -1 +  \frac{\eta^2}{\eta^2_i} \Big]^{1/2} \,. \label{ometaid}\ee

\vspace{5mm}

 \section{Analysis of  $I_2$: eqn. (\ref{I2})  }\label{app:I2}

 Consider the first term in $I_2$ eqn. (\ref{I2}),
 \be  I_{2,a}(\Lambda,k,\eta)  =    \int^\eta_{\eta_i}\Bigg[\sqrt{\frac{\omega^\phi_k(\eta')}{\omega^\phi_k(\eta_i)}}+
\sqrt{\frac{\omega^\phi_k(\eta_i)}{\omega^\phi_k(\eta')}}\, \Bigg]\,\times
\Bigg[\frac{1-\cos\Big(\int^{\eta'}_{\eta_i}
\big(\omega^\phi_k(\eta_1)-\Lambda\big)d\eta_1 \Big)}{\eta'-\eta_i}   \Bigg]\,d\eta'  \,,\label{I2a} \ee For $\Lambda \gg k,m_\phi$ the argument of the cosine becomes simply $ \Lambda\,(\eta'-\eta_i)$. Define $x = \Lambda\,(\eta'-\eta_i)$, and change integration variable to $x$, with $x_f = \Lambda\,(\eta-\eta_i)$, yielding
\be  I_{2,a}(\Lambda,k,\eta)  =    \int^{x_f}_{0}\Bigg[\sqrt{\frac{\omega^\phi_k(\eta_i+x/\Lambda)}{\omega^\phi_k(\eta_i)}}+
\sqrt{\frac{\omega^\phi_k(\eta_i)}{\omega^\phi_k(\eta_i+x/\Lambda)}}\, \Bigg]\,\times
\Bigg[\frac{1-\cos(x)}{x}   \Bigg]\,dx  \,.\label{I2xi} \ee In the limit $\Lambda \rightarrow \infty$ we find
\be I_{2,a}\rightarrow   2\big[\ln(x_f )+\gamma_E-Ci(x_f )\big]  \label{asyI2a}\ee where $\gamma_E= 0.577\cdots$ is Euler's constant and  for $x_f \gg 1$ we find $ Ci(x_f)= \sin(x_f)/x_f +\cdots  $.
    We confirmed the result (\ref{asyI2a}) numerically.  Therefore for $\Lambda \gg k, m_\phi,1/(\eta-\eta_i)$ we find
\be I_{2,a} = 2\,\ln\big[\Lambda\,e^{\gamma_E}\,(\eta-\eta_i)\big]\,.  \label{asyI2afin}\ee

Let us now consider
 \be  I_{2,b}(k,\eta)  =   - \int^\eta_{\eta_i}\Bigg[\sqrt{\frac{\omega^\phi_k(\eta')}{\omega^\phi_k(\eta_i)}}+
\sqrt{\frac{\omega^\phi_k(\eta_i)}{\omega^\phi_k(\eta')}}\, \Bigg]\,\times
\Bigg[\frac{1-\cos\Big(\int^{\eta'}_{\eta_i}
\big(\omega^\phi_k(\eta_1)-k\big)d\eta_1 \Big)}{\eta'-\eta_i}   \Bigg]\,d\eta'  \,.\label{I2b} \ee

 Using the identities obtained in appendix (\ref{app:identities}) for a radiation dominated cosmology,  we write
 \bea   \omega^\phi_k(\eta)    & = &       \sqrt{k^2+ m^2_\phi H^2_R\,\eta^2} = \Bigg\{k^2+ m^2_\phi H^2_R\,\eta^2_i + m^2_\phi H^2_R\,\eta^2_i \Big[\Big(1 + \frac{\eta-\eta_i}{\eta_i}\Big)^2 -1 \Big]\Bigg\}^{1/2}\nonumber \\ & = & \omega_i\,W[\xi] \,, \label{Wdef}\eea where we introduced the definitions
  \be W[\xi] =  \frac{1}{\gamma_i}\Bigg[\gamma^2_i-1+(1+\xi)^2 \Bigg]^{1/2}   \label{W} \,,  \ee
 \be \omega_i \equiv \omega^\phi_k(\eta_i)~~;~~   \xi =  \Big(\frac{\eta}{\eta_i}-1\Big)~~;~~ \gamma_i = \frac{\omega_i}{m_\phi\,H_R\,\eta_i}\equiv \frac{E^\phi_k(t_i)}{m_\phi}\,,  \label{defs} \ee and
  $\gamma_i$ is  the local Lorentz factor at   time   $\eta_i$.

 In terms of these variables  we find
\bea  J[\xi']\equiv \int^{\eta'}_{\eta_i}
 \Big[\omega^\phi_k(\eta_1)-k\Big]\,d\eta_1   & = &  \frac{\omega_i\,\eta_i}{2} \,\Bigg\{\big(1+   \xi' \big)W[\xi'] -1   - 2\,\frac{\xi'}{\gamma_i}\,\sqrt{\gamma^2_i-1}  \nonumber \\ & + &
 \frac{\big(\gamma^2_i -1 \big)}{  \gamma_i}\,\ln\Bigg[\frac{\gamma_i\,W[\xi']+1 + \xi'}{1+\gamma_i} \Bigg]\Bigg\}\,.  \label{Jofz} \eea
   We note that the fulfillment of the adiabatic condition at all times implies that
 \be \omega_i\,\eta_i = \frac{E^{\phi}_k(\eta_i)}{H(\eta_i)} \gg 1 \,.\label{adiaini}\ee  For $\xi' \ll 1$ it is straightforward to find that    $J[\xi']$ features the expansion
 \be J[\xi'] = \omega_i\,\eta_i \, \xi' \,\Bigg[ 1-\sqrt{1-\frac{1}{\gamma^2_i}} +\frac{1}{2}\,\frac{\xi'}{\gamma^2_i} + \cdots \Bigg]\,. \label{Jxiexp} \ee

 In terms of these variables we find that the subtracted integral $I_{2b}(k,\eta,\eta_b)$  defined by eqn. (\ref{ISubfin}) is given by
 \be I_{2,b}[k,\eta,\eta_b] = -\int^{\xi}_{\xi_b} \Bigg[\sqrt{W[\xi']}+\frac{1}{\sqrt{W[\xi']}} \Bigg]\,\Bigg[\frac{1-\cos[J(\xi')]}{\xi'} \Bigg] \,d\xi'  \,.\label{I2bfin} \ee

Consider the two contributions to this function,
\bea F_1(\xi)  & = &  \int^{\xi}_{\xi_b}   \Bigg[\sqrt{W[\xi']}+\frac{1}{\sqrt{W[\xi']}} \Bigg]\,   \frac{d\xi'}{\xi'}       \, \label{F1sub} \\
  F_2(\xi)  & = &  \int^{\xi}_{\xi_b} \Bigg[\sqrt{W[\xi']}+\frac{1}{\sqrt{W[\xi']}} \Bigg]\, \frac{\cos[J(\xi')]}{\xi'}    \,d\xi' \,.\label{F2sub}\eea

\begin{figure}[ht!]
\begin{center}
\includegraphics[height=3.5in,width=4.5in,keepaspectratio=true]{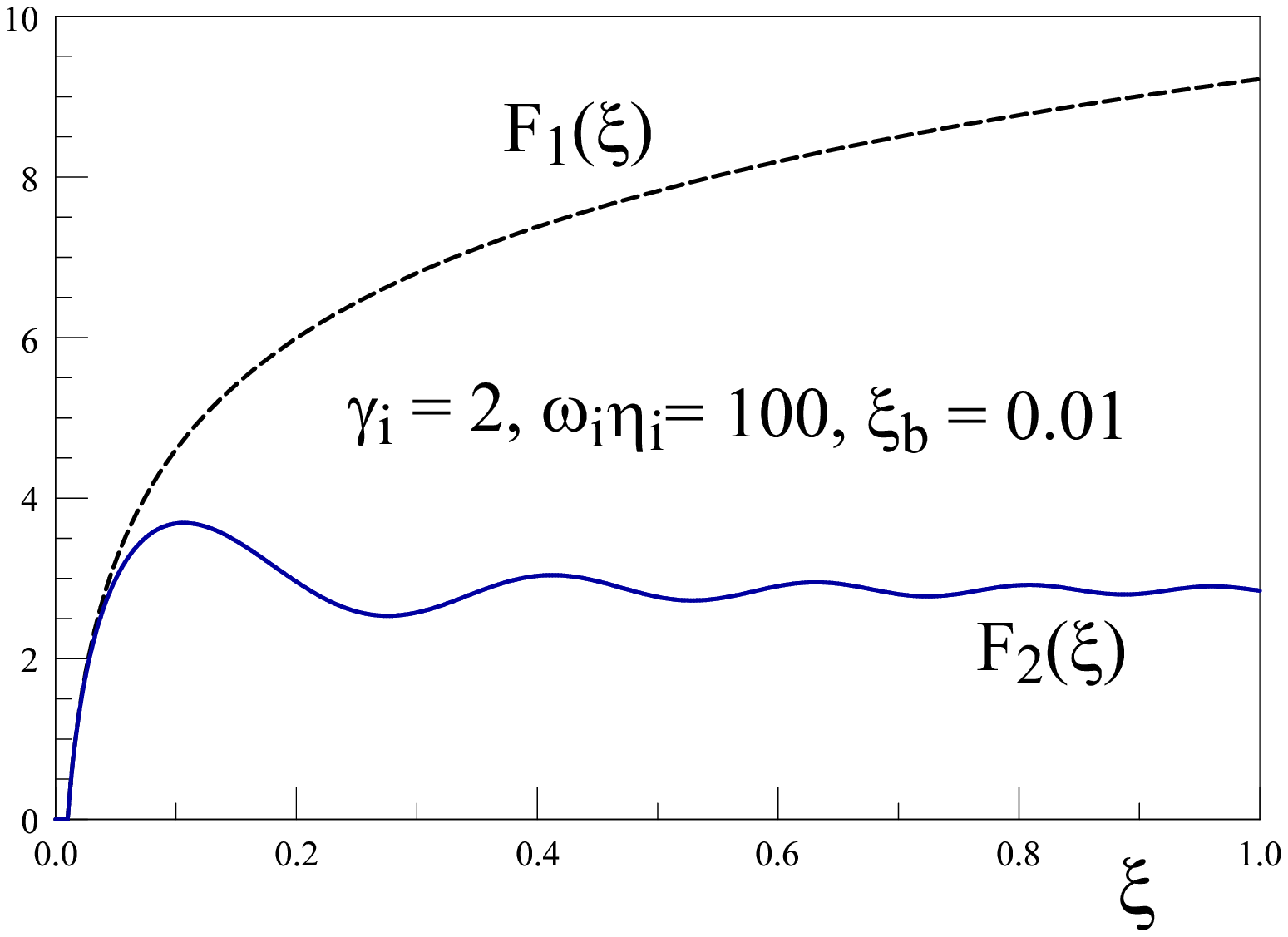}
\caption{The contributions $F_1(\xi),F_2(\xi)$, for $ {\xi_b}=0.01, \gamma_i = 2, \omega_i \eta_i =100$. }
\label{fig:f12g2}
\end{center}
\end{figure}

\begin{figure}[ht!]
\begin{center}
\includegraphics[height=3.5in,width=4.5in,keepaspectratio=true]{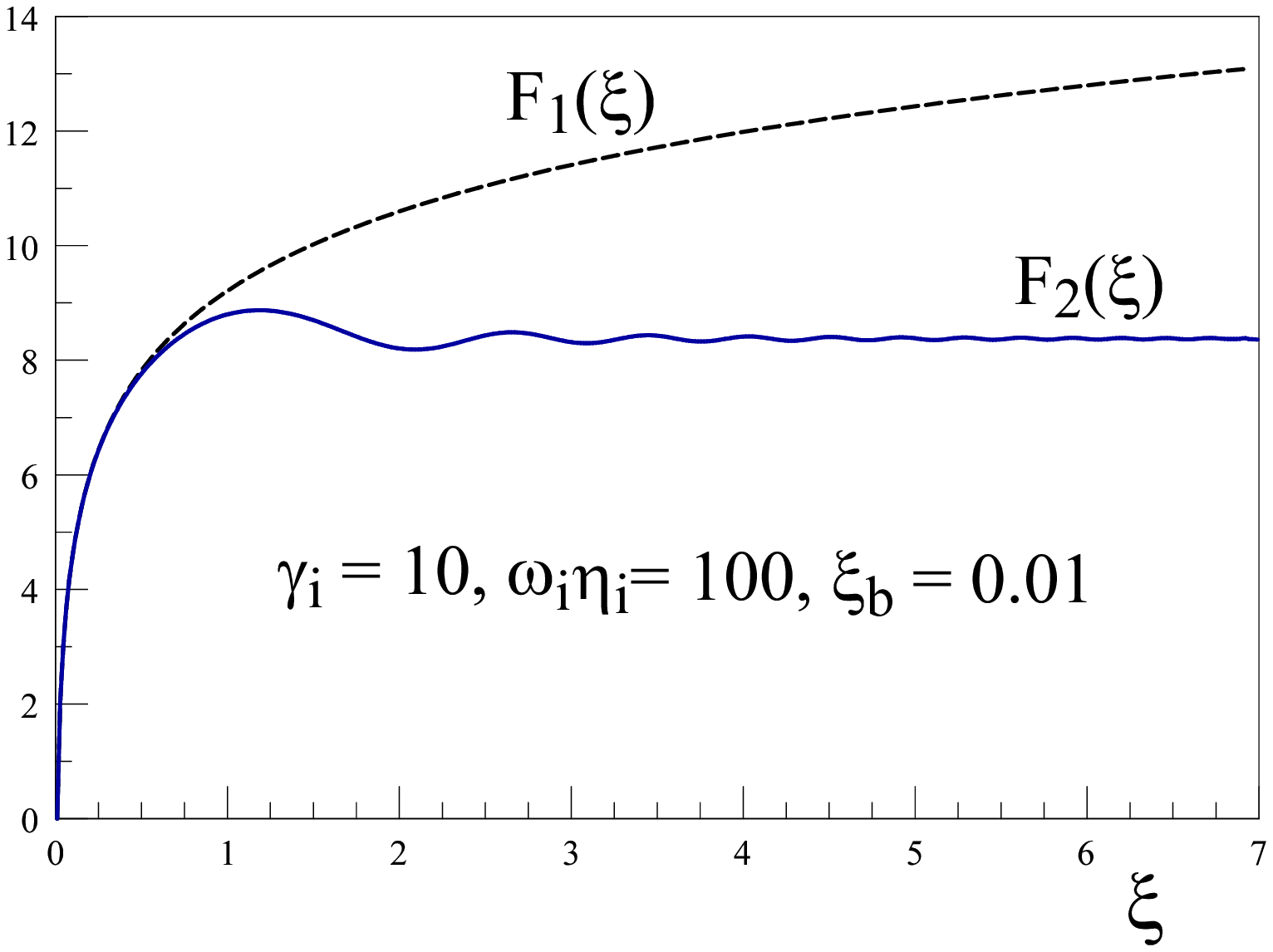}
\caption{The contributions $F_1(\xi),F_2(\xi)$, for $ {\xi_b}=0.01, \gamma_i = 10, \omega_i \eta_i =100$. }
\label{fig:f12g10}
\end{center}
\end{figure}

  During the time scale when $J(\xi') \ll  1$  the term $\cos[J(\xi')] \simeq 1$ therefore $F_2(\xi) \simeq F_1(\xi)$ and $I_{2b} \simeq 0$. Figures (\ref{fig:f12g2}, \ref{fig:f12g10}) display $F_{1,2}(\xi)$ for $\omega_i \eta_i =100$ and $\gamma_i = 2,10$ respectively for $\xi_b = 1/\omega_i\eta_i$. $F_2(\xi)$ grows up to a maximum at $\xi_m$ at which $J(\xi_m) = \pi/2$ and begins damped oscillations reaching a plateau. During the rise-time up to the maximum $F_2(\xi) \simeq F_1(\xi)$ thereby yielding $I_{2b}(k,\eta,\eta_b) \simeq 0$ during the interval $\xi_b \leq \xi \lesssim \xi_m$.

\begin{figure}[ht!]
\begin{center}
\includegraphics[height=3.5in,width=4in,keepaspectratio=true]{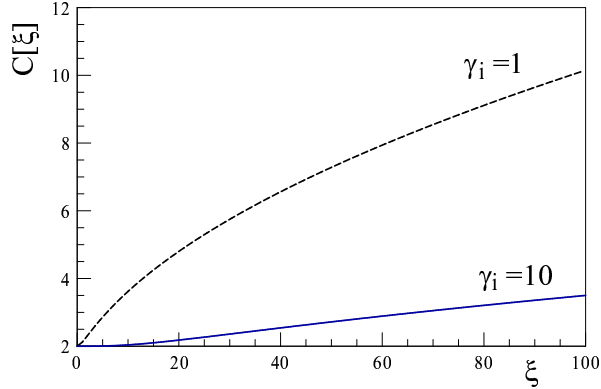}
\caption{The  function  $C[\xi]=W^{1/2}[\xi]+ 1/W^{1/2}[\xi]$ vs $\xi$, for $  \gamma_i = 1,10$. }
\label{fig:cofxi}
\end{center}
\end{figure}

  Although in general the value of $\xi_m$ must be obtained numerically, for $\omega_i \eta_i \gg 1$ there are two limits that afford an analytic estimate: \textbf{a:)} for $\omega_i \eta_i \gg 1$ and $\gamma_i \simeq 1$, we assume, self consistently that $\xi_m \ll 1$, therefore from  eqn. (\ref{Jxiexp})   we
  obtain
  \be  \xi_m \simeq  \frac{\pi}{2}\,\Bigg\{\omega_i \eta_i \Bigg[ 1-\sqrt{1-\frac{1}{\gamma^2_i}} \Bigg]\Bigg\}^{-1} ~~\mathrm{for}~~ \gamma_i \simeq 1  \,,   \label{chimax1}\ee this expression confirms the assumption that $\xi_m \ll 1$ for $\gamma_i \simeq 1$.  \textbf{b:)} for $\gamma_i \gg 1$, it is convenient to carry out the integral (\ref{Jofz}) by expanding $\omega^\phi_k(\eta_1) \simeq k + m^2_\phi\,C^2(\eta_1)/k +\cdots $ and keeping the leading order term, we find
  \be \xi_m \simeq  \Bigg\{\Bigg[1+   \frac{3\,\pi \gamma^2_i}{\omega_i\,\eta_i} \Bigg]^{\frac{1}{3}}-1 \Bigg\} ~~\mathrm{for}~~ \gamma_i   \gg 1\,.   \label{chimax2} \ee This latter expression is fairly accurate even for $\gamma_i \simeq 2,3$.   We have confirmed numerically the validity of  these approximate values of the maxima of $F_2(\xi)$  (see figures (\ref{fig:f12g2},\ref{fig:f12g10})). In both cases we find that for $\omega_i\eta_i \gg 1$ the value at the maxima fulfill $\xi_m/\gamma_i \ll 1$.  In summary, we find that during  the time interval  $\xi_b < \xi < \xi_m$   $F_1(\xi) \simeq F_2(\xi)\simeq 2 \ln[\xi/\xi_b]$ and $I_{2b}[k,\eta,\eta_b] \simeq 0$. For $\xi > \xi_m$ the contribution $F_2(\xi) \simeq F_2(\xi_m)\simeq 2\ln[\xi_m/\xi_b]$ remaining constant while $F_1(\xi)$ increases monotonically. The above analysis shows that for $\omega_i \eta_i \gg 1$ it follows that $\xi_m \ll \gamma_i$  in the whole range of $\gamma_i$, therefore during the interval $\xi_m < \xi  < \gamma_i$  and $W[\xi'] \simeq 1$, hence
  \be F_1(\xi) \simeq F_1(\xi_m) + 2\,\ln\Big[\frac{\xi}{\xi_m} \Big]~~;~~\xi_m < \xi < \gamma_i \,, \label{F1xiint}\ee with $F_1(\xi_m) \simeq F_2(\xi_m)\simeq 2\ln[\xi_m/\xi_b]$.  For $\xi \gg   \gamma_i$ the function $ \sqrt{W[\xi]}+\frac{1}{\sqrt{W[\xi]}} \geq  2$ as shown in fig.   (\ref{fig:cofxi}) hence $F_1[\xi] > 2 \,\ln[\xi]$,  with asymptotic behavior
  \be F_1[\xi] \simeq  2\ln\Big[\frac{\gamma_i}{\xi_b} \Big] + 2 \,\sqrt{\frac{\xi}{\gamma_i}}~~;~~ \mathrm{for}~\xi \gg \gamma_i \,. \label{F1asy2}\ee

  The behavior of $F_{1,2}$ in the ultrarelativistic case $\gamma_i \gg 1$ is summarized as follows:
  \bea F_1[\xi] &\simeq & F_2[\xi] \simeq 2\ln\Big[\frac{\xi}{\xi_b} \Big] ~~ \mathrm{for} ~~ \xi_b \lesssim \xi \lesssim \xi_m \,,\nonumber \\
    F_1[\xi] & \simeq &   2\ln\Big[\frac{\xi}{\xi_b} \Big] ~~;~~ F_2[\xi]\simeq 2\ln\Big[\frac{\xi_m}{\xi_b} \Big] ~~ \mathrm{for} ~~ \xi_m \lesssim \xi \lesssim \gamma_i  \,,\nonumber \\
  F_1[\xi] & \simeq &   2\ln\Big[\frac{\gamma_i}{\xi_b} \Big] +  2\sqrt{\frac{\xi}{\gamma_i} } ~~;~~ F_2[\xi]\simeq 2\ln\Big[\frac{\xi_m}{\xi_b} \Big] ~~ \mathrm{for} ~~ \xi  \gg \gamma_i  \,.\label{fif2behaviorur}\eea

\section{ Analysis of  $I_{3b}(k,\eta,\eta_b)$: eqn. (\ref{I3})}\label{app:I3}

Let us now consider the following integral in $I_{3b}$, namely the second contribution to eqn.  (\ref{I3}):
\be \mathcal{I}(\eta') =  \int^{\eta'}_{\eta_i}\,
\frac{C^2(\eta_1)}{\sqrt{\omega^\phi_k(\eta_1)}}~  \frac{\sin\Big(\int^{\eta'}_{\eta_1}
\big(\omega^\phi_k(\eta_2)-k\big)d\eta_2 \Big)}{\eta'-\eta_1} \,  d\eta_1 \,, \label{Int}  \ee
this integral is similar to the case of decay into bosonic particles studied in ref.\cite{herring}.

Following the treatment in this reference we introduce the following definitions
\be \omega^{\phi}_k(\eta')\, \eta'  = z(\eta') \gg 1 \,, \label{zetap} \ee along with
\be \omega^{\phi}_k(\eta')\,(\eta'-\eta_2) = x ~~;~~  \omega^{\phi}_k(\eta')\,(\eta'-\eta_1) = \tau \,. \label{ttau}\ee In terms of these variables, it follows that
\be \omega^\phi_k(\eta_2) = \omega^\phi_k(\eta')\,R[x;z]\,, \label{omerel}\ee with
\be R[x;z] =   \Bigg[1- \frac{2x}{z\,\gamma^2}+ \frac{x^2}{z^2\,\gamma^2}\Bigg]^{1/2} \,, \label{Rfun}\ee  where there is an implicit $\eta'$ dependence in $z$ and $\gamma$.

The argument of the sine function in (\ref{Int}) becomes
\be A(\tau,\eta') = \int^\tau_0 R[x;z]\, dx - \frac{k\,\tau}{\omega^{\phi}_k(\eta')} = \tau\,\Bigg[1- \Big(1-\frac{1}{\gamma^2}\Big)^{1/2}\,  \Bigg] + \delta_k(\tau;\eta')\,, \label{aoftau}\ee with
\be  \delta(\tau;\eta')  =   \frac{z}{2} \Bigg\{\Big(1-\frac{2\,\tau}{z} \Big)- \Big(1-\frac{\tau}{z} \Big)\, R[\tau;z]
 - \frac{(\gamma^2-1)}{\gamma} \,\ln \Bigg[\frac{\gamma\, R[\tau;z] + \big(1-\frac{\tau}{z} \big)}{1+\gamma }  \Bigg] \Bigg\}      \,, \label{deltada}\ee where $z\equiv z(\eta')~;~\gamma\equiv \gamma_k(\eta')$ .  Writing
 \be \frac{C^2(\eta_1)}{\sqrt{\omega^\phi_k(\eta_1)}} = \frac{C^2(\eta')}{\sqrt{\omega^\phi_k(\eta')}}\, {P}[\eta',\eta_1] \,,
 \label{Pfun}\ee and using (\ref{omerel}) it is straightforward to find
 \be  {P}[\tau;\eta'] = \frac{\Big[1-\frac{\tau}{z} \Big]^2}{\sqrt{R[\tau;z]}} \,.\label{poftau} \ee We finally obtain
 \be \mathcal{I}(\eta') = \frac{C^2(\eta')}{\sqrt{\omega^\phi_k(\eta')}}\, \int^{\widetilde{z}}_0 \mathcal{P}[\tau;\eta'] ~~ \frac{\sin[A(\tau,\eta')]}{\tau}\,d\tau  \, ,\label{Intv2}\ee where
 $\widetilde{z} = \omega^{\phi}_k(\eta')\, (\eta'-\eta_i) $. Gathering together this result with eqn. (\ref{I3atime}) for $I_{3a}$ we find
 \be I_3(k,\eta) = \frac{\pi}{2}\,m_\phi \int^\eta_{\eta_i} \frac{C(\eta')}{\gamma_k(\eta')}\Big[ 1+ \mathcal{S}(\eta')\Big]\, d\eta'\,, \label{totI3}\ee where
 \be \mathcal{S}(\eta') = \frac{2}{\pi}\, \int^{\widetilde{z}}_0  {P}[\tau;\eta'] ~~ \frac{\sin[A(\tau,\eta')]}{\tau}\,d\tau ~~;~~ \widetilde{z} = \omega^{\phi}_k(\eta')\, (\eta'-\eta_i) \,. \label{capS} \ee

  For $\eta'\gg \eta_i$ and $z(\eta') \gg 1$ the integral in (\ref{Intv2}) has an adiabatic expansion, for $\tau \ll z$ we find
 \be  \delta_k(\tau;\eta')   =    - \frac{\tau^2}{2\gamma^2\,z}\,   + \cdots \,,\label{delad} \ee therefore $\delta_k$ is of adiabatic order one and higher, furthermore,
\be   R[\tau;z]   =   1 - \frac{\tau}{z\,\gamma^2}+ \cdots \label{Rad} \ee and to leading (zeroth) adiabatic order we can replace $\mathcal{P}[\tau;\eta'] =1$. The $\tau$ integral in (\ref{Intv2}) is dominated by the region $\tau \simeq 0$, and the region for which $\tau \simeq z$ yields a contribution $\propto 1/z$, hence of first adiabatic order  or smaller. Therefore to leading (zeroth) adiabatic order we neglect $\delta_k$ in  (\ref{aoftau}) and replace $ {P} \rightarrow 1$ in (\ref{Intv2}).

Although the variables  (\ref{zetap},\ref{ttau}) allow an  explicit  identification of the nature of the adiabatic expansion, the    most suitable variables to merge the results for $I_{3b}$ with those of the contributions from $I_{2b}$ are those introduced in appendices (\ref{app:identities}) and (\ref{app:I2}).  We now recast the results for $I_{3b}$ in terms of these variables. Introducing
\be t = \frac{\eta'-\eta_1}{\eta_i}~~;~~ y = \frac{\eta'}{\eta_i} = 1+ \xi' \,,\label{tyvars}\ee in terms of which we find using (\ref{ome2id})
\be \omega^\phi_k(\eta')= \frac{\omega_i}{\gamma_i}\,f(y) ~~;~~ f(y) = \sqrt{\gamma^2_i-1+y^2}\,.  \label{fofy}\ee  Similarly, using (\ref{gamaid}) we obtain
\be \gamma_k(\eta') \equiv \gamma(y) = \sqrt{\frac{\big( \gamma^2_i-1\big)}{y^2} +1} = \frac{f(y)}{y}\,, \label{gamanew}\ee and  the variables $z,\tau$ introduced in eqns. (\ref{zetap})) and (\ref{ttau}) respectively  are given by
\be z(\eta')= \frac{\omega_i\,\eta_i}{\gamma_i}\,f(y)\,y ~~;~~ \tau = \frac{\omega_i\,\eta_i}{\gamma_i}\,f(y)\,t \,, \label{zetataunew}\ee which fulfill the identity
\be \frac{z(\eta')}{\gamma_k(\eta')}\,(\gamma^2_k(\eta')-1) = \frac{\omega_i \,\eta_i}{\gamma_i}\,(\gamma^2_i -1)\,. \label{idenew}\ee Using these results we find
\be  R[\tau,z] \equiv \mathcal{R}[t,y] = \frac{1}{\gamma(y)}\Bigg[\frac{(\gamma^2_i-1)}{y^2}+\Big(1-\frac{t}{y}\Big)^2 \Bigg]^{1/2}\,, \label{capR}\ee and the ratio (\ref{poftau}) becomes
\be  {P}[\tau,\eta'] \equiv \mathcal{P}[t,y] = \frac{\Big(1-\frac{t}{y}\Big)^2}{\sqrt{\mathcal{R}[t,y]}}\,,  \label{ratioP}\ee and $\delta(\tau,\eta')$ in eqn. (\ref{deltada}) becomes
\bea \delta(\tau,\eta') \equiv &   \Delta[t,y] &= \frac{\omega_i\,\eta_i}{\gamma_i} \Bigg\{
y\,f(y)\,\Bigg[\Big(1-2\frac{t}{y} \Big)- \Big(1-\frac{t}{y} \Big)\,\mathcal{R}[t,y] \Bigg]\nonumber \\ & (\gamma^2_i-1)& \,ln\Bigg[\frac{\gamma(y)\,\mathcal{R}[t,y]+\Big(1-\frac{t}{y}\Big)}{1+\gamma(y)} \Bigg] \Bigg\}\,. \label{capdelta} \eea Finally the function $A(\tau,\eta')$ given by eqn. (\ref{aoftau}), becomes
\be A(\tau,\eta') \equiv \mathcal{A}[t,y] = \mathcal{A}_0[t,y] + \Delta[t,y]\,,  \label{capA}\ee with
\be \mathcal{A}_0[t,y] = \frac{\omega_i\,\eta_i}{\gamma_i} \,t\,\Bigg[ \sqrt{(\gamma^2_i-1)+y^2}-\sqrt{(\gamma^2_i-1)}\Bigg] \label{capAzero}\,,\ee
and the integral (\ref{capS}) becomes
\be   \mathcal{S}(\eta') = \frac{2}{\pi}\, \int^{\xi'}_0  \mathcal{P}[t,y] \, \frac{\sin[\mathcal{A}(t,y)]}{t}\,dt ~~;~~ y= 1+\xi'~;~  \xi' =   (\frac{\eta'}{\eta_i}-1) \,. \label{capSnewvars} \ee

\begin{figure}[ht!]
\begin{center}
\includegraphics[height=4in,width=5in,keepaspectratio=true]{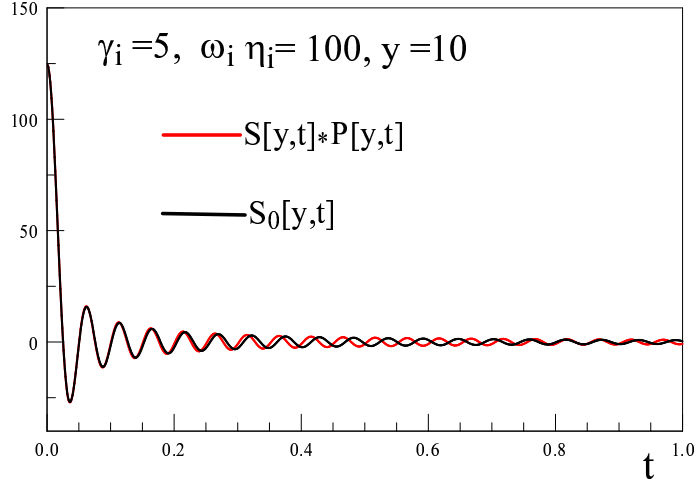}
\caption{Comparison of $S_0[y,t] = \frac{\sin[\mathcal{A}_0[y,t]]}{t}$ and $S[y,t]\times \mathcal{P}[y,t]$ with $S[y,t] = \frac{\sin[\mathcal{A}[y,t]]}{t}$ for $\gamma_i =5, \omega_i \,\eta_i = 100, y = 10$, confirming the validity of the adiabatic approximation.}
\label{fig:Scompa}
\end{center}
\end{figure}

We have argued above in this appendix that for $\omega^\phi_k(\eta')\,\eta' \gg 1$  the term $\delta \equiv \Delta$ is higher order in the adiabatic approximation and can be neglected, and that to leading order in this approximation we can set $P \equiv \mathcal{P} \rightarrow 1$. We now test this assertion numerically in terms of the new variables. Since in the new variables the product $\omega^\phi_k(\eta')\,\eta' = \frac{\omega_i\,\eta_i}{\gamma_i}\,y\,f(y) $ it follows that $\omega^\phi_k(\eta')\,\eta' \gg 1$ at all times  implies that $\omega_i\,\eta_i \gg 1$ which is precisely the statement of the validity of the adiabatic approximation at the initial time. Fig. (\ref{fig:Scompa}) compares $  \frac{\sin[\mathcal{A}_0[y,t]]}{t}$ and $  \mathcal{P}[y,t]\,\frac{\sin[\mathcal{A}[y,t]]}{t}$   for $\gamma_i =5, \omega_i \,\eta_i = 100, y = 10$, confirming the validity of the adiabatic approximation. We have explored a wide range of parameters with similar results.

Therefore to leading order in the adiabatic approximation we can replace the argument of the integral in (\ref{capSnewvars}) by $\frac{\sin[\mathcal{A}_0[y,t]]}{t}$, yielding to lowest adiabatic order
\be \mathcal{S(\xi')} = \frac{2}{\pi} \,Si\big[\alpha(\xi')\big] ~~;~~ \alpha(\xi') = \frac{\omega_i\,\eta_i}{\gamma_i} \,\xi'\,\Bigg[ \sqrt{(\gamma^2_i-1)+(1+\xi')^2}-\sqrt{(\gamma^2_i-1)}\Bigg]\,,  \label{Szeroad}\ee where $Si[x]$ is the sine-integral function, with $Si[x] \simeq x$ as $x \rightarrow 0$, reaches a maximum at $x = \pi$  and $Si[x] \rightarrow \pi/2$ for $x \gtrsim \pi$.  The maximum of $\mathcal{S}(\xi)$ occurs when
\be \alpha(\xi) = \pi \,, \label{maxS}\ee beyond which $\mathcal{S}(\xi) \simeq 1$.

\begin{figure}[ht!]
\begin{center}
\includegraphics[height=4in,width=5in,keepaspectratio=true]{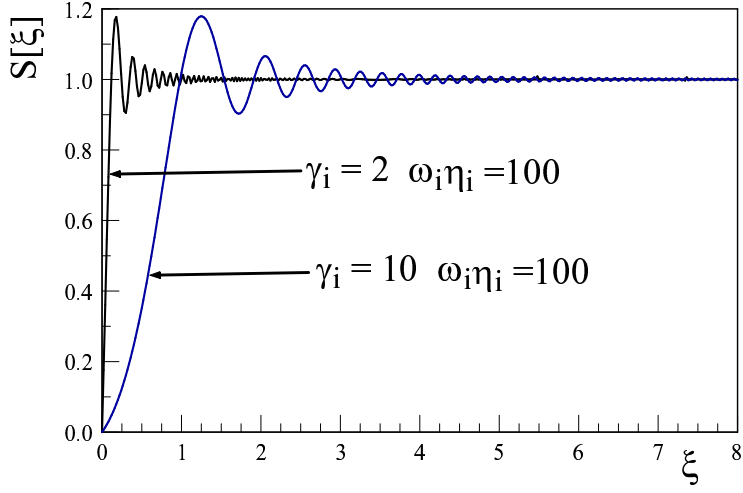}
\caption{ $S[y,t]$ for $\gamma_i = 2,10$ and $\omega_i \eta_i =100$.}
\label{fig:sofchi}
\end{center}
\end{figure}

 In particular for $\gamma_i =1$ (the particle decaying at rest) and $\omega_i \eta_i \gg 1$,  $\mathcal{S}(\xi')$ reaches a maximum at  $\xi' = \xi_s \simeq \pi/\omega_i\eta_i+\mathcal{O}(1/(\omega_i\eta_i)^2)$ with $\mathcal{S}(\xi_s) \simeq 1$. In the opposite limit, for an ultrarelativistic particle with $\gamma_i \gg 1$ and $\omega_i \eta_i \gg 1$ we find self-consistently that $S(\xi')$ reaches a maximum at $\xi_s$ with $\alpha(\xi_s) = \pi$,  where $\xi_s$ is a solution of
\be \xi_s\,(1+\xi_s)^2 = \frac{2\pi\,\gamma^2_i}{\omega_i\eta_i}\,.  \label{solchimax}\ee For $2\pi\,\gamma^2_i/\omega_i \eta_i \ll 1$ we find
\be \xi_s \simeq \Bigg[\frac{2\pi\,\gamma^2_i}{\omega_i\eta_i}\Bigg]- \Bigg[\frac{2\pi\,\gamma^2_i}{\omega_i\eta_i}\Bigg]^2 + \cdots  \,, \label{smallchim}\ee and for $2\pi\,\gamma^2_i/\omega_i\eta_i \gg 1$
  \be \xi_s \simeq \Bigg[\frac{2\pi\,\gamma^2_i}{\omega_i \eta_i } \Bigg]^{\frac{1}{3}}-\frac{2}{3} +\cdots \,.  \label{xiprilargeg} \ee In both cases we find that
 $ \frac{\xi_s}{\gamma_i} \ll 1 $ whenever $\gamma_i \gg 1$. Fig. (\ref{fig:sofchi}) displays the behavior of $\mathcal{S}(\xi)$ for $\omega_i = 100;\gamma_i = 2,10$.

Using the relations derived in appendix (\ref{app:identities}) along with the identities $C(\eta')= C(\eta_i)\,(\eta'/\eta_i) = C(\eta_i) \, (1+\xi')$ and $m_\phi C(\eta_i) = \omega_i/\gamma_i$,  it follows that  $I_3(k,\eta)$  given by (\ref{totI3}) can be written  in terms of the same variables as $I_2$, namely $\xi = \eta/\eta_i -1$ and $\eta_b = \eta_b/\eta_i-1$. We find
\be I_3(k,\xi)= \frac{\pi}{2}\, \frac{\omega_i \eta_i}{\gamma_i}\,\int^{\xi}_{\xi_b} \frac{(1+\xi')^2\,\Big[1+\mathcal{S}(\xi') \Big]}{\sqrt{(\gamma^2_i-1)+(1+\xi')^2}} \, \,d\xi'\,.\label{I3ofxi} \ee The contribution from the term with $\mathcal{S}$ in the integrand must be done numerically, however, the first term can be done analytically, yielding
\be I_{3A}(k,\xi) = \frac{\pi}{4}\,  {\omega_i \eta_i} \,\Bigg\{ (1+\xi)\,W[\xi] -1 - \frac{(\gamma^2_i-1)}{\gamma_i}\,\ln \Bigg[\frac{\gamma_i\,W[\xi]+(1+\xi)}{1+\gamma_i} \Bigg]  \Bigg\}\,, \label{I3A} \ee where we have set $\xi_b=0$ to leading adiabatic order.

The function $\mathcal{S}[\alpha(\xi)] $ has the following   behavior for $\xi \ll \xi_s$ and $\xi \gg \xi_s$,   corresponding to $\alpha(\xi) \ll \pi$ and $\alpha(\xi) \gg \pi $ respectively:

\be \mathcal{S}[\alpha(\xi)] \simeq \frac{2}{\pi} \Bigg[\alpha-\frac{\alpha^3}{18} +\frac{\alpha^5}{600} +\cdots  \Bigg] ~~;~~ \alpha \ll \pi ~~ (\xi \ll \xi_s) \label{smallalfa} \ee

\be \mathcal{S}[\alpha(\xi)] \simeq \frac{2}{\pi} \Bigg[1- \frac{\cos[\alpha]}{\alpha} - \frac{\sin[\alpha]}{\alpha^2}  +\cdots  \Bigg] ~~;~~ \alpha \gg \pi ~~ (\xi \gg \xi_s) \label{largealfa} \ee

\end{document}